\documentclass[structabstract]{aa}  
\usepackage{graphicx}
\usepackage{amsmath,amssymb}
\usepackage[applemac]{inputenc}
\usepackage[modulo, switch]{lineno}
\usepackage{xcolor}
\usepackage{txfonts}
\usepackage{natbib}
\bibpunct{(}{)}{;}{a}{}{,} % to follow the A&A style
\newcommand{\Kepler}{\textit{Kepler}}
\newcommand{\kepler}{\textit{Kepler} }

\newcommand{\Dnu}{\Delta \nu}

\newcommand{\numax}{\nu_\mathrm{max}}
\newcommand{\evid}{\mathcal{E}}
\newcommand{\model}{\mathcal{M}}
\newcommand{\muhz}{$\mu$Hz}
\newcommand{\teff}{T_\mathrm{eff}}
\newcommand{\DP}{\Delta\Pi_1}
\newcommand{\diamonds}{\textsc{D\large{iamonds}}}
\newcommand{\pa}{p_\mathrm{A}}
\newcommand{\pb}{p_\mathrm{B}}
\newcommand{\kic}{KIC~12008916}
\newcommand{\he}{He\textsc{\,\large ii}}
\def\be{\begin{equation}}
\def\ee{\end{equation}}    
\def\ba{\begin{eqnarray}}
\def\ea{\end{eqnarray}}

\begin{document}

\title{Bayesian peak bagging analysis of 19 low-mass low-luminosity red giants observed with \kepler}
\author{E. Corsaro\inst{1,2,3},
J. De Ridder\inst{1},
R. A. Garc\'{i}a\inst{2}
          }
\offprints{Enrico Corsaro\\ \email{enrico.corsaro@cea.fr}}

\institute{Instituut voor Sterrenkunde, KU Leuven, Celestijnenlaan 200D, B-3001 Leuven, Belgium
\and Laboratoire AIM, CEA/DSM -- CNRS -- Univ. Paris Diderot -- IRFU/SAp, Centre de Saclay, 91191 Gif-sur-Yvette Cedex, France
\and Instituto de Astrof\'{i}sica de Canarias, E-38205 -- Universidad de La Laguna\\Departamento de Astrof\'{i}sica, E-38206, La Laguna, Tenerife, Spain
}

   \date{Received ; accepted }

\abstract
{Non-radial oscillations, observed in thousands of red giants by the space missions CoRoT and \Kepler, allow to greatly deepen our understanding about stellar structure and evolution in cool low-mass stars. The currently available \kepler light curves contain an outstanding amount of information but a detailed analysis of the individual oscillation modes in the observed power spectra, also known as peak bagging, is computationally demanding and challenging to perform on a large number of targets.} 
{Our intent is to perform for the first time a peak bagging analysis on a sample of 19 low-mass low-luminosity red giants observed by \kepler for more than four years. This allows us to provide high-quality asteroseismic measurements that can be exploited for an intensive testing of the physics used in stellar structure models, stellar evolution and pulsation codes, as well as for refining existing asteroseismic scaling relations in the red giant branch regime.}
{For this purpose, powerful and sophisticated analysis tools are needed. We exploit the Bayesian code \diamonds, using an efficient nested sampling Monte Carlo algorithm, to perform both a fast fitting of the individual oscillation modes and a peak detection test based on the Bayesian evidence.}
{We find good agreement for the parameters estimated in the background fitting phase with those given in the literature. We extract and characterize a total of 1618 oscillation modes, providing the largest set of detailed asteroseismic mode measurements ever published. We report on the evidence of a change in regime observed in the relation between linewidths and effective temperatures of the stars occurring at the bottom of the RGB. We show the presence of a linewidth depression or plateau around $\numax$ for all the red giants of the sample. Lastly, we show a good agreement between our measurements of maximum mode amplitudes and existing maximum amplitudes from global analyses provided in the literature, useful as empirical tools to improve and simplify the future peak bagging analysis on a larger sample of evolved stars.}
{}

% 5 {} token are mandatory
%
%\abstract
%{ }
\titlerunning{Bayesian peak bagging analysis of 19 low-mass low-luminosity red giants}
      \authorrunning{E. Corsaro et al.}
\maketitle
%
%==========================================================================
\section{Introduction}
\label{sec:intro}
Since the first detection of non-radial oscillations in cool giant stars, the asteroseismology of field red giants (RGs) has encountered a substantial increase in output, especially thanks to the advent of the space-based photometric missions \textit{CoRoT} \citep[e.g.][]{DeRidder09,Kallinger10CoRoT,Mosser11universal,Mosser11mixed} and \kepler \citep[e.g.][]{Borucki10,Koch10,Bedding10,Huber10,Kallinger10Kepler,Kallinger12}, where the latter allowed also for the study of RGs in open clusters \citep[e.g.][]{Stello11membership,Miglio12,Corsaro12}, and in eccentric binary systems \citep{Beck14eccentric}. 

The discovery of so-called mixed modes \citep{Beck11Science,Mosser11mixed}, i.e. modes with p-mode as well as g-mode characteristics, only observed in stars that have passed the main sequence phase, has led to a significant improvement of our understanding of the internal structure and evolution of RGs. The characteristic period spacing of mixed modes frequencies provides a direct way to disentangle H-shell and He-core burning RGs \citep{Bedding11Nature,Mosser12} and has already been used to classify about 13,000 targets observed in the \kepler field of view \citep{Stello13}. Additional studies of the effect of rotation as seen from the splitting of mixed modes show us that the RG cores rotate faster than their convective envelope \citep{Beck12Nature,Deheuvels12,Mosser12spin}, thus opening the possibility to probe their internal rotation rates with direct observations of the oscillations at their surface. RGs are also being used as distance and age indicators to study stellar populations and trace the formation and evolution of the Galaxy \citep[e.g. see][]{Miglio09,Miglio13,Miglio15}.

Nonetheless, several problems and challenges arise in the analysis and physical interpretation of the asteroseismic properties of the RGs. Among the most important topics currently under investigation, we mention: the forward modeling and inversion of the mode frequencies aimed at probing the stellar structure, constrain the evolution, examine the physical mechanisms responsible for the transport of angular momentum inside the star \citep[e.g.][and references therein]{Tayar13,Benomar14,Deheuvels14}; the excitation and damping mechanism of the mixed modes \citep[e.g.][]{Grosjean14}, and the asymptotic behavior of their oscillation frequencies \citep{Jiang14}; the underlying physics responsible for the damping rates of the $p$ modes in a temperature range cooler than that of G-F type main-sequence and subgiant stars \citep[e.g.][]{Chaplin09,Baudin11temp,Belkacem12,Corsaro12}; the analysis of regions of sharp-structure variation inside the convective zone aimed at refining existing stellar models and at retrieving reliable helium abundances \citep[e.g.][]{Miglio10,Broomhall14,Vrard14,Corsaro15letter}. Therefore, RGs are among the most interesting and useful types of stars to test thoroughly the physics implemented in both stellar structure models and in evolution and pulsation codes, but high-precision asteroseismic measurements of individual oscillation modes are highly needed to accomplish each one of the issues listed before.

With the present four-years long datasets available from \kepler for many RGs, we have the possibility to obtain detailed asteroseismic properties such as frequency, amplitude, and lifetime of each individual oscillation mode observed, with an unprecedented level of accuracy and precision. The required analysis, usually known as peak bagging \citep[e.g.][]{App03PB}, represents a key-step to exploit the full potential of the high-quality power spectra of the stars. However, due to the large number of modes populating the power spectrum of a red giant (on the order of $\sim$100), the peak bagging usually turns into a computationally demanding analysis \citep[e.g. see][]{Benomar09,Gruberbauer09,Kallinger10CoRoT,Handberg11,Corsaro14}. For this purpose, we exploit \diamonds\,\,\citep[][hereafter CD14]{Corsaro14}, a new code well suited for model comparison and inference of high-dimensional problems in a Bayesian perspective.

In this paper we show how to adequately extend to the case of RGs the method used by CD14 for main-sequence stars, hence we perform the peak bagging analysis on a sample of low-mass low-luminosity red giant branch (RGB) stars observed by \kepler for more than four years. We thus provide and discuss our results for the background signal, the asteroseismic parameters, the mode damping and amplitude of all the stars in our sample, stressing their relevance in the light of the open questions related to the physics of the red giant stars.

\section{Observations and data} 
\label{sec:data}
To select the sample of stars investigated in this work we started from the original sample of RGs observed in long cadence (LC) by \kepler \citep{Jenkins10} and studied by \cite{Huber11} and by \cite{Corsaro13}, consisting in a total of 1111 stars. In this work we focus on low-mass low-luminosity RGB stars (hereafter LRGs, for the sake of brevity), a particular type of RGs that represents a small population of stars (about 5\,\% of the entire population of RGs observed by \Kepler). 

Our choice of analyzing LRGs, amongst the variety of RGs observed, is motivated by the following reasons:
\begin{itemize}
\item The clean oscillation pattern caused by the high values of the characteristic frequency separations, providing a more reliable fitting and identification of most of the observed modes.
\item The large number of modes (both $p$ modes and mixed modes) and radial orders observed thanks to the high frequency of maximum oscillation power $\numax$ (hence broader power excess compared to more evolved RGs), which allow for more stringent frequency inversions and studies of the seismic signatures of Helium inside the star.
\item The clear presence of rotation in the oscillations with different mixture level between pressure and gravity modes, essential to investigate the physics of angular momentum transport and its evolution in different layers inside the star.
\item The location in a short-lived phase at the bottom of the RGB, making these stars rare to observe and important to refine stellar evolution theory.
\item The covered range in stellar effective temperatures, laying just below that of the most evolved low-mass subgiant stars, filling the existing observational gaps that are crucial for our comprehension of the damping mechanism in the transition between the subgiant and the RGB phase of the stellar evolution.
\end{itemize} 

Therefore, from the original sample we only considered those stars with power spectra having an oscillation envelope with $\numax \gtrsim 110$\,\muhz, according to the stellar population study done by \cite{Miglio09} (see also \citealt{Bedding10}), with $\numax$ values provided by \cite{Huber11}. This selection reduced the number of targets to 56, which we further skimmed by visually choosing the best stars in terms of number of oscillations and SNR ($\gtrsim 1.3$ as measured from the PSD around $\numax$). We then checked for available measurements of their g-mode period spacing $\Delta\Pi_1$, as measured by \cite{Mosser12}, to have confirmation that the RGs considered are indeed settled in the RGB phase of stellar evolution, and as an auxiliary input for the mode identification process (see Sect.~\ref{sec:mode_id}). This cross-match led to 21 stars, two of which were removed due to a bad module sequence in three \kepler observing quarters (Q). The final sample obtained thus consists of 19 LRGs observed by \kepler from Q0 till Q17.1, namely a total of $\sim$1470 days, resulting in a frequency resolution of $\delta \nu_\mathrm{bin}\simeq 0.008\,$\muhz\,\,and $\sim$36,000 data bins for each power spectral density (PSD). The original PDC-SAP \kepler light curves \citep{Thompson13} were corrected following the methods described in \cite{Garcia11} and the final data were high-pass filtered ensuring a 100\,\% transmission for frequencies above 0.2\,$\mu$Hz for all stars. To minimize the effect of the \kepler regular window function due to the spacecraft angular momentum dump and the Earth downlink pointing \citep{Garcia14}, we have interpolated all the gaps of sizes smaller than 2 days using an {\it inpainting} algorithm \citep{Mathur10,Pires14}.

\begin{figure}
   \centering
   \includegraphics[width=9.1cm]{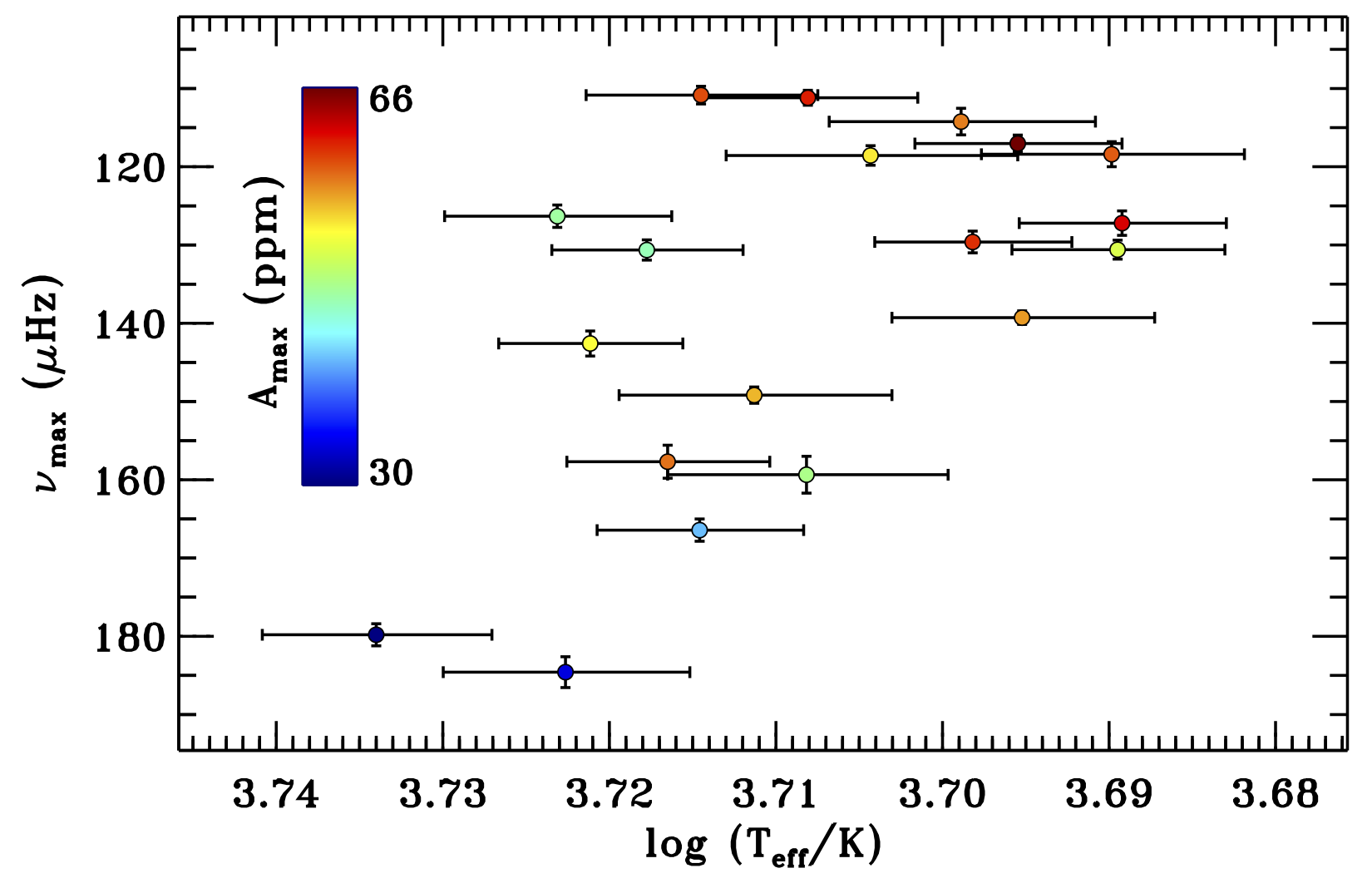}
      \caption{Asteroseismic HRD $\numax$-$\teff$ for the sample of 19 LRGs investigated. Amplitudes of maximum oscillation power, $A_\mathrm{max}$, as measured by \cite{Huber11} are indicated in a colored scale. All $\numax$ values are provided by \cite{Huber11}, while temperatures are taken from \cite{Pin12}. 1-$\sigma$ error bars are indicated for both coordinates.}
    \label{fig:hrd}
\end{figure}

Fig.~\ref{fig:hrd} illustrates the asteroseismic Hertzsprung-Russel diagram (HRD), $\numax$-$\teff$, for the sample of the 19 LRGs. The stars span a temperature range of about $4800$-$5500$\,K, and a range in $\numax$ of $\sim$75\,\muhz\,\,starting from $\sim110$\,\muhz\,\,upwards, with amplitudes at maximum power from 30 to $\sim$70\,ppm.

\section{Peak bagging analysis}
\label{sec:pb}

\subsection{Background fitting}
\label{sec:bkg}
The first step for the peak bagging analysis of the stars is to estimate the background signal in their PSD. We follow the statistical analysis done on a large sample of stars by \cite{Kallinger14}, who found that two separate granulation components with a characteristic frequency close to the region containing the oscillations, together with a third component related to long-term variations, represent best the background signal of the RGs analyzed in this work. In particular, the super-Lorentzian (or Harvey-like) profiles associated to each background component have an exponent set to four, again according to the findings by \cite{Kallinger14}. Thus, the background model that we adopt can be expressed as
\begin{equation}
P_\mathrm{bkg} \left(\nu \right) = W + R \left( \nu \right) \left[ B\left( \nu \right) + G \left( \nu \right) \right] \, ,
\label{eq:overall_bkg}
\end{equation}
where we assume $W$ to be a flat noise level since we only study RGs with high $\numax$, and with $R\left(\nu\right)$ the response function that considers the sampling rate of the observations for LC \kepler data,
\begin{equation}
R\left( \nu \right) = \mbox{sinc}^2 \left( \frac{\pi \nu}{2 \nu_\mathrm{Nyq}} \right) \, ,
\label{eq:resp}
\end{equation}
with $\nu_\mathrm{Nyq} = 283.212\,\mu$Hz the Nyquist frequency. The super-Lorentzian components are given by
\begin{equation}
B\left(\nu\right) = \sum^3_{i=1} \frac{\zeta a^2_i / b_i}{1 + \left( 2 \pi \nu / b_i \right)^4} \, ,
\label{eq:bkg}
\end{equation}
with $a_i$ the amplitude in ppm, $b_i$ the characteristic frequency in \muhz, and $\zeta = 2\sqrt{2}/\pi$ the normalization constant for a super-Lorentzian profile with exponent set to four \citep[e.g. see][]{Karoff13,Kallinger14}.
The power excess containing the oscillations is as usual described as
\begin{equation}
G\left(\nu\right) = H_\mathrm{osc} \exp \left[ - \frac{ \left( \nu - \nu_\mathrm{max} \right)^2}{2 \sigma_\mathrm{env}^2} \right] \, ,
\label{eq:env}
\end{equation}
and is only considered when fitting the background model to the overall PSD of the star.

\begin{figure}
   \centering
   \includegraphics[width=9.1cm]{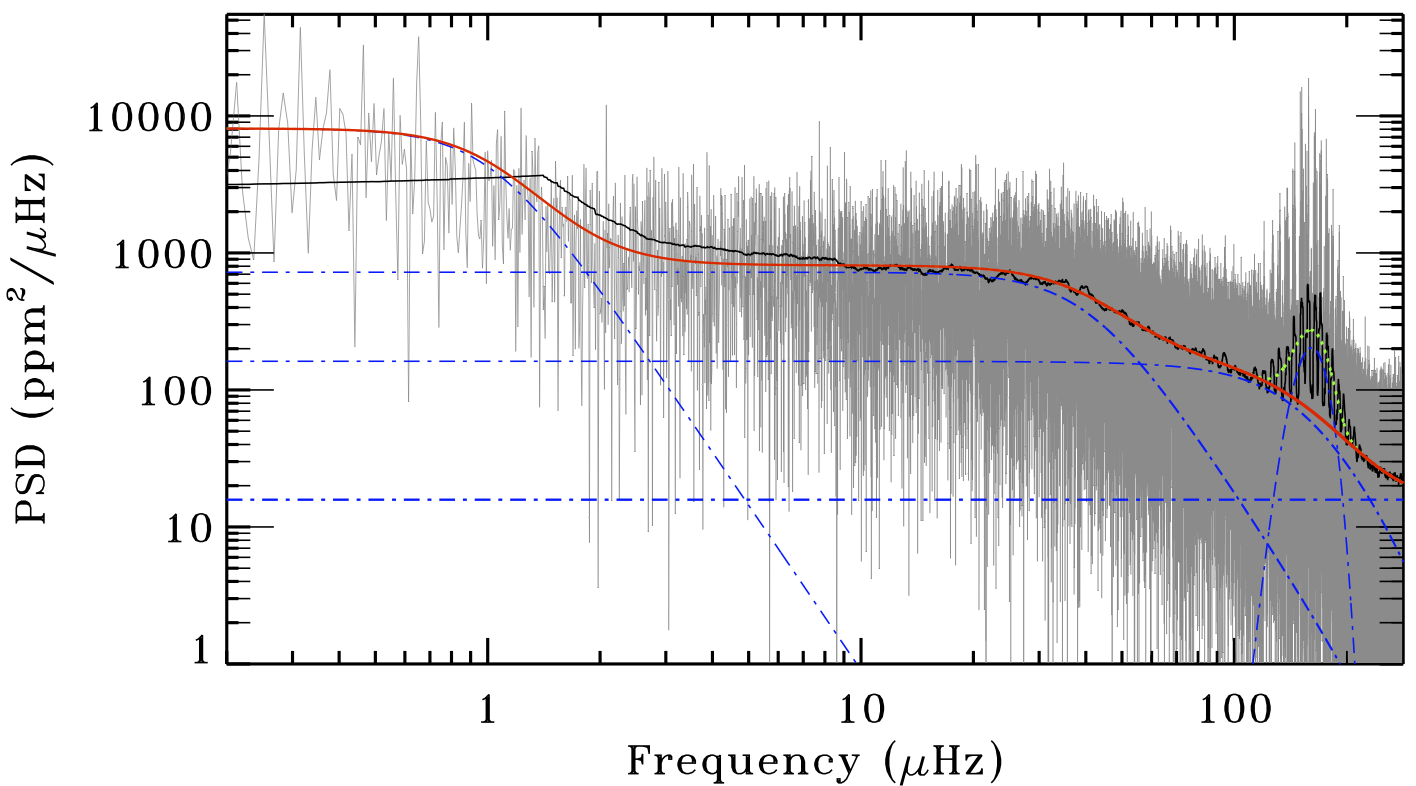}
      \caption{Background fit of the star \kic, as derived by \diamonds. The original PSD is shown in gray, whilst a smoothed version with boxcar width set to $\Delta\nu/5$, with $\Delta\nu$ taken from \cite{Mosser12Cat} (see also Table~\ref{tab:asymp_parameters}), is shown as a black line to guide the eye. The red thick line represents the background model without the Gaussian envelope and computed with the values listed in Table~\ref{tab:bkg1}. The green dotted line accounts for the additional Gaussian component using the values listed in Table~\ref{tab:bkg2}. The individual components of the background model as given by Eq.~(\ref{eq:overall_bkg}) are shown by blue dot-dashed lines.}
    \label{fig:bkg_case}
\end{figure}

The results from the fit done by means of \diamonds\,\,for all the stars of our sample are presented in Appendix~\ref{sec:bkg_results}, in which both the parameter values (Tables~\ref{tab:bkg1} and \ref{tab:bkg2}) and the plots with the background fit are given (Fig.~\ref{fig:bkg_tot}). To illustrate the results obtained with the method presented in this paper, we shall from now on refer to the star \kic, randomly chosen from our sample. The resulting background fit for \kic\,\,is shown in Fig.~\ref{fig:bkg_case}. 

We adopted uniform priors for all the free parameters of the background model by starting from the results obtained by \cite{Kallinger14}. Uniform priors ensure us the fastest computation and are easy to set up (see CD14). The configuring parameters of \diamonds\,\,used for the computations of the background fits are given in Appendix~\ref{sec:bkg_results}. Our estimated parameters describing the granulation signal of the stars in this sample agree on average within 4.6\,\% for the amplitudes ($a_\mathrm{gran,1}$, $a_\mathrm{gran,2}$) and within 5.1\,\% for the characteristic frequencies ($b_\mathrm{gran,1}$, $b_\mathrm{gran,2}$) with those measured by \cite{Kallinger14}, who used \kepler light curves about 360 days shorter than ours (up to Q13).

\subsection{Extraction of the oscillation mode parameters}
\label{sec:extraction}
The subsequent step in analyzing the stellar PSDs is to adopt an adequate fitting model for the oscillation pattern contained in the region of the power excess. In this work we adopt a method similar to that used by CD14 (see Sect.~6.3), who applied it to the case of an F-type main sequence star. We restrict our frequency range of analysis to the region containing the oscillations, which we identify as $\numax \pm 3.5 \sigma_\mathrm{env}$, with $\numax$ and $\sigma_\mathrm{env}$ listed in Table~\ref{tab:bkg2}, empirically chosen to ensure all observable modes to be included. We thus take into account some important features characterizing the oscillation pattern in RGs, as explained in the following. 

In the oscillation pattern of all the RGs, one can observe forests of modes between consecutive quadrupole and radial modes (also referred to as $\ell = 2$ and $\ell = 0$, respectively, $\ell$ being the angular degree), which are known as dipole ($\ell = 1$) mixed modes \citep{Beck11Science,Bedding11Nature,Mosser11mixed}. Since $g$ modes have the highest mode inertia because they propagate in high-density regions, their lifetime $\tau$ is significantly longer than that of pure pressure modes \citep[e.g. see][]{CD04}. As a consequence, even with an observing time $T_\mathrm{obs} > 4$ years now made available from \Kepler, mixed modes with a g-dominated character may still appear unresolved, whilst others having a more $p$ mode-like character can be partially or sometimes even fully resolved. 

\begin{figure}
   \centering
   \includegraphics[width=9.1cm]{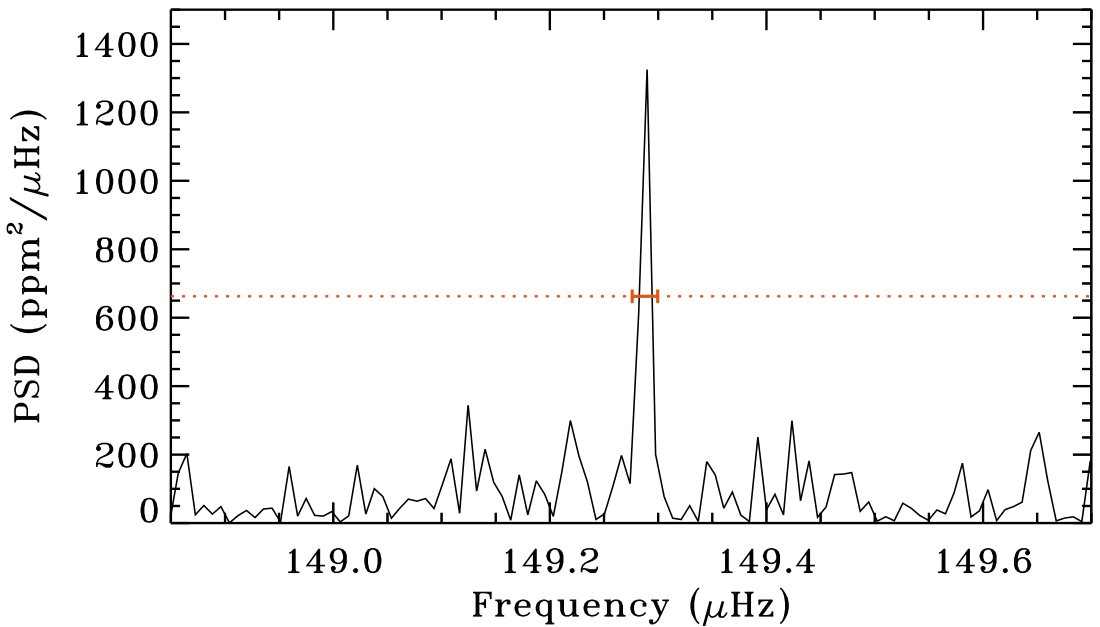}
      \caption{Example of an unresolved dipole mixed mode for \kic. The FWHM of the oscillation peak spans less than three data bins, as indicated by the red segment having a 3-bins width. The red dotted line represents the half maximum level of the peak.}
    \label{fig:unresolved_peak}
\end{figure}

In this work we distinguish between resolved or partially resolved peaks, for which $T_\mathrm{obs} \gtrsim \tau$, and those that are unresolved, for which $T_\mathrm{obs} \ll \tau$ \citep[e.g. see also][who used a similar approach]{Baudin11temp}. In the former case, the oscillation peak profile that we adopt is that of a Lorentzian \citep[e.g. see][]{Kumar88,Anderson90} and it is given by
\begin{equation}
\mathcal{P}_{\mathrm{res},0} \left( \nu \right) = \frac{A_0^2 / \left( \pi \Gamma_0 \right)}{1 + 4 \left( \frac{\nu - \nu_{0}}{\Gamma_0} \right)^2} \, ,
\label{eq:resolved_profile}
\end{equation}
where $A_0$, $\Gamma_0$, $\nu_0$ are the amplitude in ppm, the linewidth in \muhz\,\,and the frequency in \muhz, respectively, and represent the three free parameters to be estimated during the fitting process \citep[see][for more details about the relation between amplitude and linewidth of the peak]{Baudin05}.
According to the Fourier analysis, an oscillation peak that is not resolved has a profile represented as \citep{CD04}:
\begin{equation}
\mathcal{P}_{\mathrm{unres},0} \left(\nu \right) = H_0 \, \mbox{sinc}^2 \left[ \frac{\pi \left(\nu - \nu_0 \right)}{\delta \nu_\mathrm{bin}} \right] \, ,
\label{eq:unresolved_profile}
\end{equation}
where $H_0$ and $\nu_0$ are the height in PSD units and the central frequency in \muhz\,\,of the oscillation peak, respectively, and must be estimated during the fitting process, while $\delta \nu_\mathrm{bin}$ is fixed as the frequency resolution introduced in Sect.~\ref{sec:data}. Fitting the height of an unresolved mode is preferred since it is an observable and the linewidth of the peak is not a fitting parameter in this case. However, one can easily derive the corresponding amplitude for the unresolved peak as $A_0 = \sqrt{H_0\delta\nu_\mathrm{bin}/2}$. An example of this type of oscillation peak is shown in Fig.~\ref{fig:unresolved_peak} for the star \kic, where the peak width spans only a few data bins in frequency. In this case, a Lorentzian profile would represent an inadequate model because fitting the linewidth is neither meaningful nor needed for an unresolved mode. Moreover, using a Lorentzian profile to fit unresolved modes destabilizes considerably the fitting process because the linewidth tends to assume values smaller than the frequency resolution itself. Whereas this problem could be partially solved by setting a lower limit to the linewidth free parameter by means of an adequate prior, using a Lorentzian profile for the unresolved modes slows down the computation in any case since each Lorentzian profile requires one more free parameter to be fit with respect to the case of a $\mbox{sinc}^2$ profile. 

According to the description given in this section and still following CD14, we now fix the background parameters corresponding to the white noise, $W = \overline{W}$, and the super-Lorentzian profiles, $B \left( \nu \right) = \overline{B} \left( \nu \right)$, to the values listed in Table~\ref{tab:bkg1}. Then, the final peak bagging model can be represented as
\begin{equation}
P \left( \nu \right) = \overline{W} + R \left(\nu \right) \left[ \overline{B} \left( \nu \right) + P_\mathrm{osc} \left( \nu \right)\right] \, , 
\label{eq:general_pb_model}
\end{equation}
where the model representing the oscillation peaks is now given by a mixture of Lorentzian and sinc$^2$ functions
\begin{equation}
P_\mathrm{osc} \left( \nu \right) = \sum^{N_\mathrm{res}}_{i=1} \mathcal{P}_{\mathrm{res},i} \left( \nu \right) + \sum^{N_\mathrm{unres}}_{j=1} \mathcal{P}_{\mathrm{unres},j} \left( \nu \right) \, ,
\label{eq:pb_model}
\end{equation}
with $N_\mathrm{res}$ and $N_\mathrm{unres}$ the number of resolved and unresolved peaks to be fit, respectively. In an empirical Bayesian approach as that used by CD14, the prior distributions on each free parameter can easily be set by visual inspection of the PSD of the stars. Likewise done for the case of the background fitting, we adopt uniform prior distributions for all the free parameters of the peak bagging model (see CD14). An example of the resulting fit by using the peak bagging model given by Eq.~(\ref{eq:general_pb_model}) is depicted in Fig.~\ref{fig:chunk} for a chunk of PSD of \kic, where all the oscillation modes for which a Lorentzian profile was used are marked by a shaded colored band (see the figure caption for more description).

\subsection{Peak significance test}
\label{sec:test}
Testing the significance of an oscillation peak in the PSD of a star is a crucial aspect that has to be considered for providing a reliable set of modes that can be used for the modeling of the oscillations, hence to investigate the stellar structure and the evolution of the star. For this purpose, following CD14 (Sect.~6.5) we consider a detection probability for a single peak, $\pb$, defined as
\begin{equation}
\pb \equiv \frac{\evid_\mathrm{B}}{\evid_\mathrm{A} + \evid_\mathrm{B}} = 1 - \pa
\label{eq:detection_probability}
\end{equation}
where $\evid_\mathrm{A}$ and $\evid_\mathrm{B}$ are the Bayesian evidences for the models excluding and including, respectively, the oscillation peak to be tested. The probability $\pa$ is the complementary probability stating the non-detection of the oscillation peak. For clarity, we shall refer only to $\pb$ for the remainder of the paper as it represents a direct measure of the reliability of the oscillation peak.

In the Bayesian mindset, one has to provide all the peaks tested with their corresponding $\pb$ value, no matter if their detection probability is low (e.g. below 50\,\%, see also CD14, Sect.~6.5). However, we have performed a test using simulations to calibrate the method and gain a more reliable understanding about which threshold in $\pb$ is the most suitable to follow for properly selecting the oscillation modes. In particular, we have simulated 1000 chunks of PSD of a possible RG in the range $20$-$40$\,\muhz\,\,by means of the simulator provided by \cite{DeRidder06}, with either flat noise only or flat noise and a single oscillation peak. This was done by including a randomly generated flat noise component in each of the 1000 simulations and by adding an oscillation peak corresponding to a resolved $p$ mode (having the typical Lorentzian shape) in 100 randomly chosen simulations out of the initial 1000. This gives us a probability of 10\,\% to select one simulation containing a peak. 

\begin{figure}
   \centering
   \includegraphics[width=9.1cm]{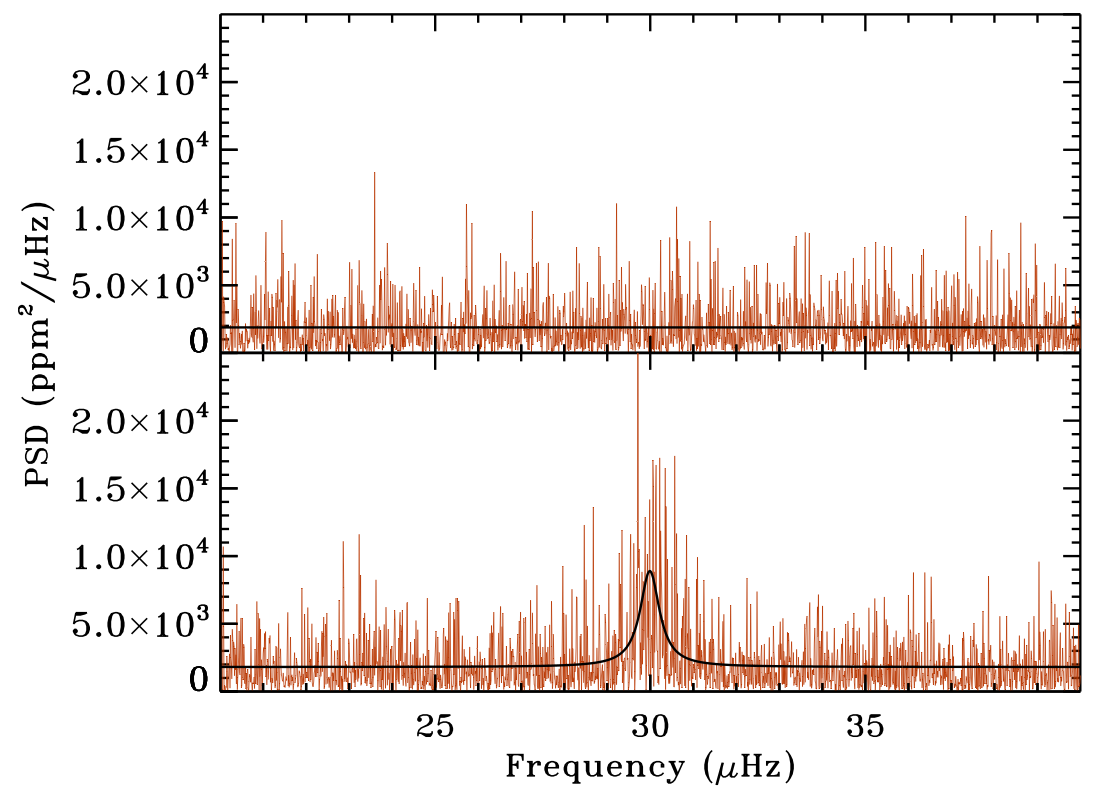}
      \caption{Example of simulations of a chunk of PSD for the peak significance test in a red giant. The top panel shows the case of a simulation containing only a flat noise component, whilst the bottom panel represents a simulation including an oscillation peak with its characteristic Lorentzian shape. The black thick line represents the resulting fit as computed by \diamonds\,\,(using $\model_\mathrm{A}$ in the top panel and $\model_\mathrm{B}$ in the bottom panel), with median values for each parameter. }
    \label{fig:sim}
\end{figure}

In each simulation containing a peak, both the amplitude, the linewidth and the central frequency of the peak were varied across the different simulations. We then performed a blind exercise by fitting the entire set of 1000 simulations without knowing which of them was containing a peak and in what order. To do the test we used two models: (i) a model, $\model_\mathrm{A}$, including only a flat noise component, represented by a single free parameter; (ii) a model, $\model_\mathrm{B}$, including both the flat noise component and the Lorentzian profile, given by Eq~(\ref{eq:resolved_profile}), for a total of four free parameters. The two models considered are therefore fit to all the individual chunks, no matter if they contain or not the oscillation peak. An example of the two fitting models is shown in Fig.~\ref{fig:sim} for two different simulations, with and without the oscillation peak.

We computed 2000 Bayesian evidences, $\evid_\mathrm{A,i}$ and $\evid_\mathrm{B,i}$ with $i$ ranging from 1 to 1000, namely one for each of the two models in each simulation. By computing the detection probabilities $p_\mathrm{B,i}$ for each simulation according to Eq.~(\ref{eq:detection_probability}), all the simulations containing the oscillation peak could be identified when adopting the threshold $\pb \gtrsim 0.99$ (meaning that the model containing the oscillation peak is significantly more likely than the one containing only the flat noise component). This probability threshold corresponds to the limit of a strong evidence condition in the Jeffreys' scale of strength \citep{Jeffreys61}. This shows that, despite $\model_\mathrm{B}$ is disfavored because it contains 3 more free parameters than $\model_\mathrm{A}$ --- according to the Bayesian perspective of Occam's razor --- it provides a much better fit than that of a single noise component when the oscillation peak is present, thus resulting in a considerably larger Bayesian evidence (see also CD14, Sect. 2.1 for more discussion). We therefore now have a method to detect peaks based on the Bayesian model comparison that works well and it is straightforward to conduct.

We can easily extend this analysis to the case of an unresolved peak by changing the function that describes the peak profile in the model $\model_\mathrm{B}$ with that given by Eq.~(\ref{eq:unresolved_profile}). Thus, following Approach 1 described in CD14, for each star in our sample we fit chunks of PSD with a frequency range containing one radial order per time, and apply the peak significance test as described before to all the ambiguous detections occurring in the selected chunk. 

\begin{figure*}[t!]
   \centering
   \includegraphics[width=18.4cm]{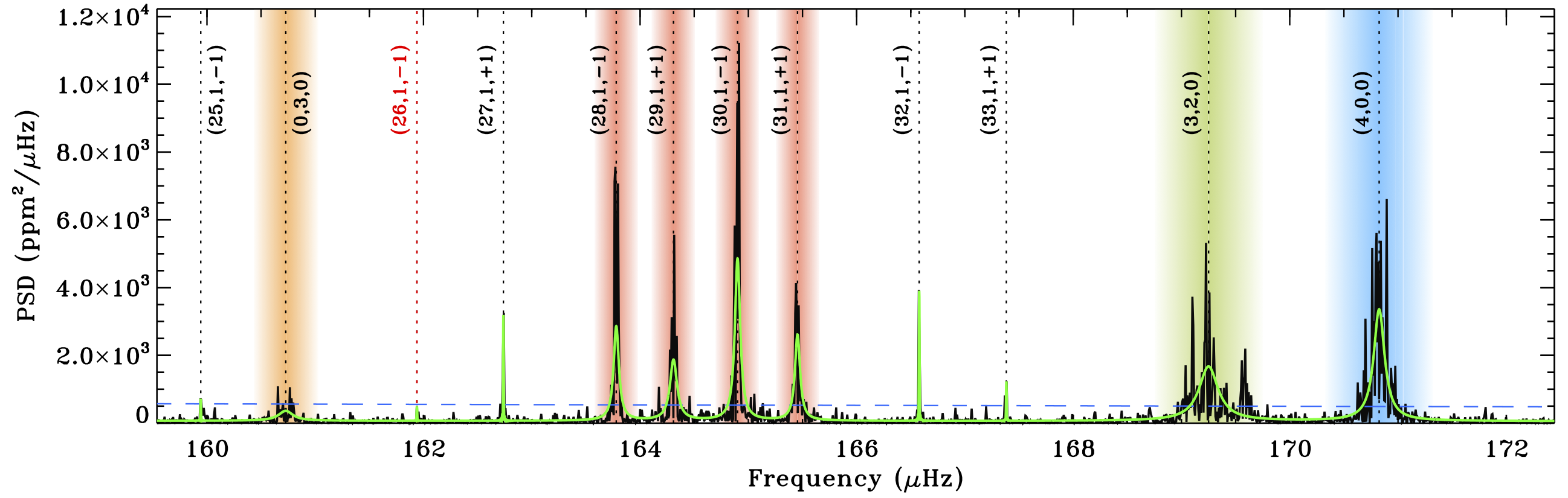}
      \caption{Example of a fit chunk for \kic\,\,derived by \diamonds. The PSD is shown in black, whilst the resulting fit using the parameters listed in Tables~\ref{tab:12008916m} and \ref{tab:12008916p} is in indicated by a green solid line. The level of eight times the background is shown with a blue dashed line as a reference for the peak significance. The colored vertical bands underline the oscillation modes modeled by a Lorentzian profile, Eq.~(\ref{eq:resolved_profile}), with $\ell = 0$ in blue, $\ell = 2$ in green, $\ell = 3$ in orange, and $\ell = 1$ mixed modes in red. The vertical dotted lines mark the central frequency $\nu_0$ for each of the fit modes, with the corresponding mode identification (Peak \#, $\ell$, $m$) also labeled, according to the description given in Appendix~\ref{sec:pb_results}. The case of the unresolved mixed mode $(26,1,-1)$ is shown in red because the peak has a detection probability ($p_\mathrm{B} = 0.002$) below the threshold suggested in Sect.~\ref{sec:test}.}
    \label{fig:chunk}
\end{figure*}

We note that while performing the peak significance test for the 19 LRGs, we found that a 8$\sigma$ level, with $\sigma$ representing the background level fit given in Sect.~\ref{sec:bkg} and used in the peak bagging model, Eq.~(\ref{eq:general_pb_model}), is a raw indication of the 99\,\% threshold in $p_\mathrm{B}$ for all the targets, and can thus be used as a guideline to select peaks to be tested with our method based on the Bayesian model comparison. Fig.~\ref{fig:chunk} illustrates an example of peak significance test performed for \kic, where $p_\mathrm{B} < 0.99$ (the peak marked by the vertical dotted red line), with a $8\sigma$ background level overlaid (dashed blue line) to show that the peak height of the tested peak is well comparable. 

\subsection{Mode identification}
\label{sec:mode_id}
The last step of the peak bagging analysis involves the mode identification. In the case of pure acoustic $\ell = 1,2,3$ modes this is accomplished by computing their position relative to that of the closest radial mode, which in turn can usually be easily identified in all the stars due to its high amplitude and linewidth. We therefore adopt the asymptotic relation for $p$ modes \citep{Vandakurov68,Tassoul80}, obtaining for each radial order $n$:
\begin{equation}
\nu_{n,\ell = 1} = \nu_{n,\ell = 0} + \Dnu/2 - \delta\nu_{01} \, ,
\end{equation} 
\begin{equation}
\nu_{n,\ell = 2} = \nu_{n,\ell = 0} + \Dnu - \delta\nu_{02} \, ,
\end{equation} 
\begin{equation}
\nu_{n,\ell = 3} = \nu_{n,\ell = 0} + \Dnu/2 - \delta\nu_{03} \, ,
\end{equation}
where $\Dnu$ is the mean large frequency separation. The mean small frequency spacings $\delta\nu_{01}$ and $\delta\nu_{02}$ are derived through the relations provided by \cite{Corsaro12}:
\begin{equation}
\delta\nu_{01} \simeq b_1 \,\Delta\nu - 0.063 \, ,
\end{equation}
\begin{equation}
\delta\nu_{02} \simeq b_2 \,\Delta\nu + 0.035 \, ,
\end{equation}
with mass-dependent slopes 
\begin{equation}
b_1 \simeq -0.073 + 0.044 \, \left( M/M_\odot \right) \, ,
\end{equation}
and
\begin{equation}
b_2 \simeq 0.138 - 0.014 \, \left( M/M_\odot \right) \, ,
\end{equation}
valid for RGB stars like in our sample.
The mean small frequency spacing $\delta\nu_{03}$ is given by the relation \citep{Huber10}:
\begin{equation}
\delta\nu_{03} \simeq 0.282 \, \Dnu + 0.16 \, .
\end{equation}
For the $\ell = 1$ mixed modes we instead adopt the asymptotic relation provided by \cite{Mosser12} (their Eq.~(9)). The mode identification for the rotationally split modes ($m = +1$,$-1$, $m$ being the azimuthal order) is instead derived from the position of the central components since the modes are in general well separated in LRGs.   
All the values for $\Dnu$, mass, the period spacing of $g$ modes $\DP$ and the coupling factor $q$, are taken from \cite{Mosser12Cat} and are listed in Table~\ref{tab:asymp_parameters} for a summary.

Whereas the asymptotic values for $p$ modes proved to be quite accurate for all the stars, in the case of mixed modes a correction for $\DP$ and $q$ from the values listed in \cite{Mosser12Cat} was required for some targets (indicated in Table~\ref{tab:asymp_parameters}), especially those showing variations in $\DP$ with the radial order (see Sect.~\ref{sec:results} for more details). The correction was made by an auxiliary fit to the mixed mode frequencies in a bi-dimensional grid $\left(\DP,q \right)$ spanning $\pm 10\,s$  and $\pm 0.05$ units, respectively, from the literature values. The auxiliary fit has the only purpose of providing a more reliable proxy for the frequency position of the dipole mixed modes, hence used for the mode identification process. 

The stars KIC~8475025, KIC~9267654, and KIC~11353313, in particular, were found to show mixed modes with a peculiar frequency position because the frequency splitting due to the rotation of the star is very similar to the frequency spacing between adjacent mixed modes. This generates a misleading mode identification at first glance. A cautious inspection of all the rotational split components for these stars was therefore needed in order to solve the ambiguity \citep[e.g. see][for more discussion on mode identification in these difficult cases]{Beck14eccentric}. For the remainder stars of the sample, the mode identification was straightforward even when rotational effects are present.

\begin{table}
\caption{Literature values for $\Dnu$, mass, $\DP$, and $q$ used for the mode identification of the stars.}             % title of Table
\centering                         
\begin{tabular}{l r c c c }       
\hline\hline
\\[-8pt]
KIC ID & \multicolumn{1}{c}{$\Dnu$} & Mass & $\DP$ & $q$\\ [1pt]
 & \multicolumn{1}{c}{($\mu$Hz$$)} & ($M_{\odot}$) & (s) & \\ [1pt]   
\hline
\\[-8pt]
03744043 &   9.90 &   1.3 &  75.98 &   0.16\\[1pt]
06117517 &  10.16 &   1.3 &  76.91 &   0.14\\[1pt]
06144777\tablefootmark{a} &  11.01 &   1.1 &  69.91 &   0.22\\[1pt]
07060732\tablefootmark{a} &  10.94 &   1.3 &  72.78 &   0.18\\[1pt]
07619745 &  13.13 &   1.5 &  79.17 &   0.15\\[1pt]
08366239\tablefootmark{a} &  13.70 &   1.9 &  86.77 &   0.17\\[1pt]
08475025\tablefootmark{a,b} &   9.66 &   1.4 &  74.80 &   0.13\\[1pt]
08718745 &  11.40 &   1.1 &  79.45 &   0.11\\[1pt]
09145955 &  11.00 &   1.4 &  77.01 &   0.16\\[1pt]
09267654\tablefootmark{a,b} &  10.34 &   1.3 &  78.41 &   0.13\\[1pt]
09475697 &   9.88 &   1.5 &  75.70 &   0.12\\[1pt]
09882316 &  13.68 &   1.6 &  80.59 &   0.15\\[1pt]
10123207 &  13.67 &   1.0 &  83.88 &   0.17\\[1pt]
10200377 &  12.47 &   1.1 &  81.58 &   0.16\\[1pt]
10257278 &  12.20 &   1.4 &  79.81 &   0.18\\[1pt]
11353313\tablefootmark{a,b} &  10.76 &   1.5 &  76.00 &   0.14\\[1pt]
11913545 &  10.18 &   1.3 &  77.84 &   0.12\\[1pt]
11968334 &  11.41 &   1.4 &  78.10 &   0.13\\[1pt]
12008916\tablefootmark{a} &  12.90 &   1.3 &  80.47 &   0.14\\[1pt]
\hline                                
\end{tabular}
\tablefoot{All the values are available from \cite{Mosser12Cat}.\\
\tablefoottext{a}{Stars for which an auxiliary fit for ($\DP$,\,$q$) was required.}
\tablefoottext{b}{Stars with a large rotational splitting, of the same magnitude of the frequency spacing between adjacent mixed modes. A careful mode identification of the split components was required for this case.}
}
\label{tab:asymp_parameters}
\end{table}

\section{Results}
\label{sec:results}
All the results for oscillation frequencies, amplitudes, linewidths for the resolved modes, and oscillation frequencies and heights for the unresolved ones are presented in Appendix~\ref{sec:pb_results}, with detection probabilities for the peaks that required a significance test as explained in Sect.~\ref{sec:test}, and with a mode identification as explained in Sect.~\ref{sec:mode_id}. A total of 1618 oscillation modes was fit for the peak bagging analysis presented in this work, with an average of $\sim85$ modes per star. We performed a peak significance test for a total of 612 peaks ($\sim$38\,\% of the entire number of oscillation modes), of which 380 ($\sim$62\,\% of tested peaks) gave a detection probability $p_\mathrm{B} \geq 0.99$. This results in a $\sim$14\,\% of the total number of peaks that did not fulfill the suggested detection threshold. The configuring parameters of \diamonds\,\,used during the computations are specified in the corresponding appendix. 

The 68\,\% Bayesian credible intervals are provided for each measurement as computed from the individual marginal probability distributions of the model parameters, according to the description given by CD14. We note that for some oscillation modes, the quoted error bars of the corresponding frequencies can be smaller than the formal frequency resolution of the PSD. Although not intuitive at first glance, this may happen because the reported error bars only refer to the capability of the adopted model to reproduce the observations, an aspect that is not uniquely depending on the resolution of the dataset, in our case represented by the formal frequency resolution obtained from the Fourier analysis \citep[e.g. see][for more discussion on the presence of error bars smaller than the resolution given by the data, as obtained from another astrophysical application of the Bayesian inference]{Froehlich12}. This effect is evident for those oscillation modes for which a sinc$^2$ profile is used, whose characteristic small FWHM (few data bins) and the well appropriate profile shape at reproducing spike-like peaks (see Fig.~\ref{fig:unresolved_peak}), leads to a more precise estimate of the frequency centroid than in the case of the Lorentzian profile fitted to resolved or partially resolved modes, in this case instead averaging for the stochastic nature of the oscillation peak over many data bins (as a result from our analysis the error bar on the frequency of the unresolved mode is on average 10 times smaller than that of the resolved modes). Therefore for clarity to the reader, we stress that in general the precision level of the error bars reported in this work is not limited by the formal frequency resolution of the PSD, because it represents the adequacy of the model presented in Sect.~\ref{sec:extraction} to fit the data.

As reported in Table~\ref{tab:asymp_parameters}, seven stars out of the 19 analyzed required a refitting of $\DP$ during the mode identification process, with deviations from the literature values found up to $\sim9$\,s for KIC~6144777, and $\sim4$\,s for KIC~7060732. The stars KIC~6144777 and \kic, in particular, show $\DP$ varying on the order of seconds with varying radial order. This caused the mode identification process to be more complex than for other stars of the sample, although still well feasible with an appropriate refitting of $\DP$ as explained in Sect.~\ref{sec:mode_id}. For completeness, we notice that in the newer catalogue provided by \cite{Mosser14}, the measurement of $\DP$ for the stars KIC~6144777 and KIC~8366239 was corrected, thus corresponding to the refitted value used in this work.

We firmly detect the signature of rotation through the presence of rotationally split modes in all the LRGs except KIC~9145955, KIC~9882316, KIC~10123207, and KIC~10200377. However, KIC~6117517 shows only the presence of two significant rotationally split modes in the highest SNR region of the PSD, which makes a detailed study of rotation for this star difficult. The stars KIC~7619745, KIC~8366239, KIC~8475025, KIC~9267654, KIC~10257278, KIC~11913545, and \kic, instead, show very few or even no oscillation modes corresponding to the zonal component $m=0$. The mode identification was possible in most of the cases, except for those low SNR oscillation peaks in the wings of the PSD for which no clear azimuthal order could be assigned, or for oscillation peaks with double mode identification. The latter case is observable in the stars KIC~8475025, KIC~9267654, and KIC~11353313, in which the confusion limit due to the large rotational splitting is present, as mentioned in Sect.~\ref{sec:mode_id}. The double mode identification for some of the peaks in the stars mentioned before can be explained by the overlap of the split component $m = +1$ with the adjacent $m = -1$ when the rotational splitting is the same as the frequency spacing between two adjacent mixed modes of different $g$ mode. 

The star KIC~7060732, despite the presence of the largest number of fit modes among the 19 LRGs (123 in total), shows the largest number of non-identified modes (40), all placed in the wings of the PSD, because of the confusion generated by the low SNR of the peaks coupled with the large rotational splitting observed. KIC~9882316 has instead the lowest number of fit modes (49) but with a clear and straightforward mode identification.

\begin{figure}
   \centering
   \includegraphics[width=9.0cm]{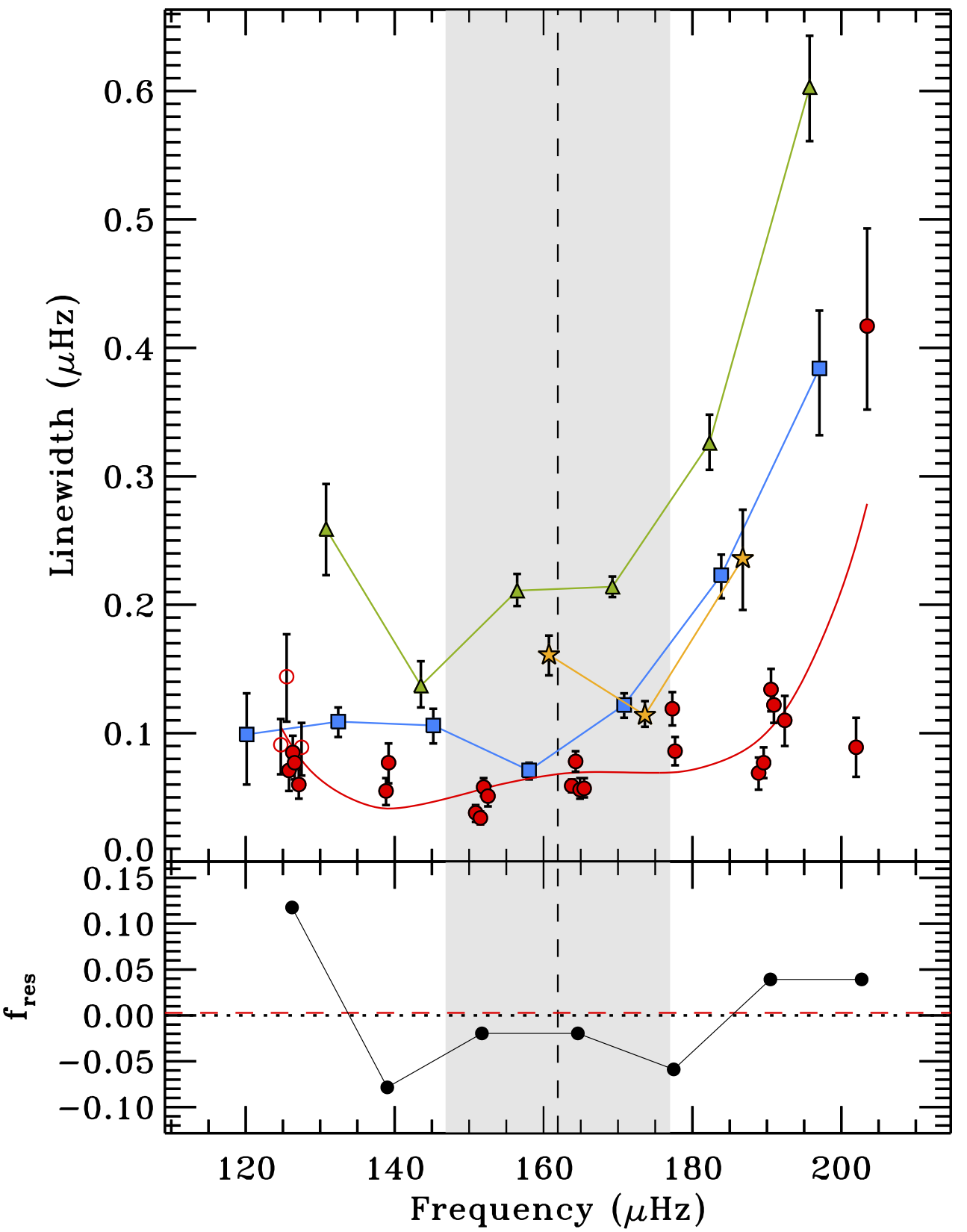}
      \caption{Mode linewidths for \kic\,\,as a function of the corresponding oscillation frequencies. \textit{Top panel}: linewidth measurements as defined by Eq.~(\ref{eq:resolved_profile}) for each angular degree ($\ell = 0$ blue squares, $\ell = 2$ green triangles, $\ell = 3$ yellow stars, and resolved $\ell = 1$ mixed modes red circles). Open symbols represent modes with detection probability under the suggested threshold (see Sect.~\ref{sec:test}). The 68\,\% credible intervals for the linewidths as derived by \diamonds\,\,are shown for each data point. The red solid line represents a polynomial fit to the linewidths of the $\ell = 1$ mixed modes, included to emphasize the trend with frequency. The shaded region represents the range $\numax \pm \sigma_\mathrm{env}$, with $\numax$ from Table~\ref{tab:bkg2} indicated by the dashed vertical line. \textit{Bottom panel}: the normalized fraction of resolved mixed modes with respect to unresolved ones, $f_\mathrm{res}$ (black dots), defined by Eq.~(\ref{eq:fraction_resolved}). The frequency position of each point is the average frequency of the resolved dipole mixed modes falling in each radial order (or that of the unresolved mixed modes if no resolved mixed modes are present). The horizontal dotted line represents the limit of resolved-dominated regime, as defined in Sect.~\ref{sec:fwhm}, whilst the horizontal dashed red line marks the average $f_\mathrm{res}$ given by Eq.~(\ref{eq:average_fraction}).}
    \label{fig:fwhm}
\end{figure}

\subsection{Mode linewidths}
\label{sec:fwhm}
The first result related to the asteroseismic measurements of individual oscillation modes is represented by the behavior of mode linewidths with increasing frequency in the PSD. The so-called linewidth depression has been observed in several CoRoT stars \citep{Benomar09gamma,Barban09,Deheuvels10,Ballot11} and more recently in a larger sample of main-sequence and subgiant stars observed with \kepler \citep{Benomar13,App14}. Theoretical studies of solar data by \cite{Belkacem11} have confirmed that $\numax$ is intimately related to the frequency position of the linewidth depression, albeit a full interpretation of this result requires more investigation.

The top panel of Fig.~\ref{fig:fwhm} shows all the linewidth measurements of \kic\,\,for the angular degrees $\ell = 0,2,3$, and the resolved (or partially resolved) $\ell = 1$ mixed modes, as a function of the corresponding oscillation frequencies in the PSD. As visible in the region marked by the vertical gray band, the linewidths either flatten or even decrease when the frequency of the oscillation peak is close to $\numax$. This is more clear for $\ell = 0,2$ modes, whilst for $\ell = 3$ the number of modes is not sufficiently large to let us conclude on a possible trend. For the case of $\ell = 1$ mixed modes we have included a low-degree polynomial fit to visualize the global trend, which also shows a depression in linewidth close to $\numax$. We note, however, that the measured linewidths of the $\ell = 2$ modes are not the physical linewidths and are only used to show the presence of the linewidth depression in comparison to the other modes with different angular degree. This is because by fitting a single Lorentzian profile to the region of the $\ell = 2$ modes we are not taking into account the possible effects arising from both the rotational splitting and from the presence of quadrupole mixed modes, thus resulting in measured linewidths that are broader than the real ones.

To investigate the behavior of the mode linewidths in a star more thoroughly we computed the normalized fraction of resolved $\ell =1$ mixed modes with respect to unresolved ones, which can be expressed for each radial order as
\begin{equation}
f_{\mathrm{res},n} = \frac{N_{\mathrm{res},n} - N_{\mathrm{unres},n}}{\sum_n \left( N_{\mathrm{res},n} + N_{\mathrm{unres},n} \right) }
\label{eq:fraction_resolved}
\end{equation}
where $N_{\mathrm{res},n}$ and $N_{\mathrm{unres},n}$ are the number of resolved and unresolved mixed modes, respectively, fit in each radial order $n$. The denominator of Eq.~(\ref{eq:fraction_resolved}) allows us to normalize by the total number of mixed modes fit in the star, hence to compare the value among different stars in the sample. The resulting normalized fraction, hereafter $f_\mathrm{res}$ for simplicity, as a function of the average frequency position of the dipole mixed modes falling in each radial order, is shown in the bottom panel of Fig.~\ref{fig:fwhm} for \kic. The horizontal dotted line separates the regime of resolved mixed modes fitting (upper part, $f_\mathrm{res} > 0$) from that of unresolved mixed modes fitting (lower part, $f_\mathrm{res} < 0$). We do not compute $f_\mathrm{res}$ for radial orders in which no mixed modes are fit. 

For radial orders where the mode linewidths encounter a depression, we therefore expect $f_\mathrm{res}$ to decrease as well, since the number of fit resolved mixed modes is likely to be lower than in the case of radial orders in which the modes have a larger linewidth. This is shown again in the bottom panel of Fig.~\ref{fig:fwhm}, where the decrease in $f_\mathrm{res}$ is sited in the region close to $\numax$. This behavior is evident for all the other stars of the sample as well (see the figures in Appendix~\ref{sec:pb_results}).
To use $f_\mathrm{res}$ as an indicator of the resolving level of the mixed modes fit in the star, we can compute an average value as
\begin{equation}
\langle f_{\mathrm{res}} \rangle = \frac{\sum_n f_{\mathrm{res},n}}{N_\mathrm{radial}}
\label{eq:average_fraction}
\end{equation}
where $N_\mathrm{radial}$ is simply the total number of radial orders where the mixed modes are fit. The value $\langle f_{\mathrm{res}} \rangle$ for \kic\,\,is indicated in the bottom panel of Fig.~\ref{fig:fwhm} by a horizontal dashed red line, and similarly for all the other stars in the corresponding figures provided in Appendix~\ref{sec:pb_results}. For \kic\,\,we find that $\langle f_{\mathrm{res}} \rangle = 0.003$, namely the highest value in the entire sample and the only one falling in the resolved-dominated regime: the number of resolved (or partially resolved) mixed modes per radial order in this star is on average larger than that of the unresolved ones. For KIC~3744043, instead, we find the minimum value of the sample, $\langle f_{\mathrm{res}} \rangle = -0.057$, laying in the unresolved-dominated regime. The highest values of $\langle f_{\mathrm{res}} \rangle$ correspond to stars in which no rotation is detected, or in most cases to stars for which the central component of dipole mixed modes, $m=0$, is missing.

\begin{figure}
   \centering
   \includegraphics[width=9.1cm]{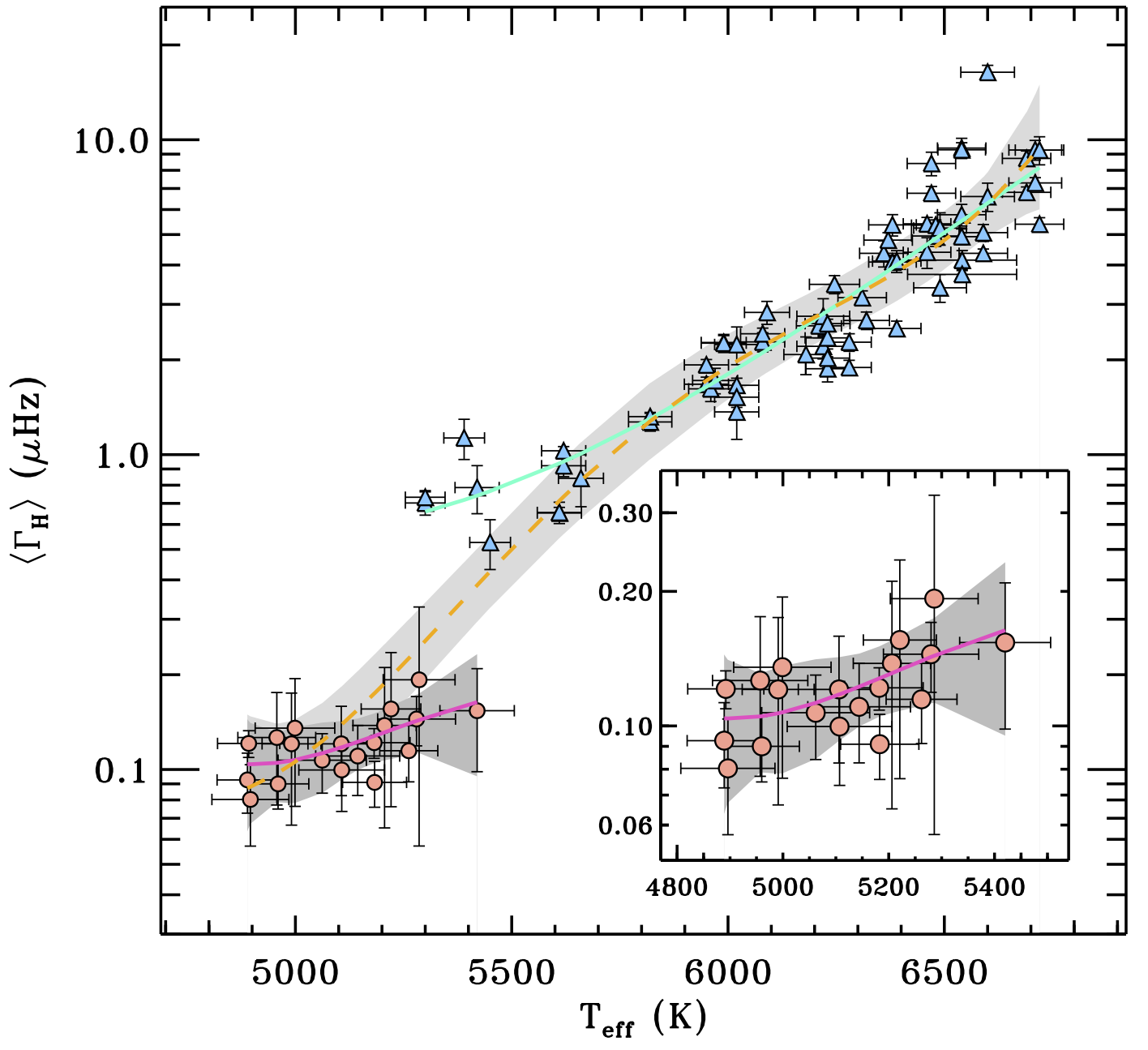}
      \caption{Average mode linewidth at maximum mode height, $\langle \Gamma_\mathrm{H} \rangle$, for the LRGs (red circles) and the sample of main-sequence and subgiant stars analyzed by \cite{App12fwhm} (blue triangles), as a function of the stellar effective temperature from \cite{Pin12}, in a linear-log scale. 1-$\sigma$ error bars are shown on both coordinates. The solid pink line shows a polynomial fit to the sample of RGs, while the solid teal line represents the law proposed by \cite{App12fwhm} for their sample of stars. The dashed yellow line is a polynomial fit to the total sample of stars. The 3-$\sigma$ regions coming from the polynomial fits are also shown for comparison. The inset displays a zoom-in of our sample of red giant stars.}
    \label{fig:fwhm_temp}
\end{figure}

An often investigated quantity related to the linewidth measurements, as already mentioned in Sect.~\ref{sec:intro} and \ref{sec:data}, is their correlation with the stellar effective temperature \citep[e.g. see][for more results and details]{Hekker10,Baudin11temp,App12fwhm,Belkacem12,Corsaro12}. A theoretical modeling of the damping rates of radial modes based on the existing observations from CoRoT and \kepler \citep{Belkacem12} has proposed a common physical description of the linewidth-temperature correlation from main-sequence to low-temperature giant stars (down to $\sim 4000$\,K), although it did not exploit linewidth measurements in the critical temperature range used in this paper.

Following \cite{Baudin11temp} and \cite{App12fwhm}, we have computed for each star in our sample an average linewidth at maximum mode height, $\langle \Gamma_\mathrm{H} \rangle$, by taking the average linewidth of radial modes in three radial orders around the radial mode with maximum height, which is more immune to systematic effects as discussed by e.g. \cite{App12fwhm}. The resulting $\langle \Gamma_\mathrm{H} \rangle$ values are shown in Fig.~\ref{fig:fwhm_temp} (red circles) as a function of $T_\mathrm{eff}$ from \cite{Pin12}, where the error bars on each $\langle \Gamma_H \rangle$ are the standard deviations from the three linewidths used in the calculation. This allows us to compare the result with the existing set of measurements for a sample of main-sequence and subgiant stars investigated by \cite{App12fwhm} (blue triangles). 
We observe a steeper temperature gradient of $\langle \Gamma_H \rangle$ for the hotter main-sequence stars with respect to the cooler RGs, in agreement with previous studies \citep{Baudin11temp,Belkacem12}. This change in the slope of the correlation between $T_\mathrm{eff}$ and linewidths appears more evident in the temperature range around $T_\mathrm{eff} \sim 5400$\,K, and is emphasized by the polynomial fits overlaid to our measurements and by the confidence regions marked with gray-shaded bands, in comparison to the power law fit derived by \cite{App12fwhm}.

\begin{figure}
   \centering
   \includegraphics[width=9.1cm]{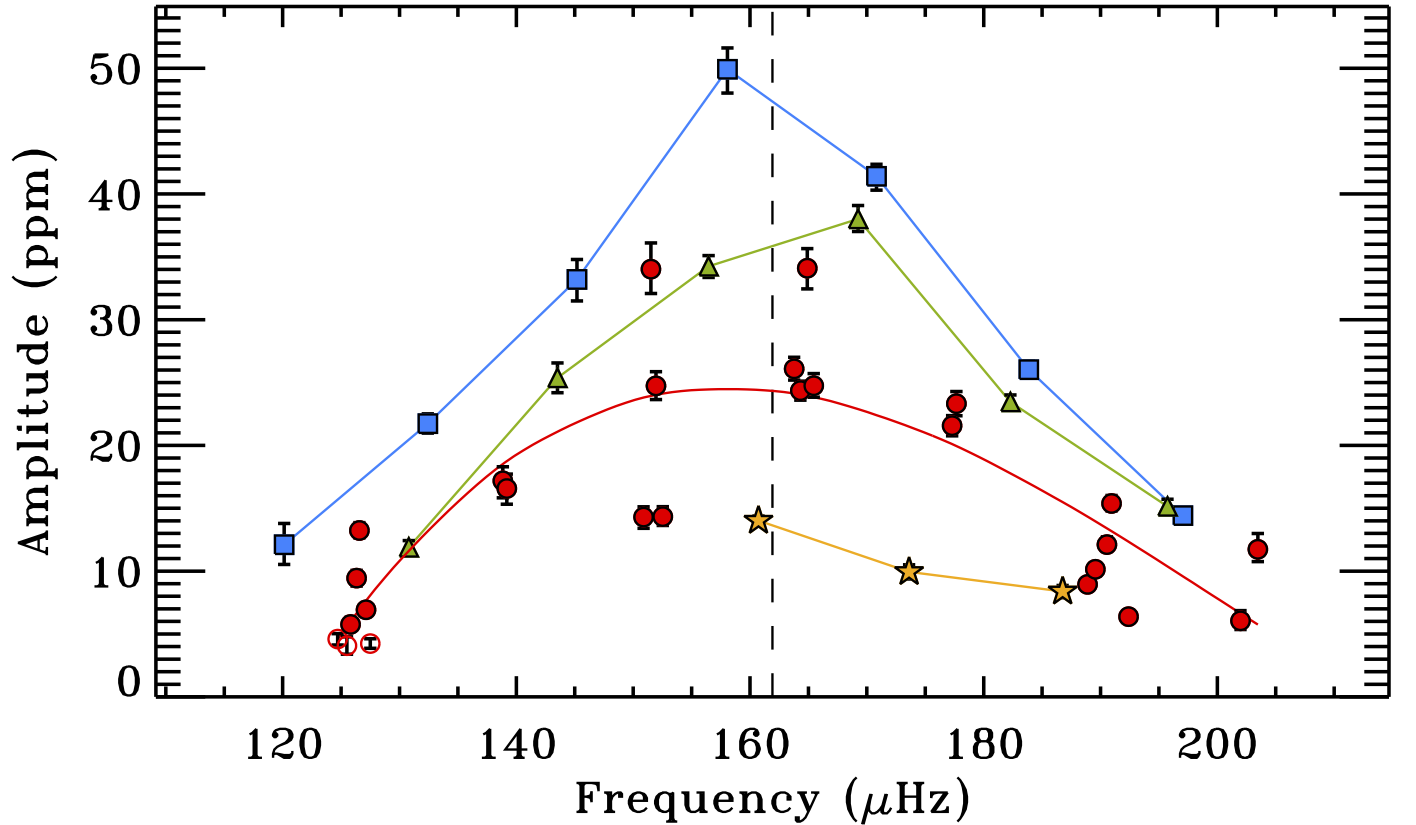}
      \caption{Mode amplitudes for \kic\,\,as a function of the corresponding oscillation frequencies. Amplitude measurements as defined by Eq.~(\ref{eq:resolved_profile}) for each angular degree ($\ell = 0$ blue squares, $\ell = 2$ green triangles, $\ell = 3$ yellow stars, and resolved $\ell = 1$ mixed modes red circles). Open symbols represent modes with detection probability under the suggested threshold (see Sect.~\ref{sec:test}). The 68\,\% credible intervals for the amplitudes as derived by \diamonds\,\,are shown for each data point. The solid red line represents a polynomial fit to the amplitudes of the $\ell = 1$ mixed modes, included to emphasize the trend with frequency. The dashed vertical line indicates the $\numax$ value listed in Table~\ref{tab:bkg2}.}
    \label{fig:amplitude}
\end{figure}

\subsection{Mode amplitudes}
\label{sec:amplitudes}
Mode amplitudes are the primary information to study the detectability of the oscillations and the physical process responsible for their excitation mechanism, especially in the less known case of the mixed modes \citep[e.g.][]{Grosjean14}. Retrieving empirical relations that allow us to predict amplitudes in other RGs stars \citep[e.g.][and references therein]{Corsaro13} is certainly useful to facilitate future detailed asteroseismic analyses on a larger sample of targets.

An example of the resulting mode amplitudes as a function of the corresponding oscillation frequencies for \kic\,\,is given in Fig.~\ref{fig:amplitude} for the different angular degrees, including the resolved mixed modes for which a polynomial fit to enhance the trend was added. The plots for all the other stars can be found in Appendix~\ref{sec:pb_results}. By computing an average ratio (or visibility factor) of the amplitudes between $\ell = 0$ modes, $\langle A_0 \rangle$, and $\ell = 2$, resolved $\ell = 1$, and $\ell = 3$ modes, $\langle A_2 \rangle$, $\langle A_1 \rangle$, $\langle A_3 \rangle$, respectively, we find that $\langle A_0 \rangle / \langle A_2 \rangle = 1.19 \pm 0.10$, $\langle A_0 \rangle / \langle A_1 \rangle = 1.84 \pm 0.26$, and $\langle A_0 \rangle / \langle A_3 \rangle = 2.77 \pm 0.41$. The error bars are computed from the dispersion in the ratios coming from the sample analyzed. These visibilities are useful for setting up prior distributions on mode amplitudes in other red giant stars (see Sect.~\ref{sec:discussion} for more discussion), while individual mode amplitudes can also be used to compare visibilities of LRGs with stars in different evolutionary stages \citep[e.g.][]{Ballot11visibility,Benomar13}.

\begin{figure}
   \centering
   \includegraphics[width=9.0cm]{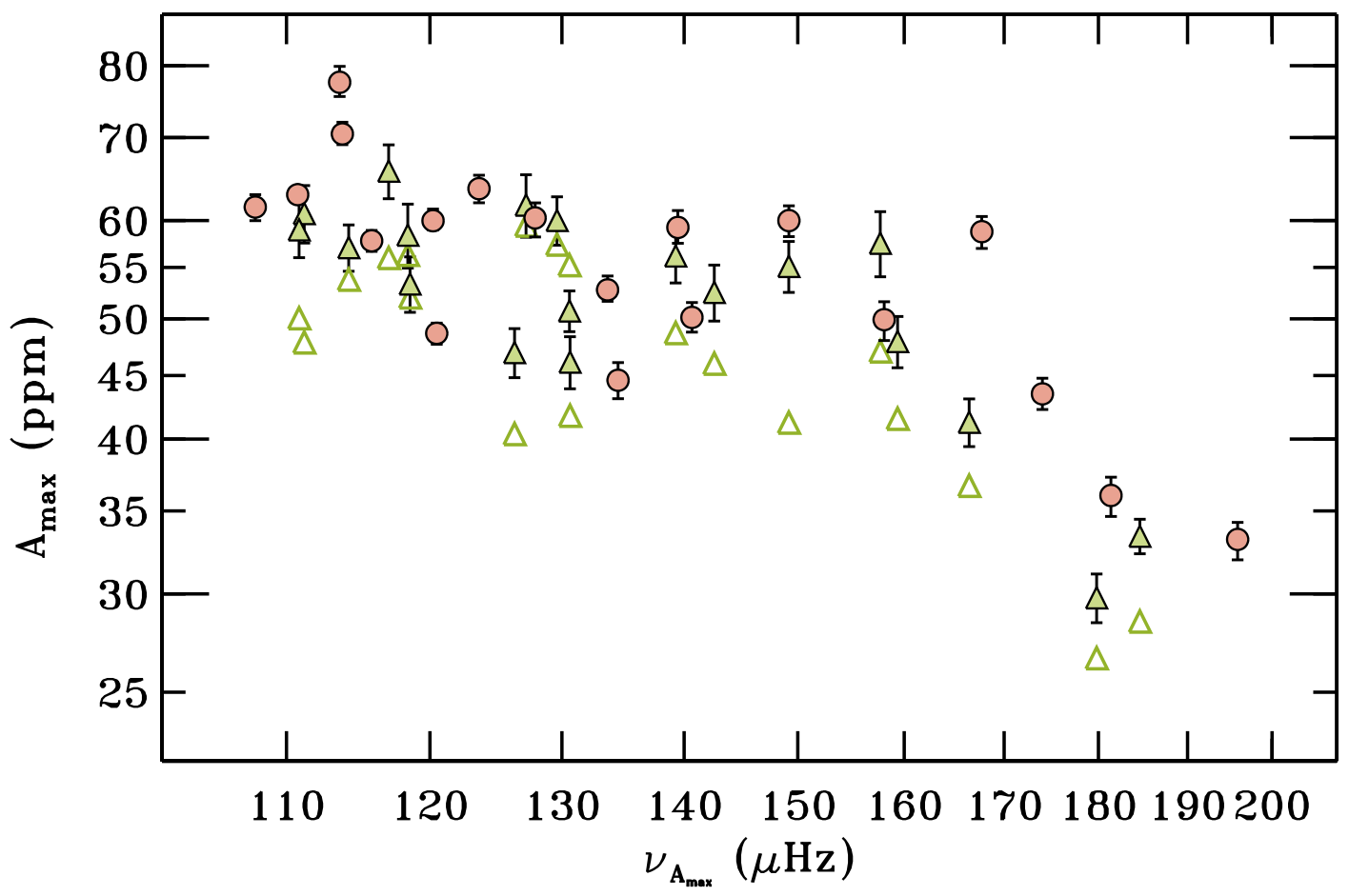}
      \caption{Maximum radial mode amplitudes for the LRGs (red circles) as a function of the corresponding radial mode frequency $\nu_\mathrm{A_\mathrm{max}}$ in a log-log scale, with 68\,\% Bayesian credible intervals overlaid. Included, the amplitude measurements by \cite{Huber11} for the same stars (filled green triangles), with their 1-$\sigma$ error bars, reported as a function of their $\numax$ values. The corresponding amplitude predictions for the measurements by \cite{Huber11}, computed from the best scaling relation model found by \cite{Corsaro13}, are also included for a comparison (open green triangles).}
    \label{fig:amax}
\end{figure}

To show how the behavior of the mode amplitude varies from star to star, we plot the maximum radial mode amplitude of each star, here indicated as $A_\mathrm{max}$, against the corresponding radial mode frequency $\nu_\mathrm{A_\mathrm{max}}$ in Fig.~\ref{fig:amax} (red circles). We note that the amplitude measurements from this work agree on average within 6\,\% with those listed by \cite{Huber11}, despite the different datasets used and the method adopted, which in the latter case provides maximum amplitudes per radial mode from a global analysis of the power excess of the stellar oscillations \citep[see][and references therein for more details]{Huber11}. The values $A_\mathrm{max}$ and $\nu_\mathrm{A_\mathrm{max}}$ from this work are all part of the results coming from a detailed peak bagging analysis (see the tables in Appendix~\ref{sec:pb_results}). The amplitude scaling relation used for computing the predictions for the measurements provided by \cite{Huber11}, indicated with open green triangles in Fig.~\ref{fig:amax}, is given by the best model $\mathcal{M}_{4,\beta}$ of \cite{Corsaro13} for the case of LC data. When comparing our $A_\mathrm{max}$ measurements with the predictions, we find an agreement on average around 19\,\%. We point out that the amplitude scaling relation adopted for the comparison is however calibrated on a set of global parameters such as $\numax$, $\Dnu$, and $A_\mathrm{max}$ from the power excess, and not on detailed measurements for an individual oscillation mode as those provided in this work.

\section{Discussion \& conclusions}
\label{sec:discussion}
Starting from the background fitting phase, we find a good agreement ($\sim 5$\,\%) between our estimated parameters of the granulation signal, Eq.~(\ref{eq:bkg}), and those available from \cite{Kallinger14}, proving that even with an additional year of observation this background signature has remained nearly unchanged for this sample of stars. This shows that we can easily retrieve prior distributions for the background free parameters of the RGs even if new datasets are used. This aspect is helpful for improving our performances in extracting oscillations from a large number of stars since the background fitting is a necessary step in the peak bagging analysis.

As shown by Fig.~\ref{fig:fwhm} for \kic, and in all the other figures in Appendix~\ref{sec:pb_results} for the entire sample of LRGs investigated, every star manifests a clear linewidth depression in the frequency region close to $\numax$, consistent with the expectations for less evolved stars. This behavior is clear for $\ell = 0, 2$ modes (although the measured linewidths of the quadrupole modes are not the real ones, as stated in Sect.~\ref{sec:fwhm}) and is less pronounced for resolved (or partially resolved) $\ell = 1$ mixed modes, although still evident (see e.g. the polynomial fit shown in red in Fig.~\ref{fig:fwhm}, top panel), whereas it is questionable for $\ell = 3$ modes because the low number of fit peaks does not fully cover the observed frequency range of the oscillations. The presence of a depression in linewidth for all the stars is also supported by our computation of the normalized fraction of resolved to unresolved mixed modes, $f_\mathrm{res}$, given by Eq.~(\ref{eq:fraction_resolved}). The average normalized fraction $\langle f_\mathrm{res} \rangle$ given by Eq.~(\ref{eq:average_fraction}) instead tells us that even with four years of nearly continuous observations with \Kepler, all the red giants in our sample are still dominated by the presence of unresolved mixed modes, except for \kic\,\,where the unresolved modes basically equal in number those that are instead resolved or partially resolved. This proves that the lifetime of the $g$ modes, clearly not lower than that of the most g-dominated mixed modes, is still not measurable. A detailed study involving the new individual linewidth measurements provided here would definitely help to tighten the constraints on the physical relation between the location of the linewidth depression, the cut-off frequency, and $\numax$ in red giants, thus shedding light to our understanding of the damping mechanism of the oscillations, especially in the case of the dipole mixed modes \citep{Grosjean14}.

Another significant result involving linewidths is their correlation with the stellar effective temperatures. In the comparison given in Fig.~\ref{fig:fwhm_temp}, we observe a temperature gradient in the linewidths that is steeper for main-sequence and subgiant stars than for RGs, thus flattening toward lower temperatures, as already found by \cite{Baudin11temp} and \cite{Belkacem12}. Conversely to the large spread in linewidths found by \cite{Baudin11temp} for the RGs observed with CoRoT, our linewidths appear to be much less dispersed, systematically lower, and quite correlated with $T_\mathrm{eff}$, despite the small temperature range available, as also highlighted by the polynomial fit, and the corresponding confidence region, to our sample shown in Fig.~\ref{fig:fwhm_temp}. This is in agreement with the theoretical predictions by \cite{Belkacem12}. We find that the change in the slope of the correlation between linewidths and temperatures from our sample of RGs to that of the main-sequence and subgiant stars observed by \kepler and analyzed by \cite{App12fwhm}, is more evident in the temperature range around $T_\mathrm{eff} \sim 5400$\,K and could be related to a change in the regime responsible for the mode damping mechanism at the beginning of the RGB. Importantly, the temperature range of the RGs analyzed in this work, 4800-5500\,K, covers the gap in observations between red giants and subgiant and main-sequence stars from previous works that were using CoRoT and \kepler data \citep[e.g.][]{Baudin11temp,Belkacem12,Corsaro12}. We therefore consider that the new results presented are important to perform a more realistic modeling of the underlying physics responsible for the observed change in gradient occurring at the bottom of the RGB.

We have noted that our measurements of maximum mode amplitude, available from Appendix~\ref{sec:pb_results} and plotted in Fig.~\ref{fig:amax}, are in a good agreement ($\sim 6$\,\%) with the amplitudes of maximum power derived by \cite{Huber11} according to the method described by \cite{K05,K08}. This shows that, at least for the range of $\numax$ investigated in this work, the existing amplitude scaling relations can in principle be used to retrieve raw predictions on maximum amplitudes of individual radial modes. By introducing the optimal amplitude scaling relation derived by \cite{Corsaro13}, we have that the amplitude predictions for our sample of stars agree within $\sim$19\,\%. By exploiting the average amplitude visibilities, $\langle A_0 \rangle / \langle A_{\ell} \rangle$, one is able to derive amplitude estimates for the different angular degrees. These tools are very useful for the preparation of prior distributions on mode amplitudes in RGs having $\numax \in \left[110, 200 \right]$\,\muhz, hence they build the base for setting up the peak bagging analysis for a large number of stars. A wider range of $\numax$ will certainly allow to better test these simple relations and eventually obtain and calibrate one or more specific scaling relations for individual mode amplitudes in a similar fashion as investigated by \cite{Corsaro13}. Our physical comprehension and modeling of the stochastic excitation mechanism of the mixed modes in RGs \citep[e.g. see][]{Grosjean14} will certainly benefit from individual mode amplitudes and corresponding linewidth measurements of the dipole mixed modes provided in this paper.

In addition, the remarkable precision level obtained on the individual mode frequencies of the $p$ modes of all the LRGs, ranging from $10^{-2}$ to $10^{-3}$\,\muhz, coupled with the large number of $p$ modes extracted from each star (spanning from six to nine different radial orders), allow us to constrain the regions of sharp structure variation caused by the \he\,\,zone inside the red giants with an unprecedented level of detail \citep[see][for more details]{Corsaro15letter}. The new frequency measurements derived here will help to improve existing stellar structure models of evolved low-mass stars settling in the RGB phase of the stellar evolution.

Through the peak bagging analysis exploiting the high-quality \kepler data used in this work, we could extract and characterize a total of more than 1600 different oscillation modes. Following the work proposed by CD14, we successfully extended the fitting procedure to the more complex case of the red giant stars, for which a large number of mixed modes is present. As shown by CD14, and now in this work, the Bayesian approach coupled with high-efficiency sampling algorithms, offers a valuable and robust way of inferring asteroseismic parameters from power spectra. This approach allows us to perform the fitting of many oscillation modes, i.e. of several tens of free parameters, in a reasonably short time (of the order of minutes). Besides, the computation of the Bayesian evidence has allowed us to quickly and easily test the peak significance of more than 600 oscillation modes, as also discussed in Sect.~\ref{sec:test} and Sect.~\ref{sec:results}.

By adopting the peak bagging model given by Eq.~(\ref{eq:pb_model}), the \diamonds\,\,code has performed well for all the computations done. This is in general explained by an easy setting up of the prior distributions for the free parameters of the model, by the absence of degeneracies in the sampling of the parameter space when adopting individual parameters for each mode and with the combination of amplitude-linewidth instead of height-linewidth in the mode profile (see Eq.~\ref{eq:resolved_profile}), and by the avoided estimation of the linewidths for the narrowest (namely unresolved) oscillation peaks (see Sect.~\ref{sec:extraction}). The latter case in particular proves that by using a mixture of two different profiles, hence distinguishing between resolved (or partially resolved) modes, and unresolved ones, considerably improves our performances for the fitting of the oscillations in RGs. The main drawbacks of this approach are related to the requirement of a visual inspection of the region of the PSD of the star containing the oscillations, thus involving the selection of the modes to be fit,  the choice of the peak profile to be adopted, and the setting up of the corresponding prior distributions. However, these steps can all be improved and made easier by means of auxiliary tools based on existing asteroseismic scaling relations, as presented in Sect.~\ref{sec:mode_id}. Since the number of mixed modes in RGs is largely dominant over that of $p$ modes, and since among the mixed modes most are still unresolved, as seen in nearly all the stars analyzed here with more than four years of observations (Sect.~\ref{sec:fwhm}), we therefore recommend the adoption of a peak bagging model along the same lines as those described in Sect.~\ref{sec:extraction} for future fitting of the oscillations in evolved low-mass stars. 

The mode identification process, presented in Sect.~\ref{sec:mode_id}, may sometimes be problematic for the case of the dipole mixed modes since having a reliable estimate of the true period spacing $\DP$ is necessary. By starting from the literature values listed in Table~\ref{tab:asymp_parameters}, we find that a refitting of $\DP$ was required for seven stars out of the 19 analyzed ($\sim37$\,\% of the sample), meaning that the extraction process for the mixed modes may need an additional search for a proper solution of $\DP$, always coupled with a visual inspection of the result. Based on the experience gained with the analysis done for the 19 LRGs, we consider the visual inspection for the mode identification of the mixed modes an important and often decisive step, especially for those stars where the confusion limit due to the rotational splitting is present (KIC~8475025, KIC~9267654, KIC~11353313) and for those with variations in $\DP$ with the radial order (KIC~ 6144777 and \kic). This aspect clearly represents a limitation to the full automatization of the peak bagging analysis in red giant stars, but it might be improved in the future by analyzing and comparing the results arising from a larger variety of targets, especially those having higher rotation rates. We however stress that, provided that the used values of $\DP$ are accurate enough, the asymptotic relation for the mixed modes works remarkably well at reproducing their frequencies for all the peaks identified in our analysis. This is in agreement with the recent theoretical study by \cite{Jiang14}, who investigated the asymptotic regime of the mixed modes and confirmed the reliability of their asymptotic relation.

Last but not least, the firm detection of rotation in 14 stars of the sample will greatly help to constrain the physical mechanisms responsible for the angular momentum transport in RGs \citep[e.g.][and references therein]{Tayar13}. In particular, by reconstructing the rotational profile of the star up to the core level by using the mixed modes with different mixture content between $p$ and $g$ modes, provided in this work, it will be possible to investigate the evolution of the angular momentum transport inside the star starting from the main-sequence phase of the stellar evolution. Thanks to the variety of masses of the stars in our sample, spanning the range $1$-$2$\,$M_\odot$ (Table~\ref{tab:asymp_parameters}), the models involving the decoupling of the convective and radiative zones in low-mass RGs, expected after the first dredge up \citep{Tayar13}, can be intensively tested and refined.

To conclude, the full set of results presented in Sect.~\ref{sec:results} and discussed in Sect.~\ref{sec:discussion}, representing the largest set of detailed asteroseismic mode measurements ever made available, with the high-precision and accuracy level achieved from four-years datasets, offers a great opportunity to perform thorough investigations and testing of the existing stellar models and stellar evolution theory, asteroseismic scaling relations, and of the underlying physics of both the damping and the excitation mechanisms of solar-like and mixed mode oscillations in red giant stars. Future works following the approach presented here to extend the peak bagging analysis to larger samples of red giants observed by \Kepler, will definitely bring an outstanding improvement to our understanding level on the stellar physics of low-mass stars.  

\begin{acknowledgements}
E.C. is funded by the European Community's Seventh Framework Programme (FP7/2007-2013) under grant agreement n$^\circ$312844 (SPACEINN).
The research leading to these results has received funding from the European
Research Council under the European Community's Seventh Framework Programme
(FP7/2007--2013) ERC grant agreement n$^\circ$227224 (PROSPERITY), 
from the Fund for Scientific Research of Flanders (G.0728.11), 
and from the Belgian federal science policy office (C90291 Gaia-DPAC).
E.C. thanks F. Baudin, P. Beck, G. Davies, T. Kallinger and A. Miglio for useful discussions.
\end{acknowledgements}

\bibliographystyle{aa} % style aa.bst
%\bibliography{biblio} % your references

\appendix
\section{Results for the background fitting}
\label{sec:bkg_results}
\begin{figure*}
   \centering
   \includegraphics[width=9.2cm]{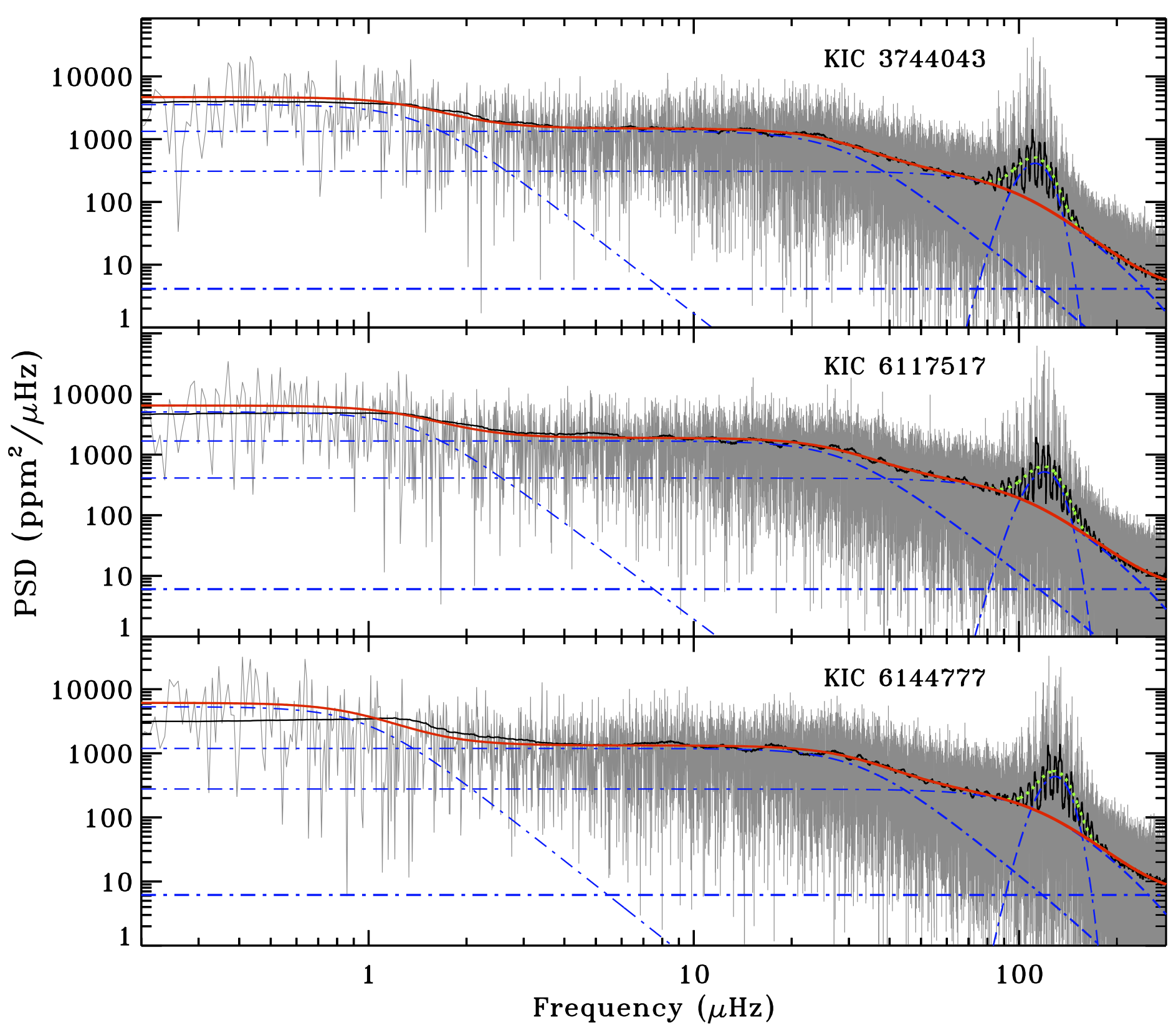}\includegraphics[width=9.2cm]{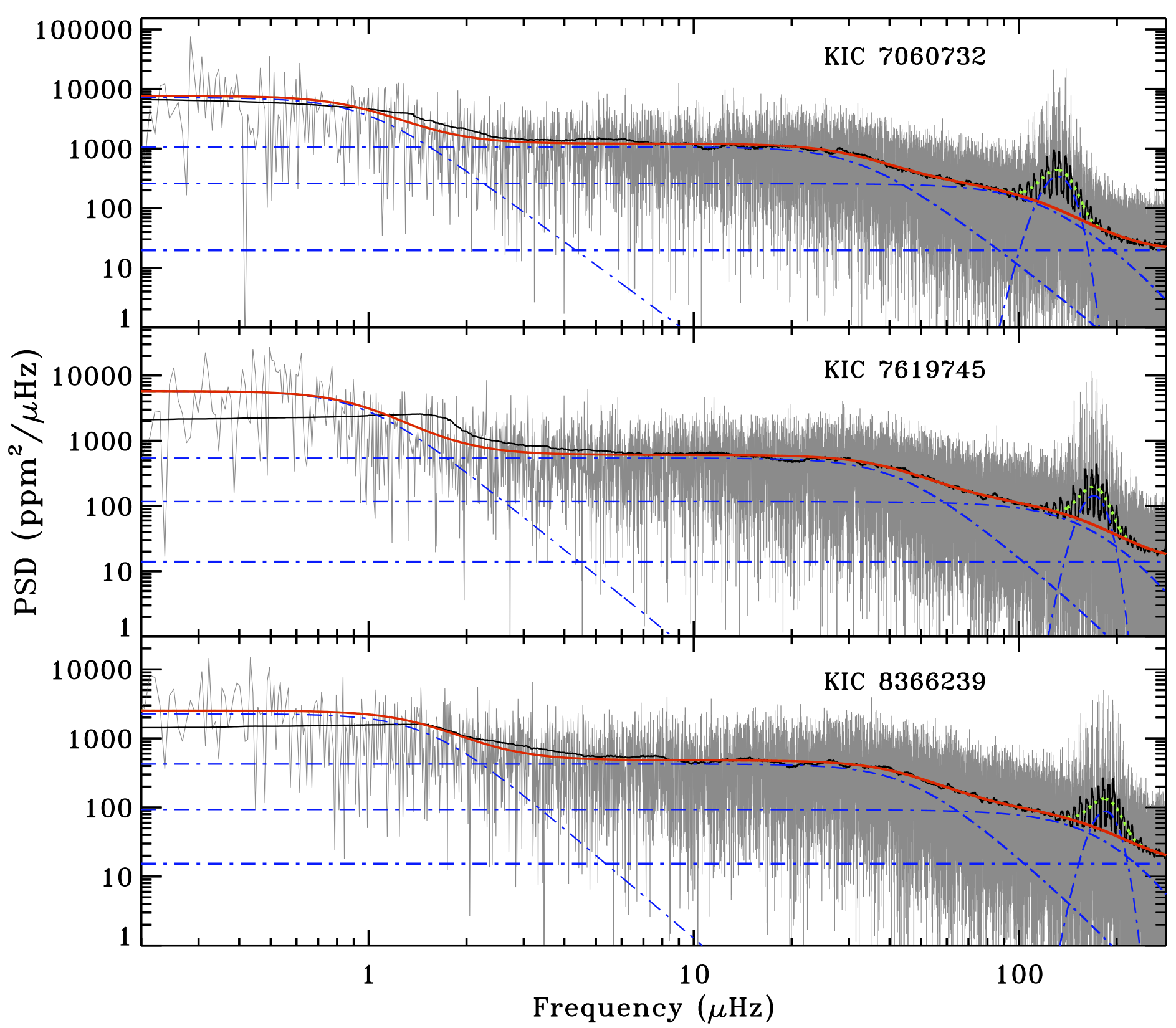}
   \includegraphics[width=9.2cm]{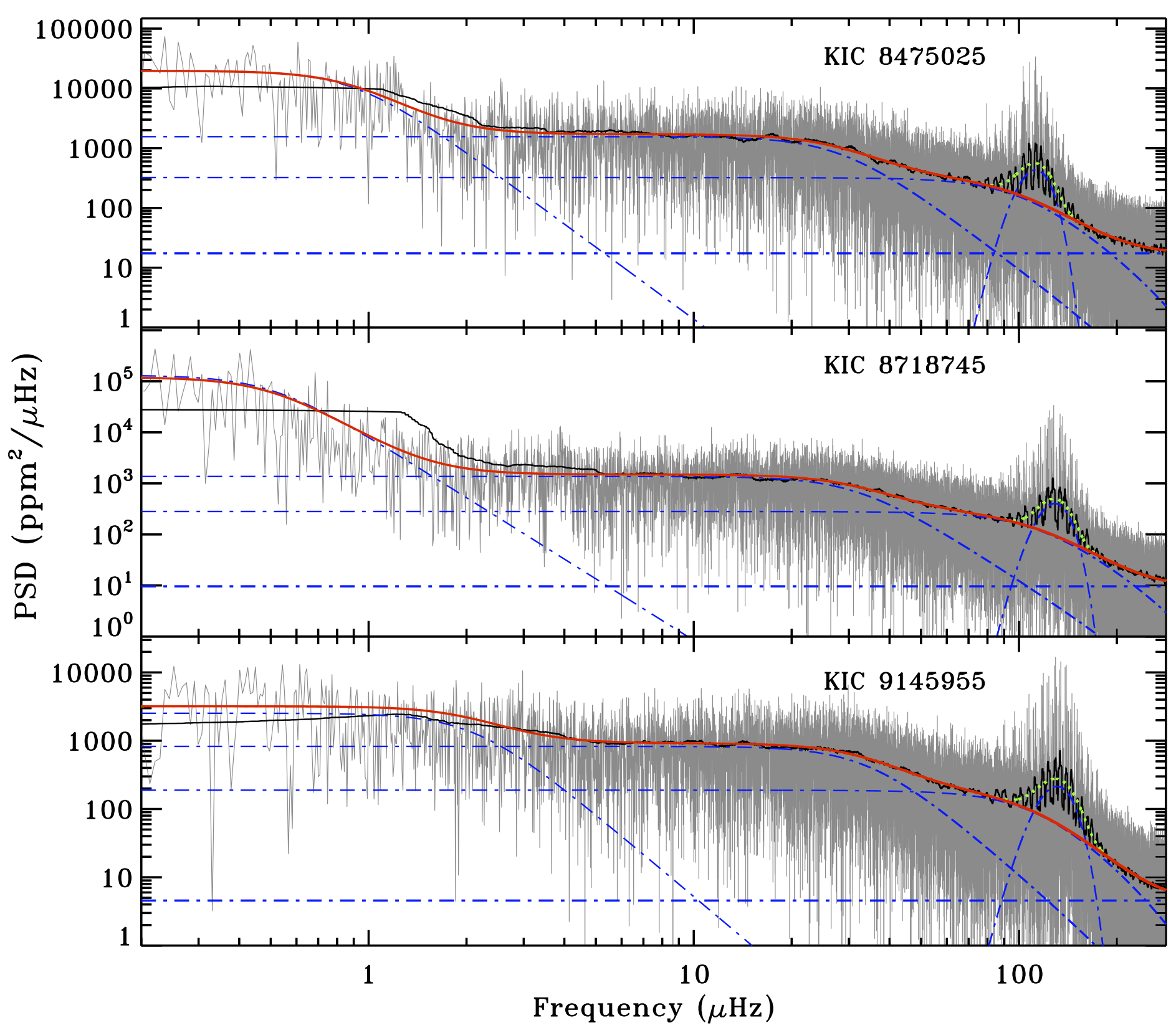}\includegraphics[width=9.2cm]{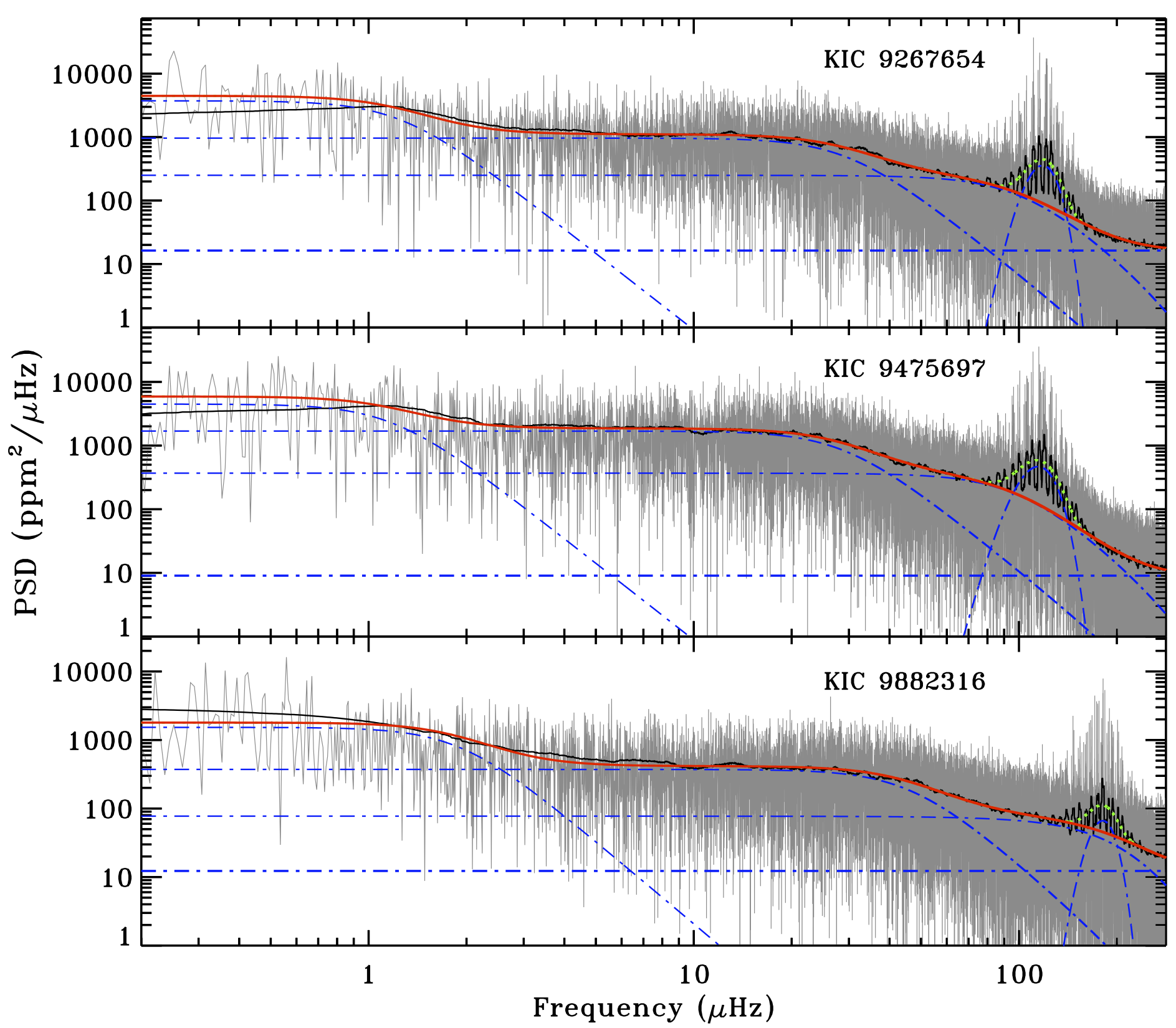}
   \includegraphics[width=9.2cm]{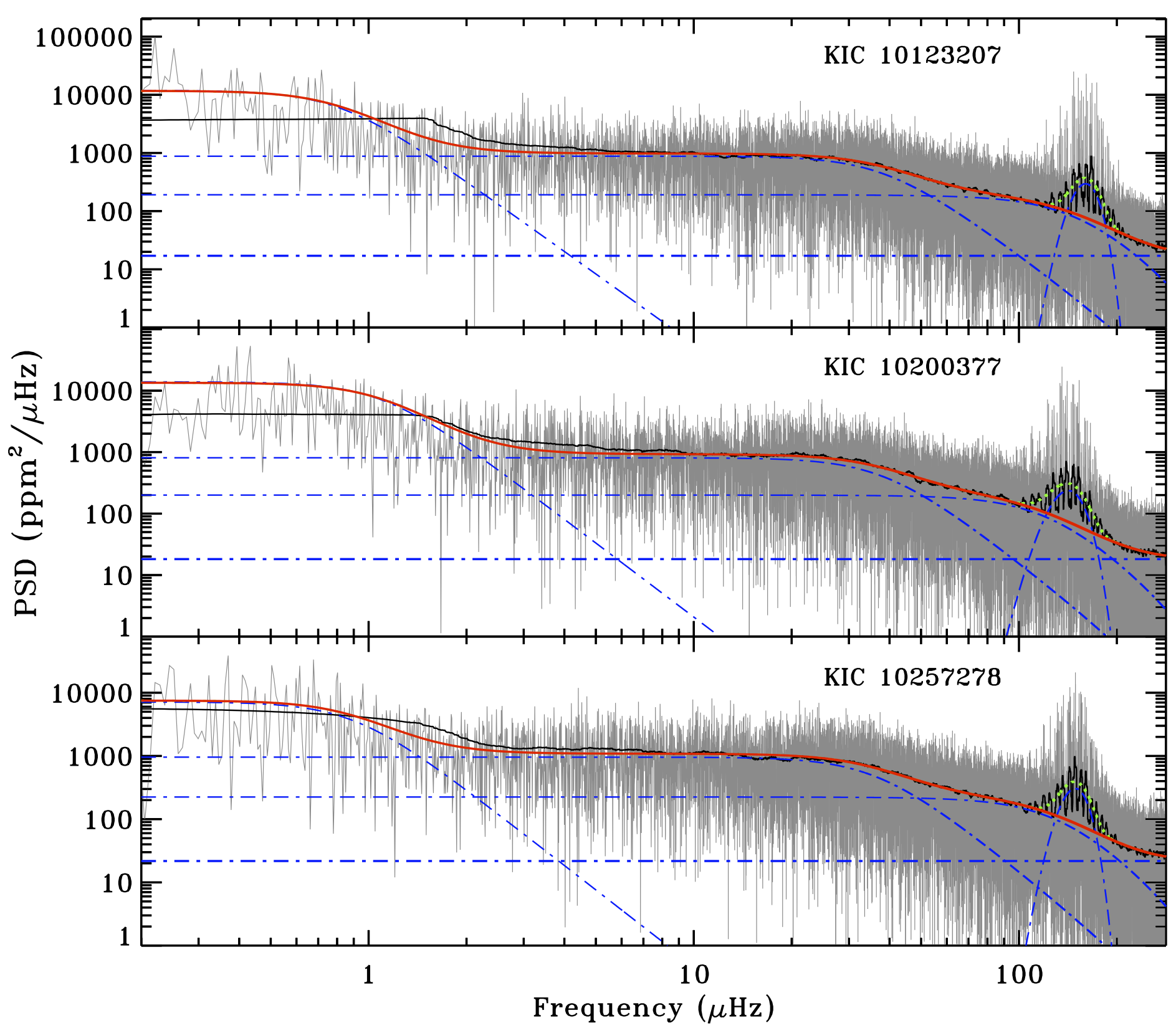}\includegraphics[width=9.2cm]{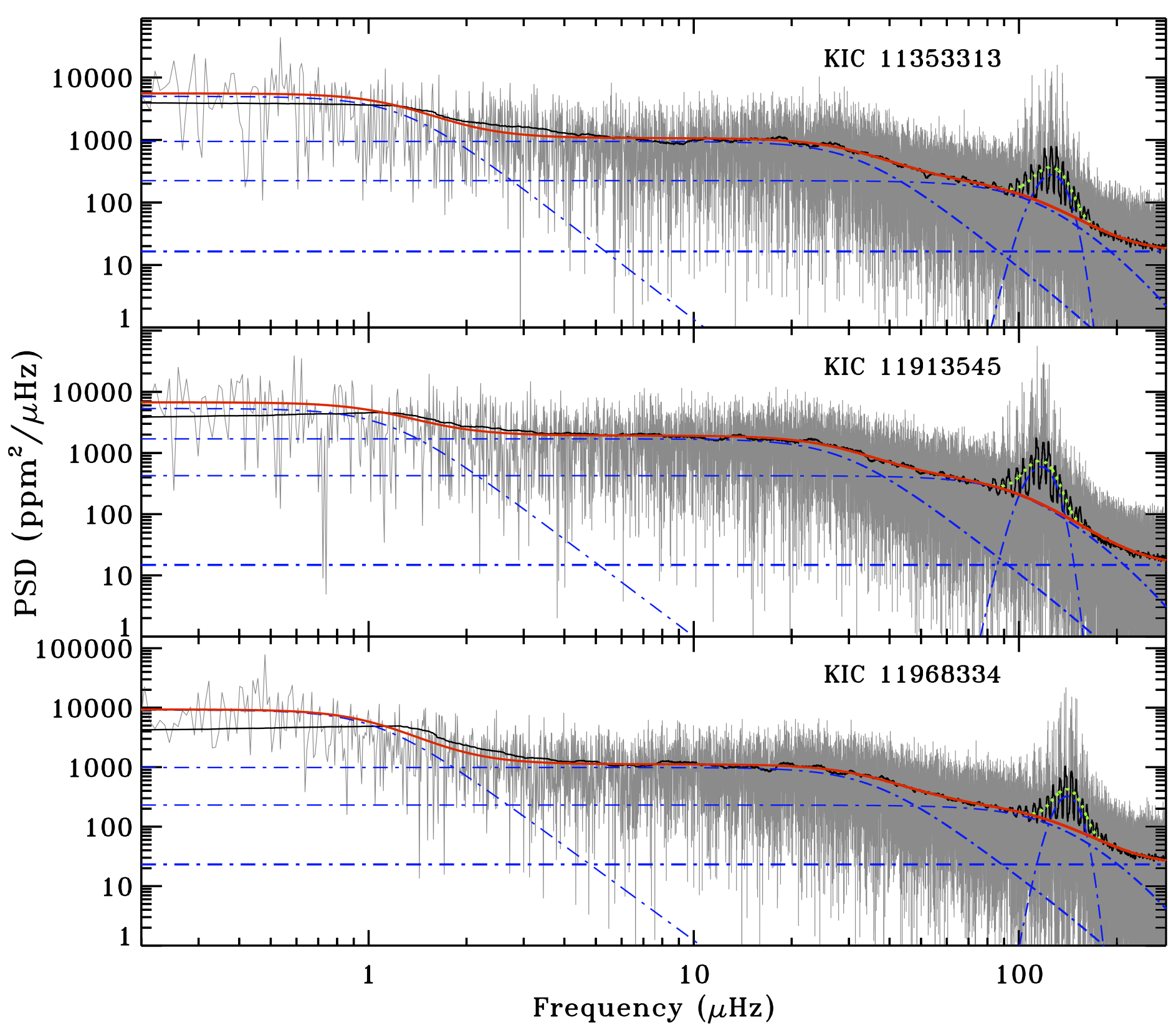}
      \caption{Same as for Fig.~\ref{fig:bkg_case} but for the remaining stars of the sample.}
    \label{fig:bkg_tot}
\end{figure*}

The parameter values derived by means of \diamonds\,\,for the entire sample of 19 LRGs analyzed in this work are listed in Table~\ref{tab:bkg1} for the Harvey-like profiles $B \left( \nu \right)$, Eq.~(\ref{eq:bkg}), and in Table~\ref{tab:bkg2} for the Gaussian envelope $G \left( \nu \right)$, Eq.~(\ref{eq:env}).
The background fits are shown in Fig.~\ref{fig:bkg_tot} for all the stars except \kic, which is instead shown in Fig.~\ref{fig:bkg_case}. 

Referring to the definitions presented by CD14, the configuring parameters of \diamonds\,\,used for all the computations are: initial enlargement fraction $1.5 \leq f_0 \leq 2.1$, shrinking rate $ 0.01 \leq \alpha \leq 0.03$, number of live points $N_\mathrm{live} = 500$, number of clusters $1 \leq N_\mathrm{clust} \leq 10$, number of total drawing attempts $M_\mathrm{attempts} = 10^4$, number of nested iterations before the first clustering $M_\mathrm{init} = 1500$, and number of nested iterations with the same clustering $M_\mathrm{same} = 50$.

% Figures of the background fit for all the stars

\begin{table*}
\caption{Median values with corresponding 68.3\,\% shortest credible intervals of the background parameters for the 19 RGs investigated, related to the white noise component and the super-Lorentzian profiles given by Eq.~(\ref{eq:bkg}), as derived by \diamonds.}             % title of Table
\centering                         
\begin{tabular}{l r r c c c c c}       
\hline\hline
\\[-8pt]         
KIC ID & \multicolumn{1}{c}{$W$} & \multicolumn{1}{c}{$a_\mathrm{long}$} & $b_\mathrm{long}$ & $a_\mathrm{gran,1}$ & $b_\mathrm{gran,1}$ & $a_\mathrm{gran,2}$ & $b_\mathrm{gran,2}$\\ [1pt]
 & \multicolumn{1}{c}{(ppm$^2$ $\mu$Hz$^{-1}$)} & \multicolumn{1}{c}{(ppm)}  & (\muhz) & (ppm) & (\muhz) & (ppm) & (\muhz)\\ [1pt]   
\hline
\\[-8pt]
03744043         & $4.13_{-0.11}^{+0.11}$   &   $72.33_{-3.80}^{+3.52}$   &   $1.47_{-0.15}^{+0.13}$   &   $194.59_{-2.09}^{+2.64}$   &   $28.45_{-0.62}^{+0.55}$   &   $173.68_{-2.10}^{+2.12}$   &   $97.81_{-1.03}^{+0.95}$ \\[1pt]
06117517         & $6.05_{-0.19}^{+0.18}$   &   $84.16_{-4.24}^{+4.48}$   &   $1.40_{-0.17}^{+0.13}$   &   $221.66_{-2.61}^{+2.95}$   &   $29.27_{-0.59}^{+0.61}$   &   $205.40_{-2.47}^{+2.33}$   &   $102.61_{-1.04}^{+1.09}$ \\[1pt]
06144777         & $6.17_{-0.22}^{+0.20}$   &   $73.19_{-3.87}^{+4.05}$   &   $1.00_{-0.12}^{+0.12}$   &   $197.86_{-1.97}^{+2.48}$   &   $32.88_{-0.67}^{+0.56}$   &   $179.43_{-1.94}^{+1.96}$   &   $116.04_{-1.26}^{+1.31}$ \\[1pt]
07060732         & $19.70_{-0.38}^{+0.42}$   &   $84.34_{-3.83}^{+3.82}$   &   $1.00_{-0.08}^{+0.07}$   &   $186.61_{-2.00}^{+2.15}$   &   $32.76_{-0.60}^{+0.62}$   &   $172.96_{-1.79}^{+2.04}$   &   $115.97_{-1.91}^{+1.85}$ \\
\hline                                
\end{tabular}
\label{tab:bkg1}
\end{table*}

\begin{table*}
\caption{Median values with corresponding 68.3\,\% shortest credible intervals of the parameters related to the Gaussian envelope, Eq.~(\ref{eq:env}), for the 19 RGs investigated, as derived by \diamonds.}             % title of Table
\centering                         
\begin{tabular}{l r c c}       
\hline\hline
\\[-8pt]         
KIC ID & \multicolumn{1}{c}{$H_\mathrm{osc}$} & $\numax$ & $\sigma_\mathrm{env}$\\ [1pt]
 & \multicolumn{1}{c}{(ppm$^2$ $\mu$Hz$^{-1}$)} & (\muhz)  & (\muhz) \\ [1pt]   
\hline
\\[-8pt]
03744043         & $470.59_{-13.49}^{+11.77}$   &   $112.52_{-0.23}^{+0.25}$   &   $12.47_{-0.24}^{+0.21}$ \\[1pt]
06117517         & $602.54_{-16.95}^{+15.48}$   &   $120.27_{-0.24}^{+0.25}$   &   $13.22_{-0.27}^{+0.26}$ \\[1pt]
06144777         & $517.92_{-14.11}^{+12.74}$   &   $129.69_{-0.24}^{+0.22}$   &   $13.16_{-0.25}^{+0.23}$ \\[1pt]
07060732         & $408.89_{-10.80}^{+11.58}$   &   $132.29_{-0.24}^{+0.24}$   &   $13.16_{-0.28}^{+0.27}$ \\
\hline
\end{tabular}
\label{tab:bkg2}
\end{table*}

\section{Results for the oscillation modes}
\label{sec:pb_results}
For listing the results from the peak bagging analysis we adopt a nomenclature similar to that used by CD14 and list the oscillation modes in ascending frequency order, divided by angular degree. Each mode can be identified uniquely by the set of numbers (Peak \#, $\ell$, $m$). An example of this nomenclature for \kic\,\,is displayed in Fig.~\ref{fig:chunk}. The missing values for amplitudes and linewidths correspond to the cases of unresolved mixed modes, were only the heights are fit. The missing values for the detection probability $p_\mathrm{B}$ represent peaks for which the significance test was not necessary due to their large height and the clear position with respect to the asymptotic relations for $p$ and mixed modes (Sect.~\ref{sec:mode_id}). 

This appendix also includes the figures for mode linewidths and mode amplitudes, for all the stars of the sample, similarly to Fig.~\ref{fig:fwhm} and Fig.~\ref{fig:amplitude}, respectively, shown for \kic.

Referring to the definitions presented by CD14, the configuring parameters of \diamonds\,\,used for all the computations are: initial enlargement fraction $0.5 \leq f_0 \leq 3.4$, shrinking rate $\alpha = 0.01$, number of live points $N_\mathrm{live} = 1000$, number of clusters $1 \leq N_\mathrm{clust} \leq 4$, number of total drawing attempts $M_\mathrm{attempts} = 10^4$, number of nested iterations before the first clustering $M_\mathrm{init} = 1000$, and number of nested iterations with the same clustering $M_\mathrm{same} = 50$.

% Figures of mode linewidths and amplitudes for all the stars
\clearpage

\begin{figure}
   \centering
   \includegraphics[width=9.0cm]{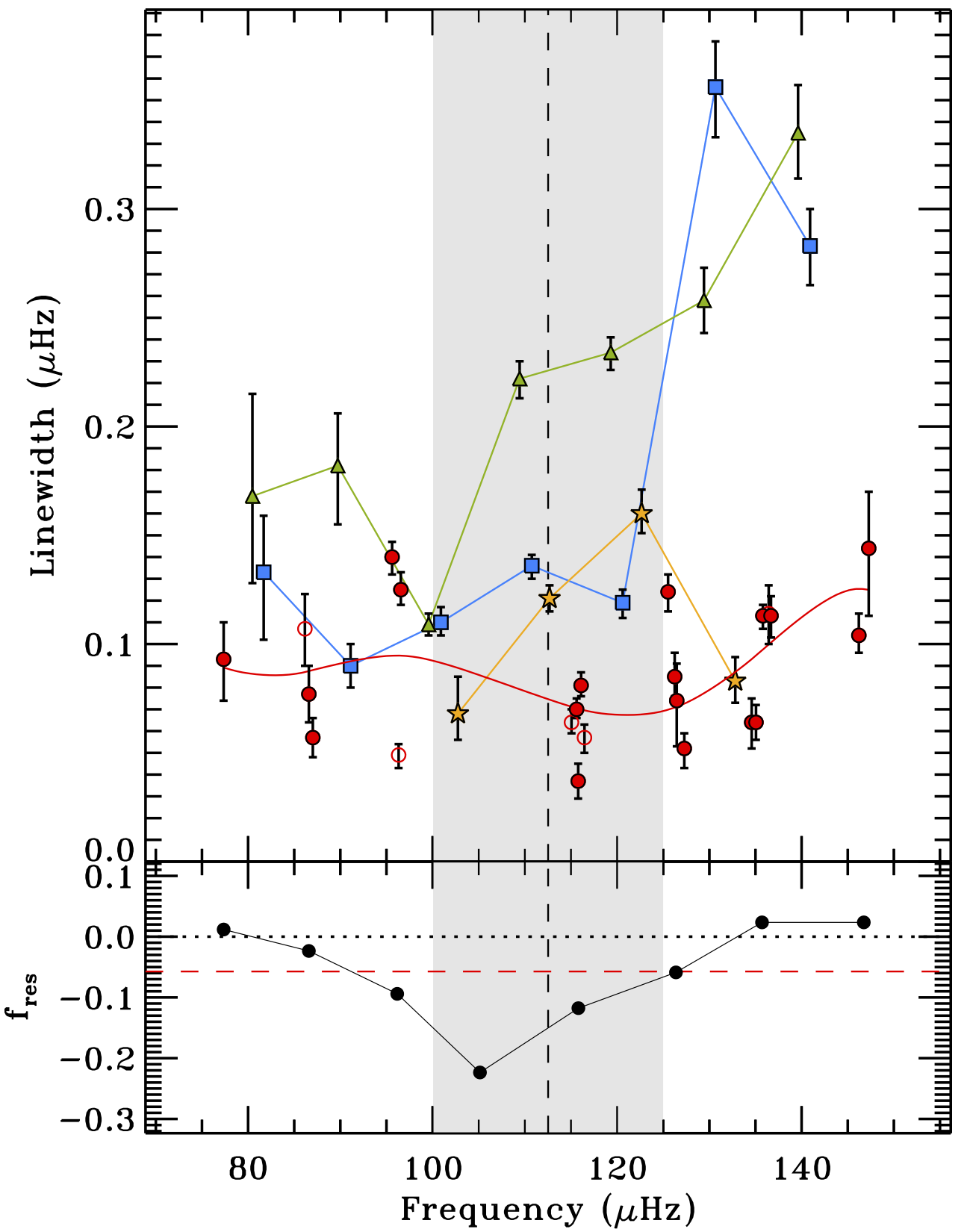}
      \caption{Mode linewidths for KIC~3744043 as a function of the corresponding oscillation frequencies. \textit{Top panel}: linewidth measurements as defined by Eq.~(\ref{eq:resolved_profile}) for each angular degree ($\ell = 0$ blue squares, $\ell = 2$ green triangles, $\ell = 3$ yellow stars, and resolved $\ell = 1$ mixed modes red circles). Open symbols represent modes with detection probability under the suggested threshold (see Sect.~\ref{sec:test}). The 68\,\% credible intervals for the linewidths as derived by \diamonds\,\,are shown for each data point. The red solid line represents a polynomial fit to the linewidths of the $\ell = 1$ mixed modes, included to emphasize the trend with frequency. The shaded region represents the range $\numax \pm \sigma_\mathrm{env}$, with $\numax$ from Table~\ref{tab:bkg2} indicated by the dashed vertical line. \textit{Bottom panel}: the normalized fraction of resolved mixed modes with respect to unresolved ones, $f_\mathrm{res}$ (black dots), defined by Eq.~(\ref{eq:fraction_resolved}). The frequency position of each point is the average frequency of the resolved dipole mixed modes falling in each radial order (or that of the unresolved mixed modes if no resolved mixed modes are present). The horizontal dotted line represents the limit of resolved-dominated regime, as defined in Sect.~\ref{sec:fwhm}, while the horizontal dashed red line marks the average $f_\mathrm{res}$ given by Eq.~(\ref{eq:average_fraction}).}
    \label{fig:3744043fwhm}
\end{figure}

\begin{figure}
   \centering
   \includegraphics[width=9.0cm]{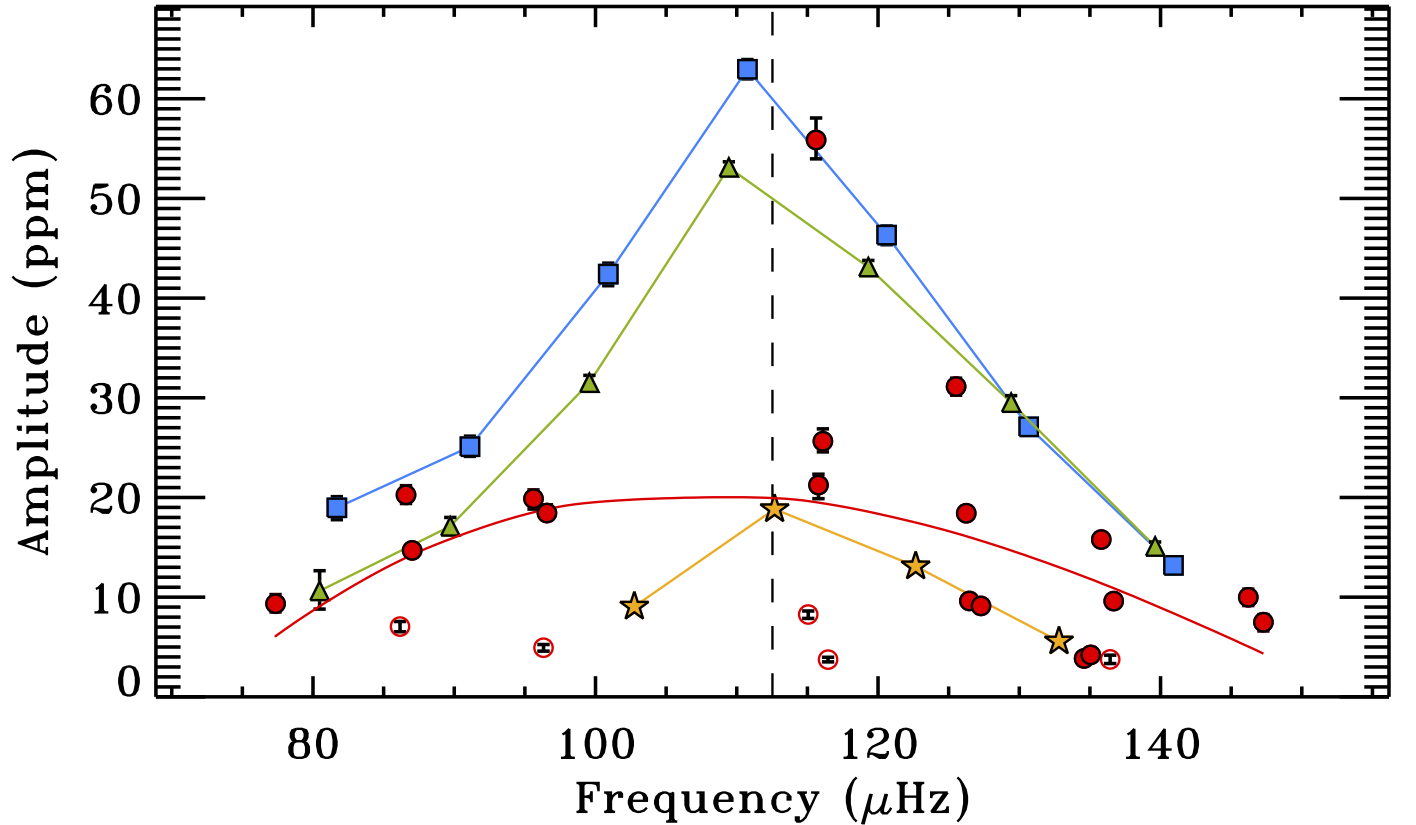}
      \caption{Mode amplitudes for KIC~3744043 as a function of the corresponding oscillation frequencies. Amplitude measurements as defined by Eq.~(\ref{eq:resolved_profile}) for each angular degree ($\ell = 0$ blue squares, $\ell = 2$ green triangles, $\ell = 3$ yellow stars, and resolved $\ell = 1$ mixed modes red circles). Open symbols represent modes with detection probability under the suggested threshold (see Sect.~\ref{sec:test}). The 68\,\% credible intervals for the amplitudes as derived by \diamonds\,\,are shown for each data point. The solid red line represents a polynomial fit to the amplitudes of the $\ell = 1$ mixed modes, included to emphasize the trend with frequency. The dashed vertical line indicates the $\numax$ value listed in Table~\ref{tab:bkg2}.}
    \label{fig:3744043amplitude}
\end{figure}
\clearpage

\begin{figure}
   \centering
   \includegraphics[width=9.0cm]{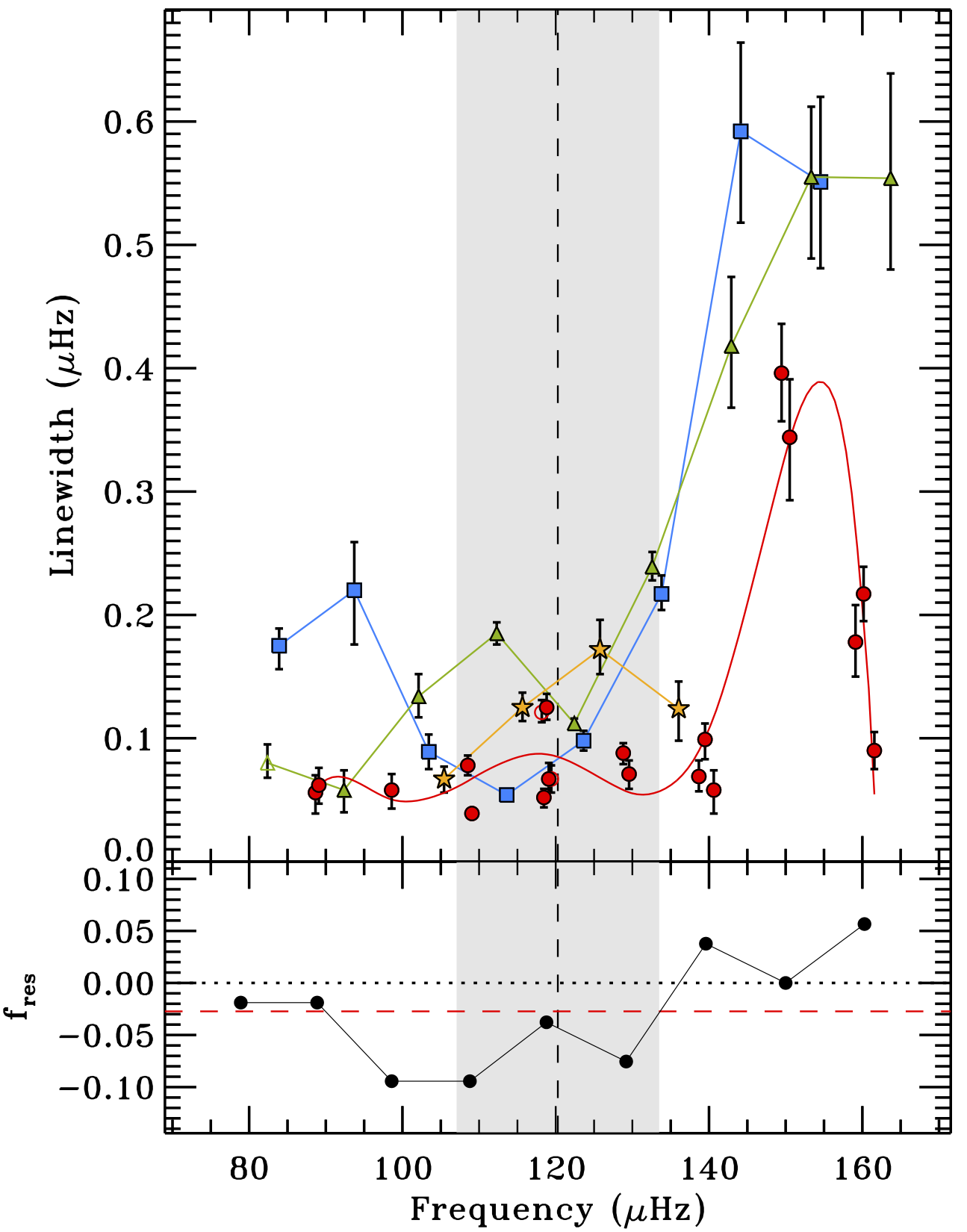}
      \caption{Mode linewidths for KIC~6117517 as a function of the corresponding oscillation frequencies. \textit{Top panel}: linewidth measurements as defined by Eq.~(\ref{eq:resolved_profile}) for each angular degree ($\ell = 0$ blue squares, $\ell = 2$ green triangles, $\ell = 3$ yellow stars, and resolved $\ell = 1$ mixed modes red circles). Open symbols represent modes with detection probability under the suggested threshold (see Sect.~\ref{sec:test}). The 68\,\% credible intervals for the linewidths as derived by \diamonds\,\,are shown for each data point. The red solid line represents a polynomial fit to the linewidths of the $\ell = 1$ mixed modes, included to emphasize the trend with frequency. The shaded region represents the range $\numax \pm \sigma_\mathrm{env}$, with $\numax$ from Table~\ref{tab:bkg2} indicated by the dashed vertical line. \textit{Bottom panel}: the normalized fraction of resolved mixed modes with respect to unresolved ones, $f_\mathrm{res}$ (black dots), defined by Eq.~(\ref{eq:fraction_resolved}). The frequency position of each point is the average frequency of the resolved dipole mixed modes falling in each radial order (or that of the unresolved mixed modes if no resolved mixed modes are present). The horizontal dotted line represents the limit of resolved-dominated regime, as defined in Sect.~\ref{sec:fwhm}, while the horizontal dashed red line marks the average $f_\mathrm{res}$ given by Eq.~(\ref{eq:average_fraction}).}
    \label{fig:6117517fwhm}
\end{figure}

\begin{figure}
   \centering
   \includegraphics[width=9.0cm]{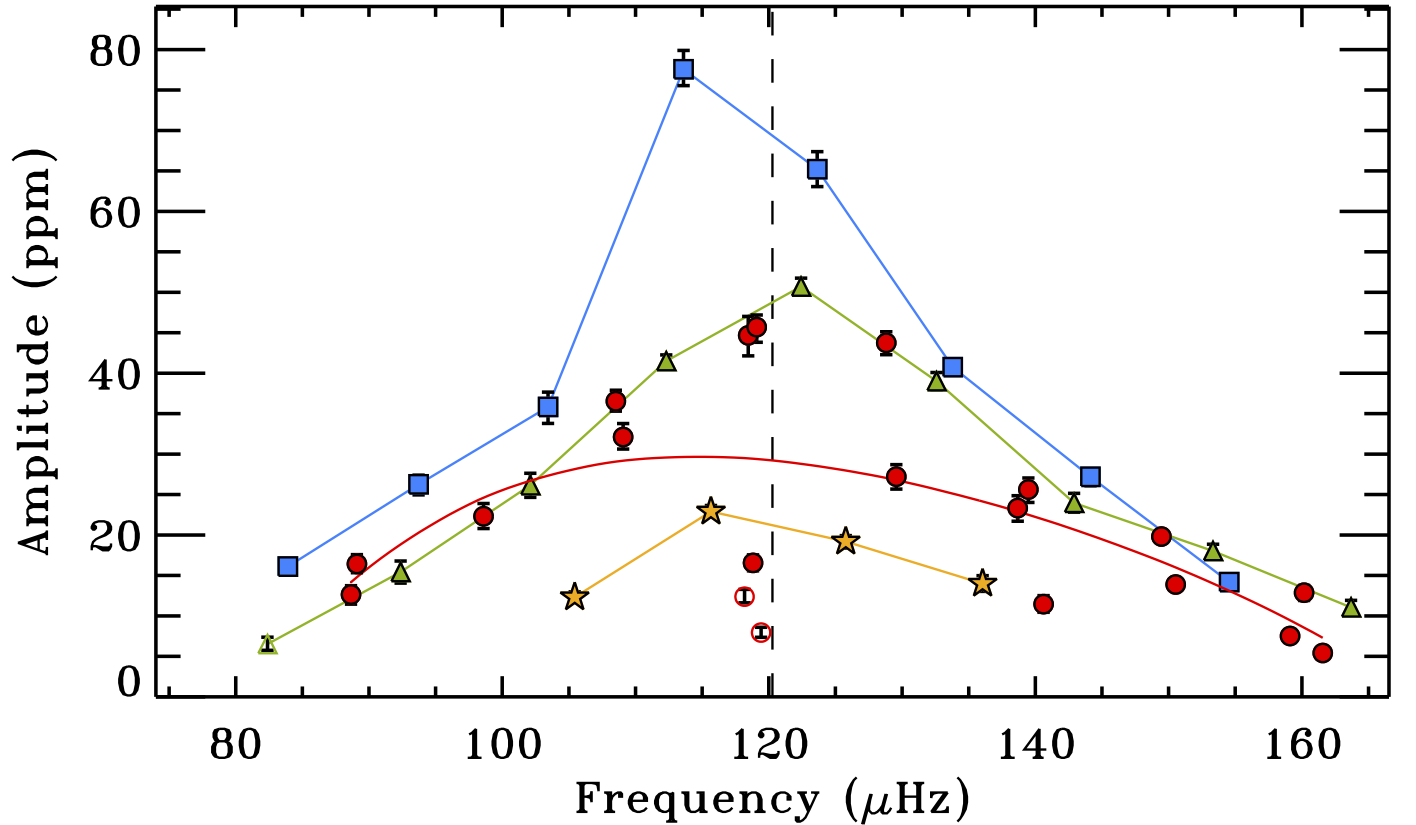}
      \caption{Mode amplitudes for KIC~6117517 as a function of the corresponding oscillation frequencies. Amplitude measurements as defined by Eq.~(\ref{eq:resolved_profile}) for each angular degree ($\ell = 0$ blue squares, $\ell = 2$ green triangles, $\ell = 3$ yellow stars, and resolved $\ell = 1$ mixed modes red circles). Open symbols represent modes with detection probability under the suggested threshold (see Sect.~\ref{sec:test}). The 68\,\% credible intervals for the amplitudes as derived by \diamonds\,\,are shown for each data point. The solid red line represents a polynomial fit to the amplitudes of the $\ell = 1$ mixed modes, included to emphasize the trend with frequency. The dashed vertical line indicates the $\numax$ value listed in Table~\ref{tab:bkg2}.}
    \label{fig:6117517amplitude}
\end{figure}
\clearpage

\begin{figure}
   \centering
   \includegraphics[width=9.0cm]{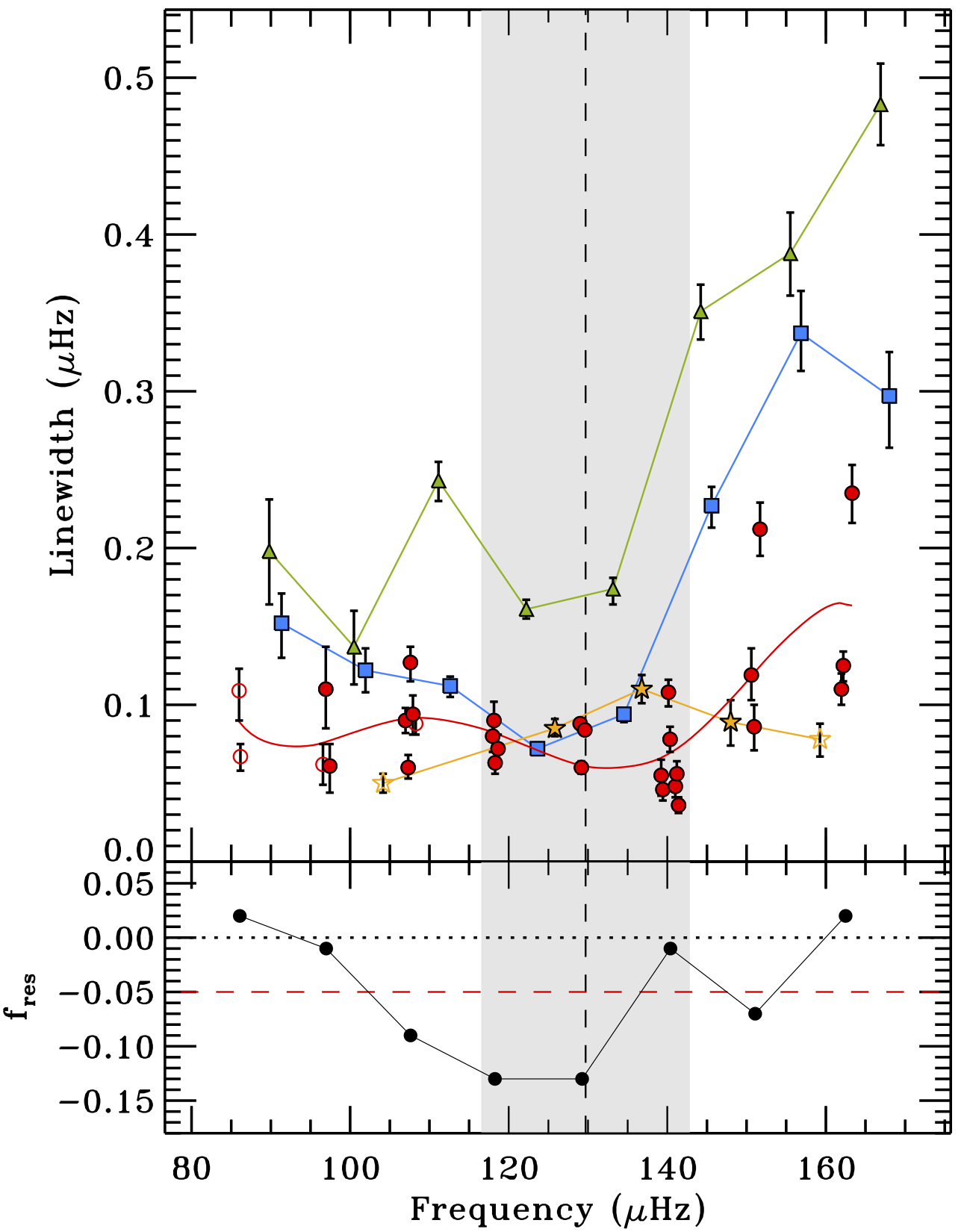}
      \caption{Mode linewidths for KIC~6144777 as a function of the corresponding oscillation frequencies. \textit{Top panel}: linewidth measurements as defined by Eq.~(\ref{eq:resolved_profile}) for each angular degree ($\ell = 0$ blue squares, $\ell = 2$ green triangles, $\ell = 3$ yellow stars, and resolved $\ell = 1$ mixed modes red circles). Open symbols represent modes with detection probability under the suggested threshold (see Sect.~\ref{sec:test}). The 68\,\% credible intervals for the linewidths as derived by \diamonds\,\,are shown for each data point. The red solid line represents a polynomial fit to the linewidths of the $\ell = 1$ mixed modes, included to emphasize the trend with frequency. The shaded region represents the range $\numax \pm \sigma_\mathrm{env}$, with $\numax$ from Table~\ref{tab:bkg2} indicated by the dashed vertical line. \textit{Bottom panel}: the normalized fraction of resolved mixed modes with respect to unresolved ones, $f_\mathrm{res}$ (black dots), defined by Eq.~(\ref{eq:fraction_resolved}). The frequency position of each point is the average frequency of the resolved dipole mixed modes falling in each radial order (or that of the unresolved mixed modes if no resolved mixed modes are present). The horizontal dotted line represents the limit of resolved-dominated regime, as defined in Sect.~\ref{sec:fwhm}, while the horizontal dashed red line marks the average $f_\mathrm{res}$ given by Eq.~(\ref{eq:average_fraction}).}
    \label{fig:6144777fwhm}
\end{figure}

\begin{figure}
   \centering
   \includegraphics[width=9.0cm]{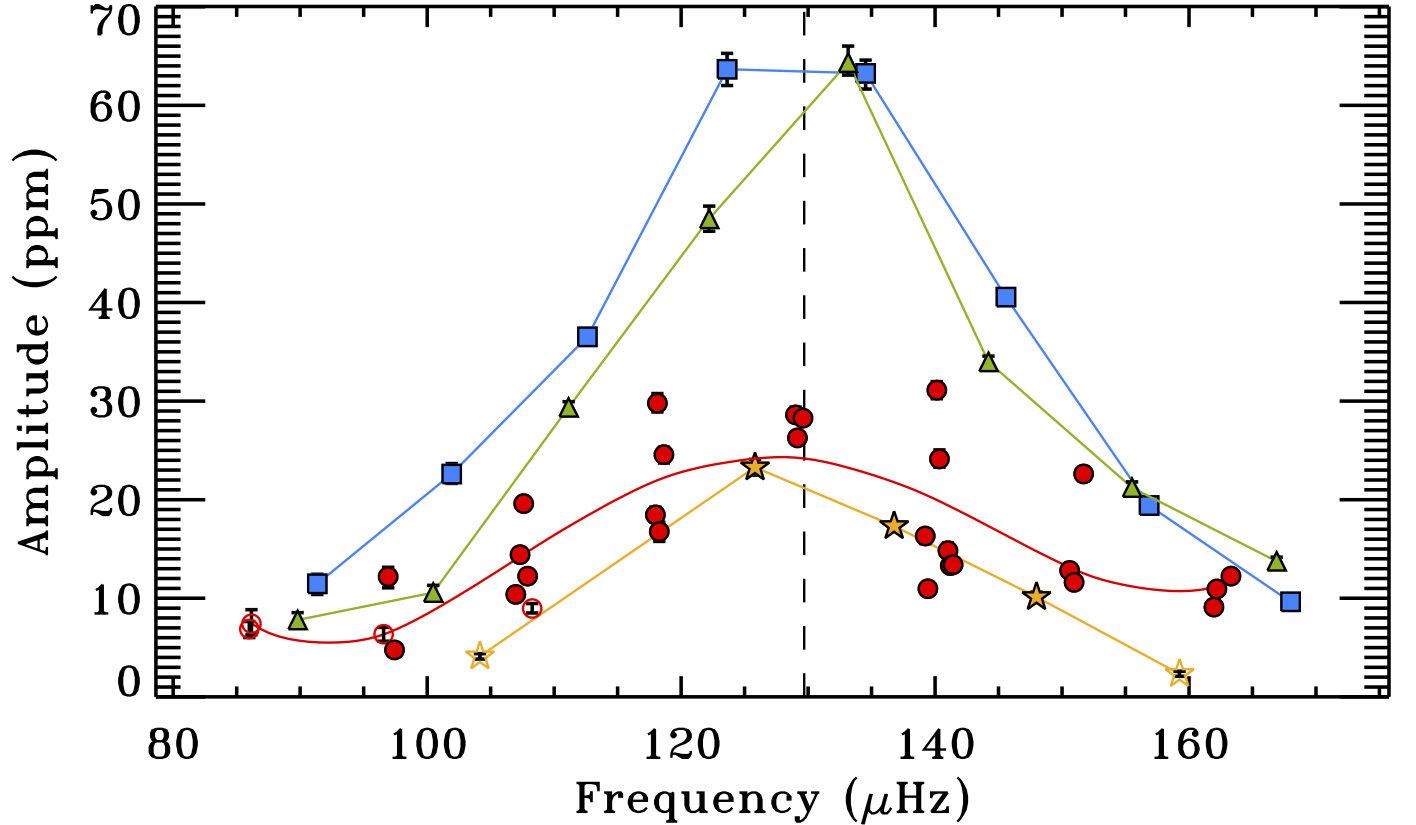}
      \caption{Mode amplitudes for KIC~6144777 as a function of the corresponding oscillation frequencies. Amplitude measurements as defined by Eq.~(\ref{eq:resolved_profile}) for each angular degree ($\ell = 0$ blue squares, $\ell = 2$ green triangles, $\ell = 3$ yellow stars, and resolved $\ell = 1$ mixed modes red circles). Open symbols represent modes with detection probability under the suggested threshold (see Sect.~\ref{sec:test}). The 68\,\% credible intervals for the amplitudes as derived by \diamonds\,\,are shown for each data point. The solid red line represents a polynomial fit to the amplitudes of the $\ell = 1$ mixed modes, included to emphasize the trend with frequency. The dashed vertical line indicates the $\numax$ value listed in Table~\ref{tab:bkg2}.}
    \label{fig:6144777amplitude}
\end{figure}
\clearpage

\begin{figure}
   \centering
   \includegraphics[width=9.0cm]{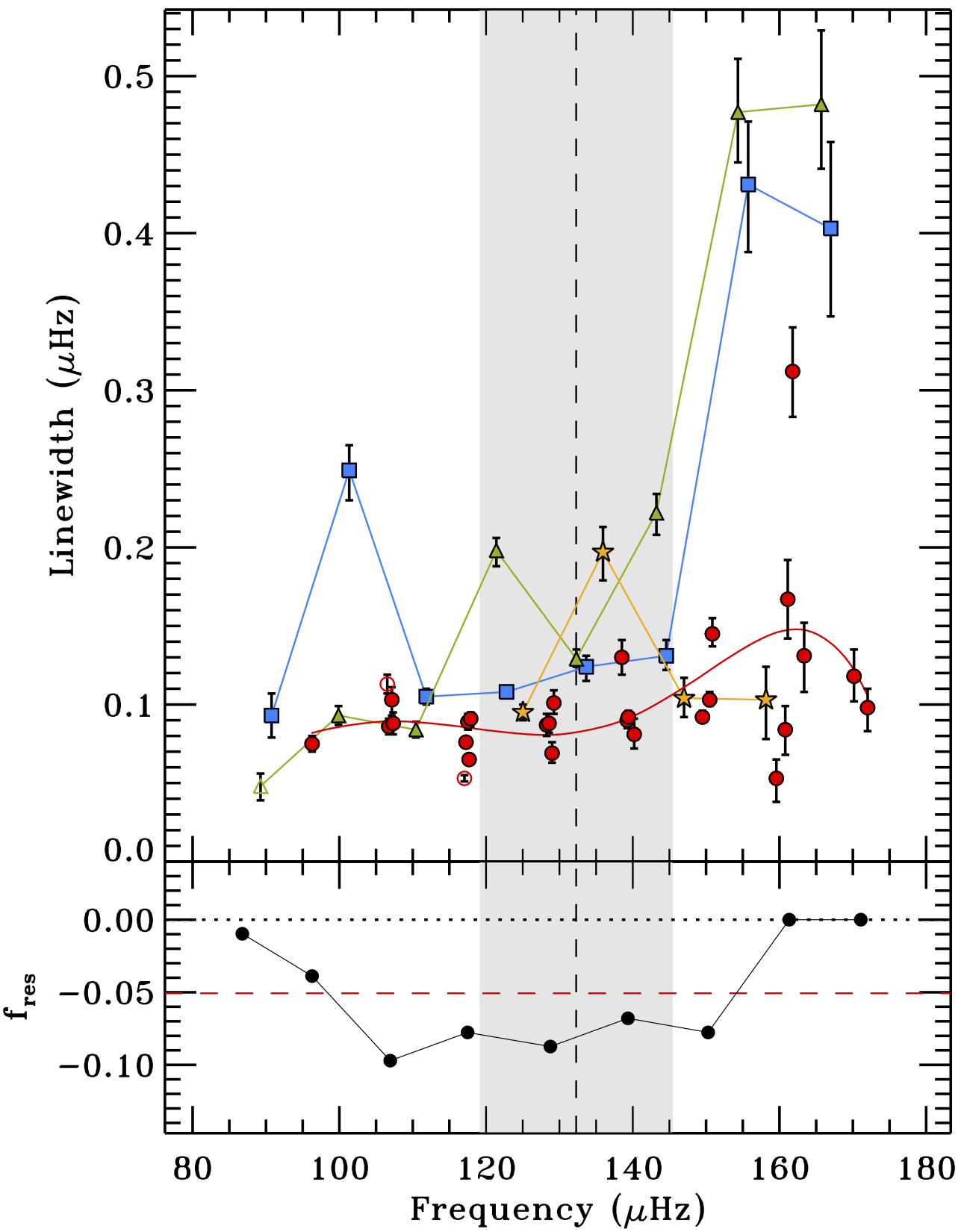}
      \caption{Mode linewidths for KIC~7060732 as a function of the corresponding oscillation frequencies. \textit{Top panel}: linewidth measurements as defined by Eq.~(\ref{eq:resolved_profile}) for each angular degree ($\ell = 0$ blue squares, $\ell = 2$ green triangles, $\ell = 3$ yellow stars, and resolved $\ell = 1$ mixed modes red circles). Open symbols represent modes with detection probability under the suggested threshold (see Sect.~\ref{sec:test}). The 68\,\% credible intervals for the linewidths as derived by \diamonds\,\,are shown for each data point. The red solid line represents a polynomial fit to the linewidths of the $\ell = 1$ mixed modes, included to emphasize the trend with frequency. The shaded region represents the range $\numax \pm \sigma_\mathrm{env}$, with $\numax$ from Table~\ref{tab:bkg2} indicated by the dashed vertical line. \textit{Bottom panel}: the normalized fraction of resolved mixed modes with respect to unresolved ones, $f_\mathrm{res}$ (black dots), defined by Eq.~(\ref{eq:fraction_resolved}). The frequency position of each point is the average frequency of the resolved dipole mixed modes falling in each radial order (or that of the unresolved mixed modes if no resolved mixed modes are present). The horizontal dotted line represents the limit of resolved-dominated regime, as defined in Sect.~\ref{sec:fwhm}, while the horizontal dashed red line marks the average $f_\mathrm{res}$ given by Eq.~(\ref{eq:average_fraction}).}
    \label{fig:7060732fwhm}
\end{figure}

\begin{figure}
   \centering
   \includegraphics[width=9.0cm]{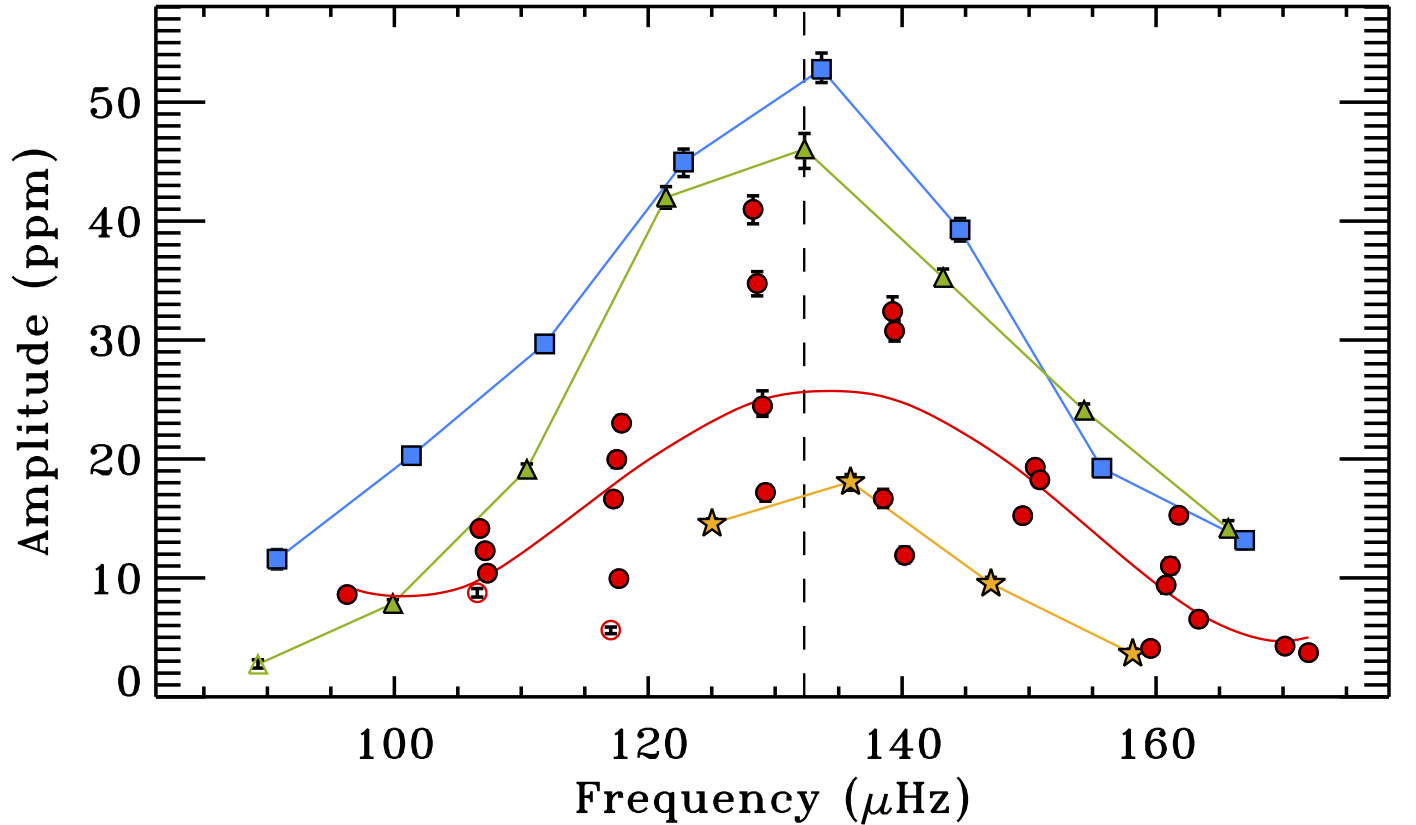}
      \caption{Mode amplitudes for KIC~7060732 as a function of the corresponding oscillation frequencies. Amplitude measurements as defined by Eq.~(\ref{eq:resolved_profile}) for each angular degree ($\ell = 0$ blue squares, $\ell = 2$ green triangles, $\ell = 3$ yellow stars, and resolved $\ell = 1$ mixed modes red circles). Open symbols represent modes with detection probability under the suggested threshold (see Sect.~\ref{sec:test}). The 68\,\% credible intervals for the amplitudes as derived by \diamonds\,\,are shown for each data point. The solid red line represents a polynomial fit to the amplitudes of the $\ell = 1$ mixed modes, included to emphasize the trend with frequency. The dashed vertical line indicates the $\numax$ value listed in Table~\ref{tab:bkg2}.}
    \label{fig:7060732amplitude}
\end{figure}
\clearpage

\begin{figure}
   \centering
   \includegraphics[width=9.0cm]{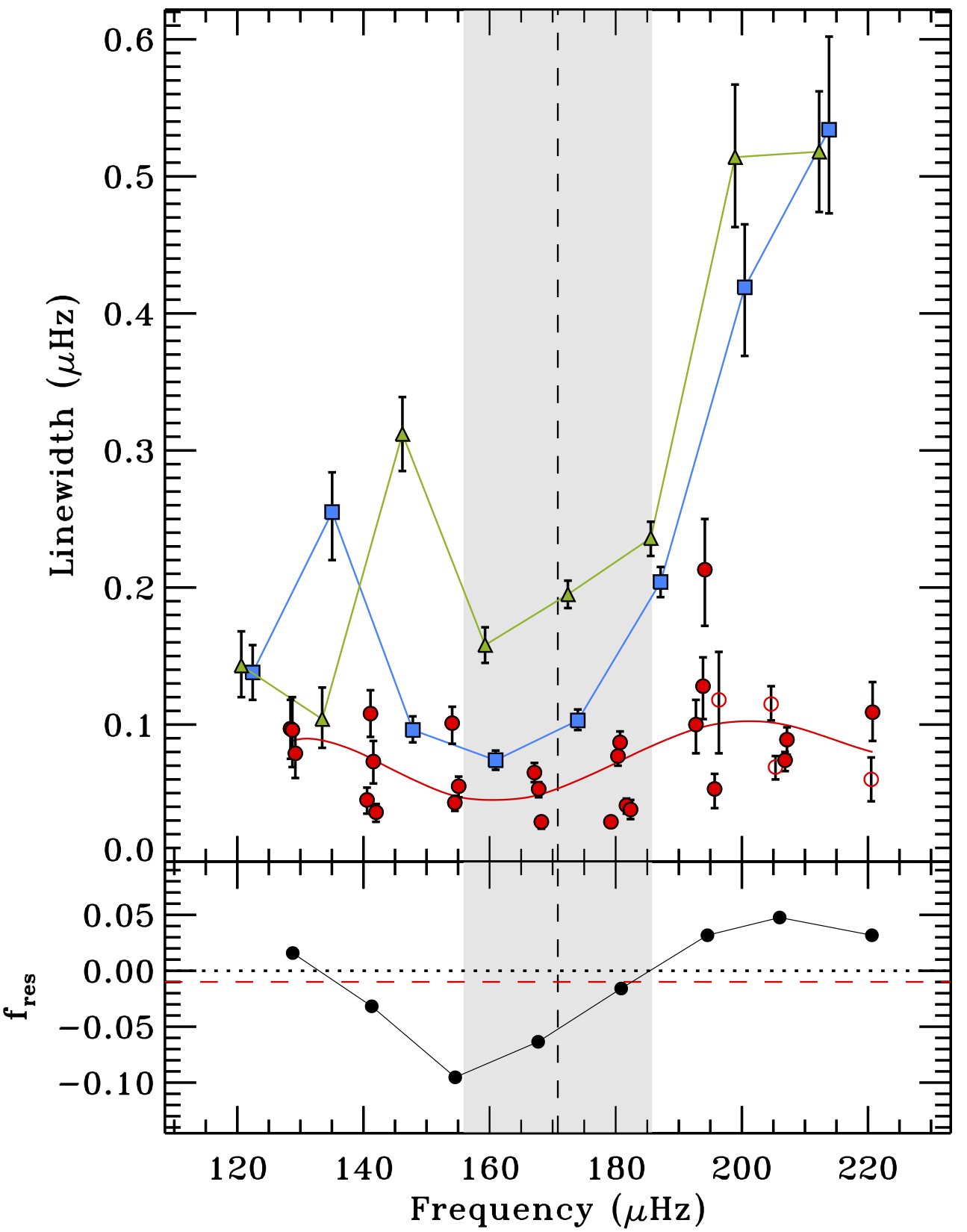}
      \caption{Mode linewidths for KIC~7619745 as a function of the corresponding oscillation frequencies. \textit{Top panel}: linewidth measurements as defined by Eq.~(\ref{eq:resolved_profile}) for each angular degree ($\ell = 0$ blue squares, $\ell = 2$ green triangles, $\ell = 3$ yellow stars, and resolved $\ell = 1$ mixed modes red circles). Open symbols represent modes with detection probability under the suggested threshold (see Sect.~\ref{sec:test}). The 68\,\% credible intervals for the linewidths as derived by \diamonds\,\,are shown for each data point. The red solid line represents a polynomial fit to the linewidths of the $\ell = 1$ mixed modes, included to emphasize the trend with frequency. The shaded region represents the range $\numax \pm \sigma_\mathrm{env}$, with $\numax$ from Table~\ref{tab:bkg2} indicated by the dashed vertical line. \textit{Bottom panel}: the normalized fraction of resolved mixed modes with respect to unresolved ones, $f_\mathrm{res}$ (black dots), defined by Eq.~(\ref{eq:fraction_resolved}). The frequency position of each point is the average frequency of the resolved dipole mixed modes falling in each radial order (or that of the unresolved mixed modes if no resolved mixed modes are present). The horizontal dotted line represents the limit of resolved-dominated regime, as defined in Sect.~\ref{sec:fwhm}, while the horizontal dashed red line marks the average $f_\mathrm{res}$ given by Eq.~(\ref{eq:average_fraction}).}
    \label{fig:7619745fwhm}
\end{figure}

\begin{figure}
   \centering
   \includegraphics[width=9.0cm]{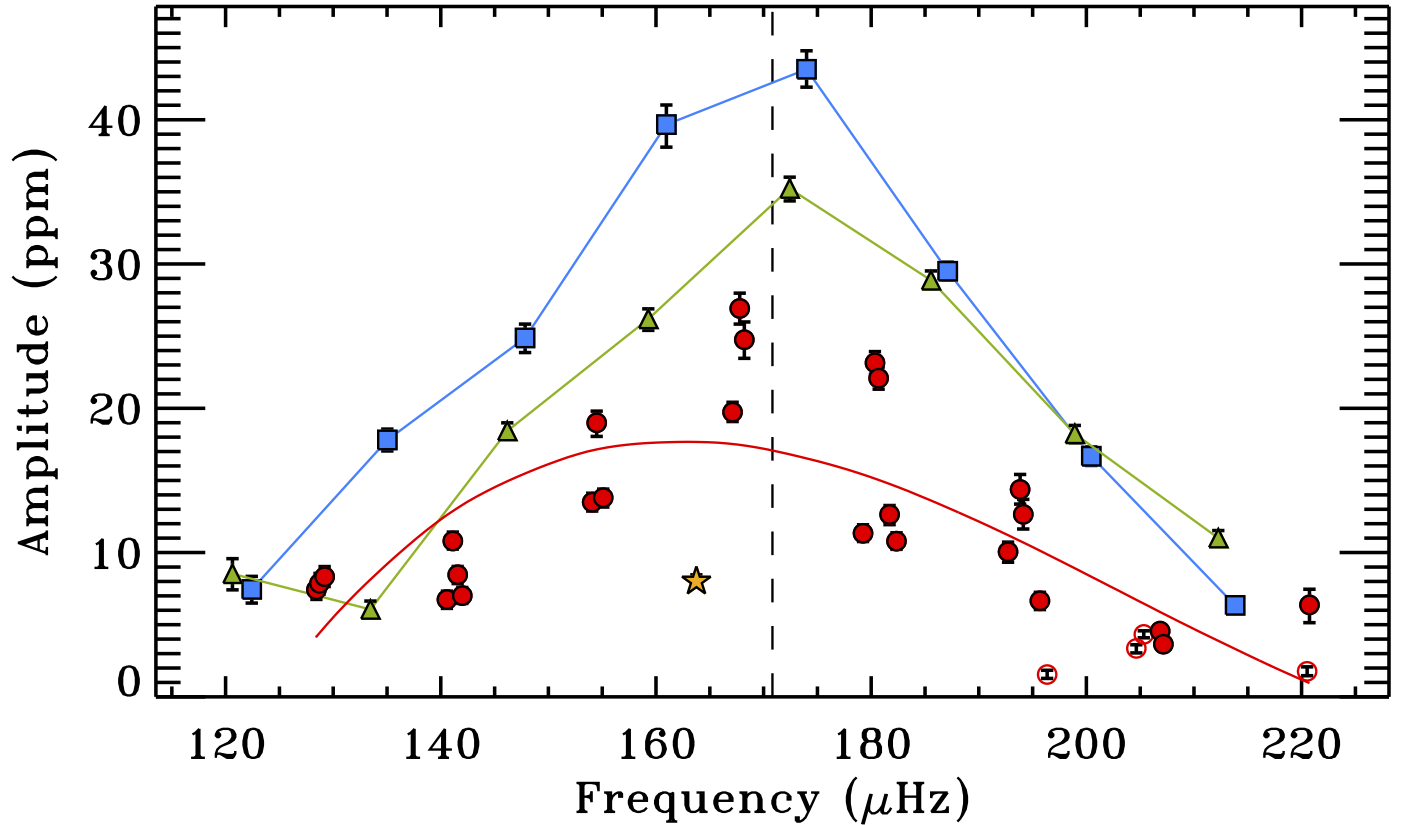}
      \caption{Mode amplitudes for KIC~7619745 as a function of the corresponding oscillation frequencies. Amplitude measurements as defined by Eq.~(\ref{eq:resolved_profile}) for each angular degree ($\ell = 0$ blue squares, $\ell = 2$ green triangles, $\ell = 3$ yellow stars, and resolved $\ell = 1$ mixed modes red circles). Open symbols represent modes with detection probability under the suggested threshold (see Sect.~\ref{sec:test}). The 68\,\% credible intervals for the amplitudes as derived by \diamonds\,\,are shown for each data point. The solid red line represents a polynomial fit to the amplitudes of the $\ell = 1$ mixed modes, included to emphasize the trend with frequency. The dashed vertical line indicates the $\numax$ value listed in Table~\ref{tab:bkg2}.}
    \label{fig:7619745amplitude}
\end{figure}
\clearpage

\begin{figure}
   \centering
   \includegraphics[width=9.0cm]{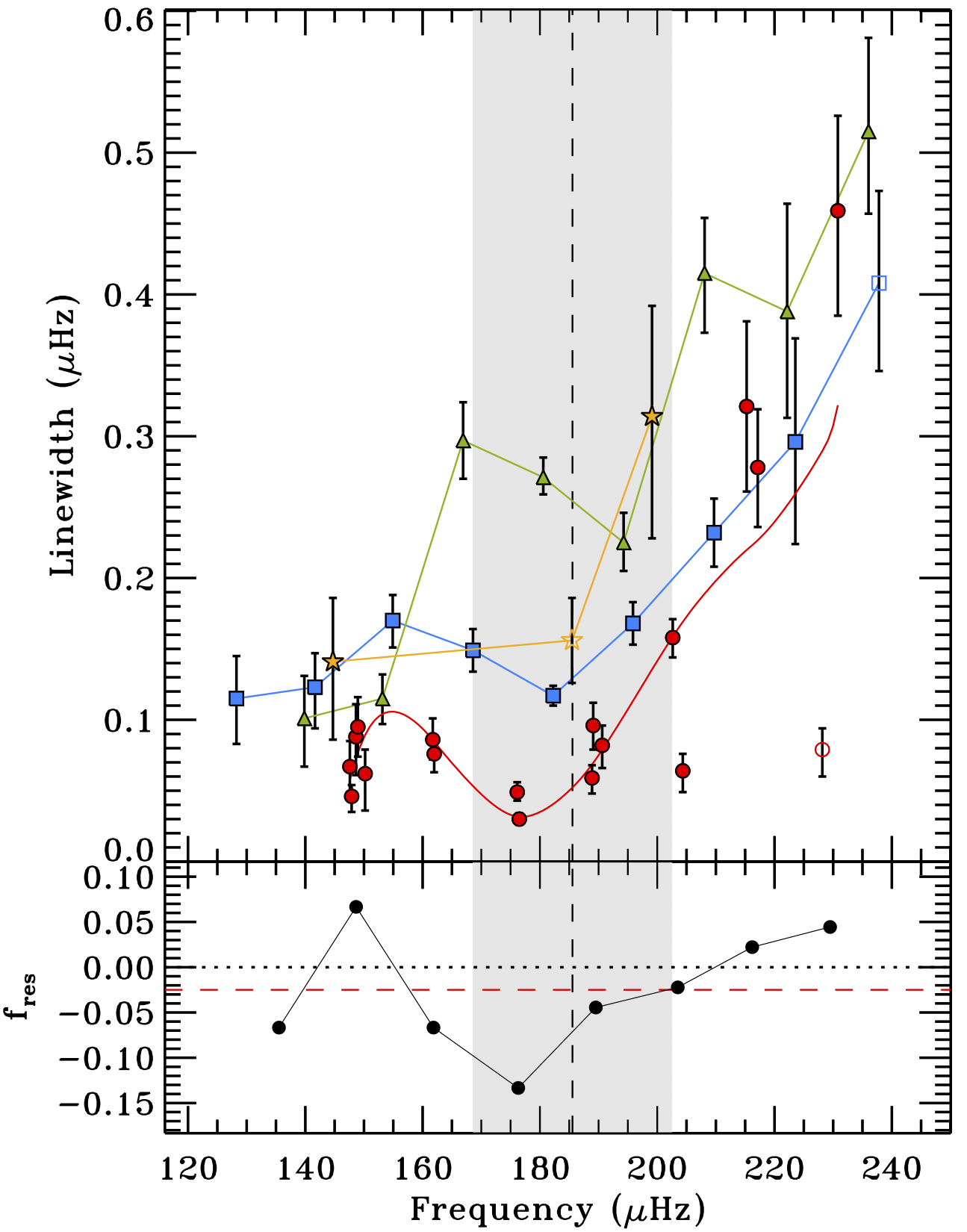}
      \caption{Mode linewidths for KIC~8366239 as a function of the corresponding oscillation frequencies. \textit{Top panel}: linewidth measurements as defined by Eq.~(\ref{eq:resolved_profile}) for each angular degree ($\ell = 0$ blue squares, $\ell = 2$ green triangles, $\ell = 3$ yellow stars, and resolved $\ell = 1$ mixed modes red circles). Open symbols represent modes with detection probability under the suggested threshold (see Sect.~\ref{sec:test}). The 68\,\% credible intervals for the linewidths as derived by \diamonds\,\,are shown for each data point. The red solid line represents a polynomial fit to the linewidths of the $\ell = 1$ mixed modes, included to emphasize the trend with frequency. The shaded region represents the range $\numax \pm \sigma_\mathrm{env}$, with $\numax$ from Table~\ref{tab:bkg2} indicated by the dashed vertical line. \textit{Bottom panel}: the normalized fraction of resolved mixed modes with respect to unresolved ones, $f_\mathrm{res}$ (black dots), defined by Eq.~(\ref{eq:fraction_resolved}). The frequency position of each point is the average frequency of the resolved dipole mixed modes falling in each radial order (or that of the unresolved mixed modes if no resolved mixed modes are present). The horizontal dotted line represents the limit of resolved-dominated regime, as defined in Sect.~\ref{sec:fwhm}, while the horizontal dashed red line marks the average $f_\mathrm{res}$ given by Eq.~(\ref{eq:average_fraction}).}
    \label{fig:8366239fwhm}
\end{figure}

\begin{figure}
   \centering
   \includegraphics[width=9.0cm]{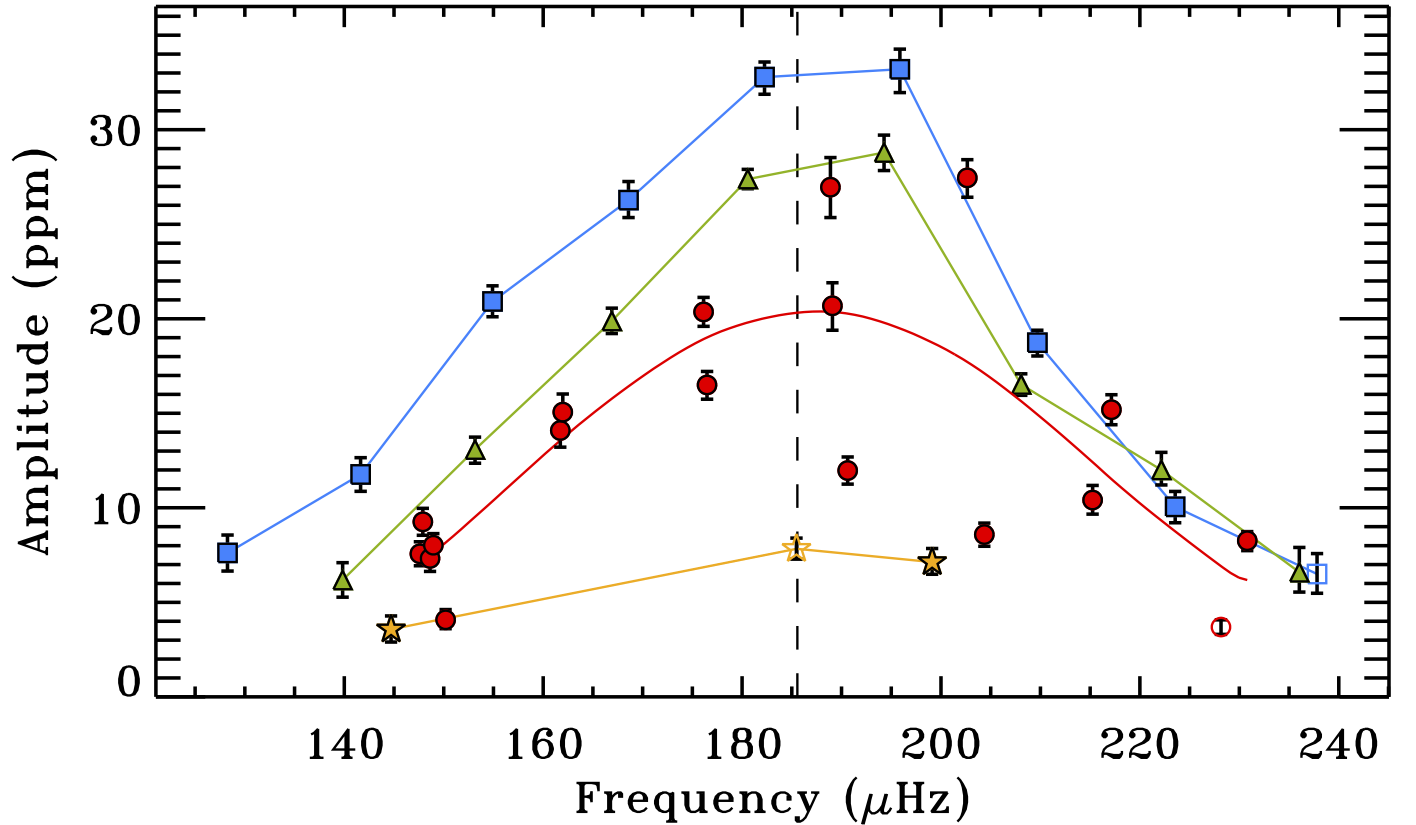}
      \caption{Mode amplitudes for KIC~8366239 as a function of the corresponding oscillation frequencies. Amplitude measurements as defined by Eq.~(\ref{eq:resolved_profile}) for each angular degree ($\ell = 0$ blue squares, $\ell = 2$ green triangles, $\ell = 3$ yellow stars, and resolved $\ell = 1$ mixed modes red circles). Open symbols represent modes with detection probability under the suggested threshold (see Sect.~\ref{sec:test}). The 68\,\% credible intervals for the amplitudes as derived by \diamonds\,\,are shown for each data point. The solid red line represents a polynomial fit to the amplitudes of the $\ell = 1$ mixed modes, included to emphasize the trend with frequency. The dashed vertical line indicates the $\numax$ value listed in Table~\ref{tab:bkg2}.}
    \label{fig:8366239amplitude}
\end{figure}
\clearpage

\begin{figure}
   \centering
   \includegraphics[width=9.0cm]{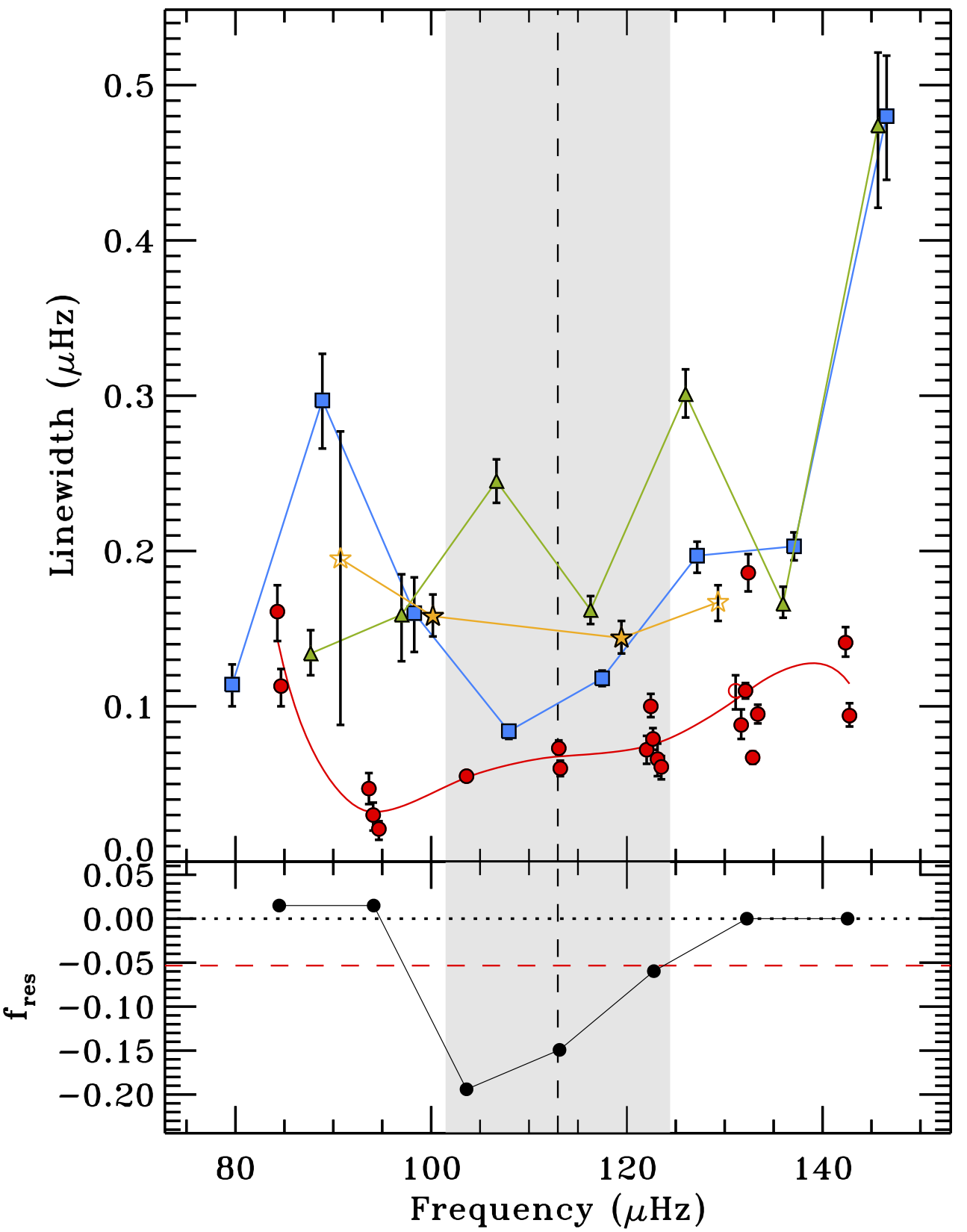}
      \caption{Mode linewidths for KIC~8475025 as a function of the corresponding oscillation frequencies. \textit{Top panel}: linewidth measurements as defined by Eq.~(\ref{eq:resolved_profile}) for each angular degree ($\ell = 0$ blue squares, $\ell = 2$ green triangles, $\ell = 3$ yellow stars, and resolved $\ell = 1$ mixed modes red circles). Open symbols represent modes with detection probability under the suggested threshold (see Sect.~\ref{sec:test}). The 68\,\% credible intervals for the linewidths as derived by \diamonds\,\,are shown for each data point. The red solid line represents a polynomial fit to the linewidths of the $\ell = 1$ mixed modes, included to emphasize the trend with frequency. The shaded region represents the range $\numax \pm \sigma_\mathrm{env}$, with $\numax$ from Table~\ref{tab:bkg2} indicated by the dashed vertical line. \textit{Bottom panel}: the normalized fraction of resolved mixed modes with respect to unresolved ones, $f_\mathrm{res}$ (black dots), defined by Eq.~(\ref{eq:fraction_resolved}). The frequency position of each point is the average frequency of the resolved dipole mixed modes falling in each radial order (or that of the unresolved mixed modes if no resolved mixed modes are present). The horizontal dotted line represents the limit of resolved-dominated regime, as defined in Sect.~\ref{sec:fwhm}, while the horizontal dashed red line marks the average $f_\mathrm{res}$ given by Eq.~(\ref{eq:average_fraction}).}
    \label{fig:8475025fwhm}
\end{figure}

\begin{figure}
   \centering
   \includegraphics[width=9.0cm]{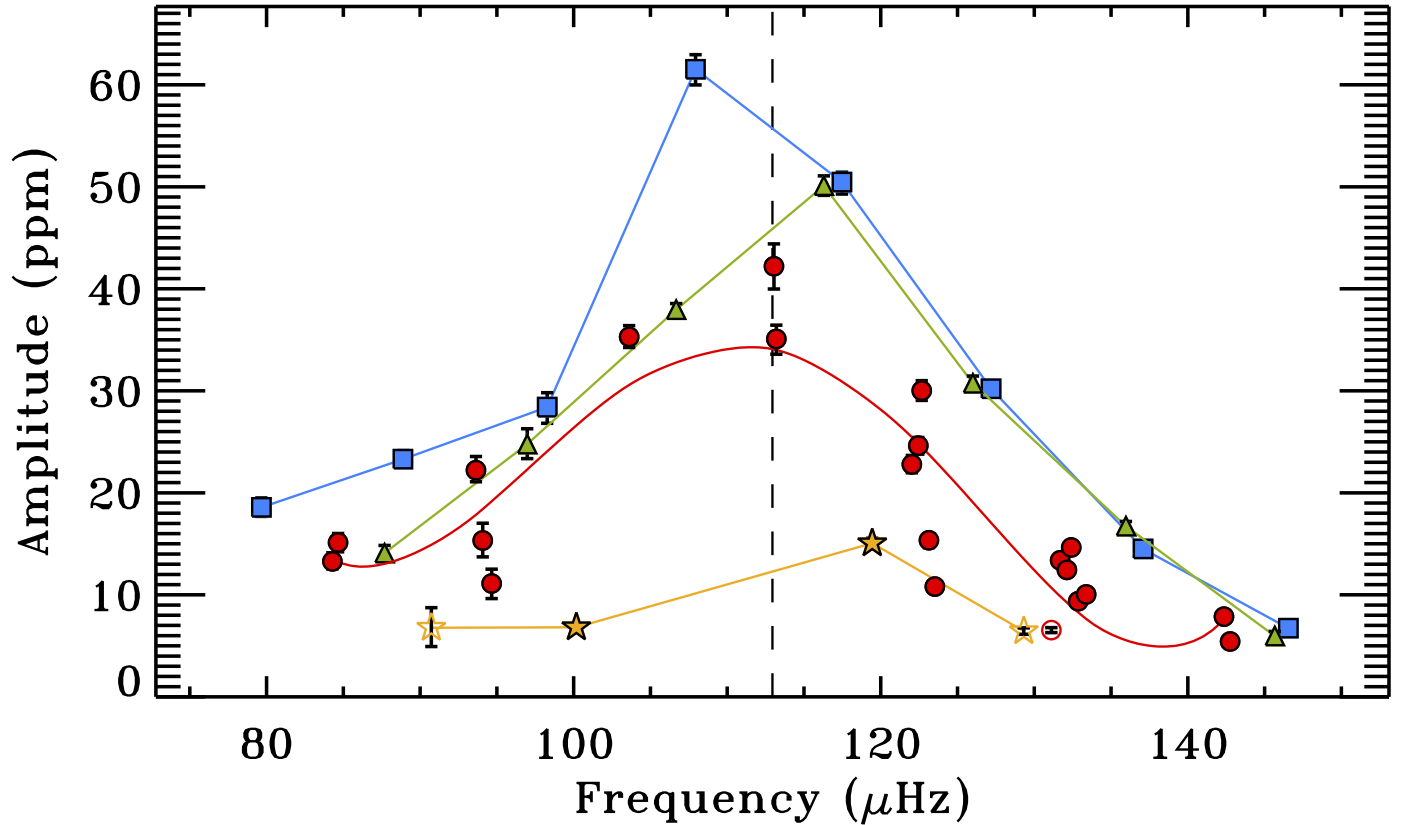}
      \caption{Mode amplitudes for KIC~8475025 as a function of the corresponding oscillation frequencies. Amplitude measurements as defined by Eq.~(\ref{eq:resolved_profile}) for each angular degree ($\ell = 0$ blue squares, $\ell = 2$ green triangles, $\ell = 3$ yellow stars, and resolved $\ell = 1$ mixed modes red circles). Open symbols represent modes with detection probability under the suggested threshold (see Sect.~\ref{sec:test}). The 68\,\% credible intervals for the amplitudes as derived by \diamonds\,\,are shown for each data point. The solid red line represents a polynomial fit to the amplitudes of the $\ell = 1$ mixed modes, included to emphasize the trend with frequency. The dashed vertical line indicates the $\numax$ value listed in Table~\ref{tab:bkg2}.}
    \label{fig:8475025amplitude}
\end{figure}
\clearpage

\begin{figure}
   \centering
   \includegraphics[width=9.0cm]{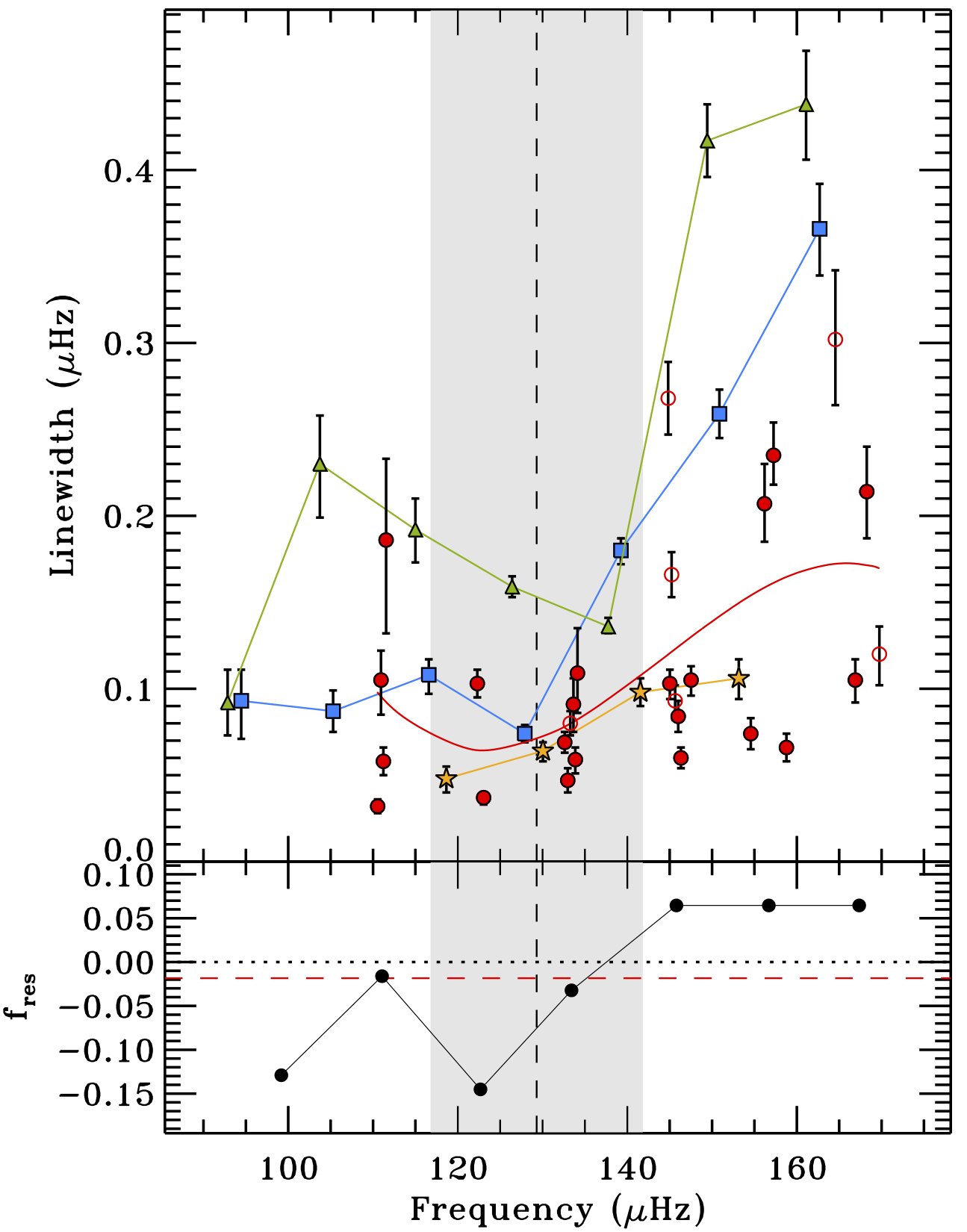}
      \caption{Mode linewidths for KIC~8718745 as a function of the corresponding oscillation frequencies. \textit{Top panel}: linewidth measurements as defined by Eq.~(\ref{eq:resolved_profile}) for each angular degree ($\ell = 0$ blue squares, $\ell = 2$ green triangles, $\ell = 3$ yellow stars, and resolved $\ell = 1$ mixed modes red circles). Open symbols represent modes with detection probability under the suggested threshold (see Sect.~\ref{sec:test}). The 68\,\% credible intervals for the linewidths as derived by \diamonds\,\,are shown for each data point. The red solid line represents a polynomial fit to the linewidths of the $\ell = 1$ mixed modes, included to emphasize the trend with frequency. The shaded region represents the range $\numax \pm \sigma_\mathrm{env}$, with $\numax$ from Table~\ref{tab:bkg2} indicated by the dashed vertical line. \textit{Bottom panel}: the normalized fraction of resolved mixed modes with respect to unresolved ones, $f_\mathrm{res}$ (black dots), defined by Eq.~(\ref{eq:fraction_resolved}). The frequency position of each point is the average frequency of the resolved dipole mixed modes falling in each radial order (or that of the unresolved mixed modes if no resolved mixed modes are present). The horizontal dotted line represents the limit of resolved-dominated regime, as defined in Sect.~\ref{sec:fwhm}, while the horizontal dashed red line marks the average $f_\mathrm{res}$ given by Eq.~(\ref{eq:average_fraction}).}
    \label{fig:8718745fwhm}
\end{figure}

\begin{figure}
   \centering
   \includegraphics[width=9.0cm]{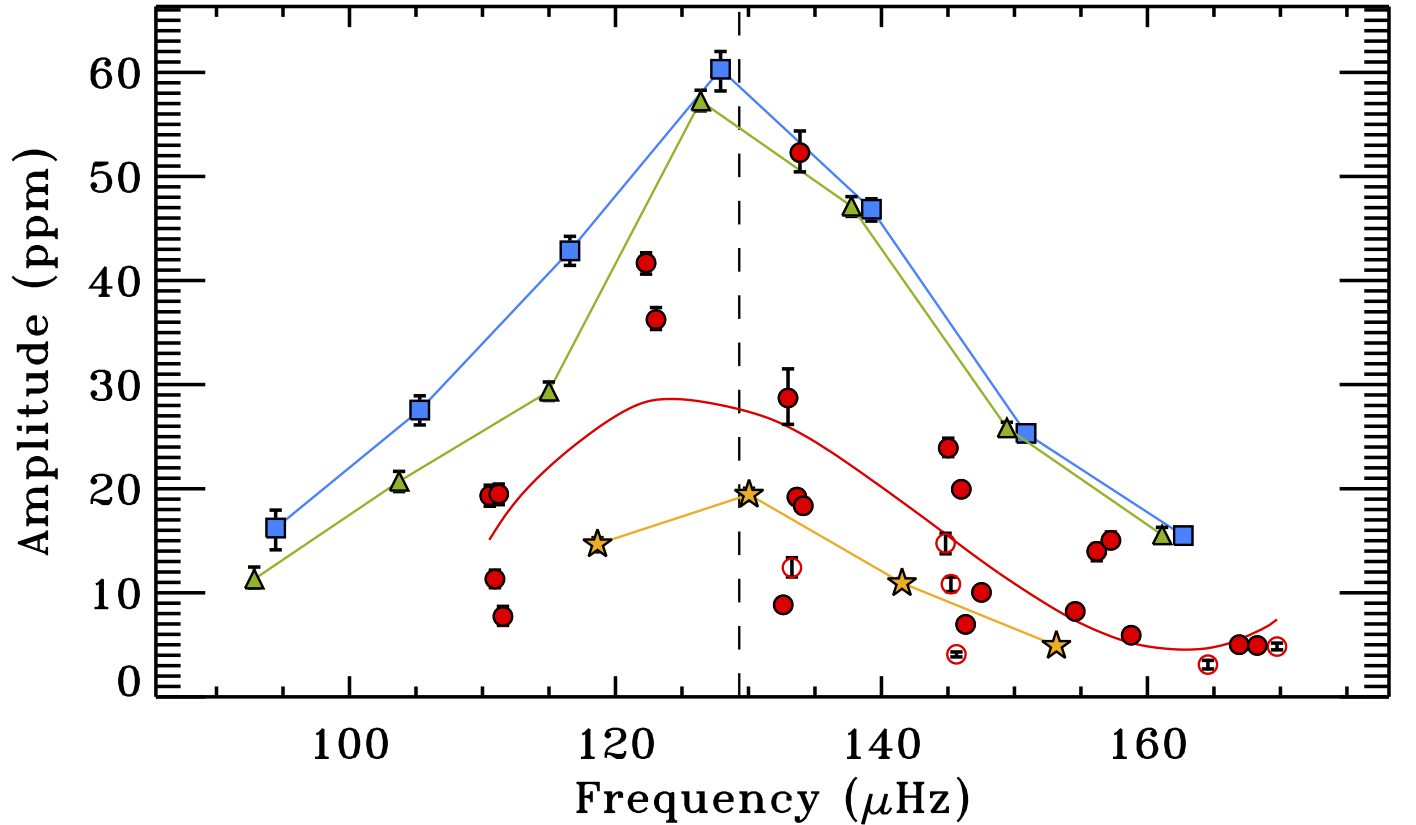}
      \caption{Mode amplitudes for KIC~8718745 as a function of the corresponding oscillation frequencies. Amplitude measurements as defined by Eq.~(\ref{eq:resolved_profile}) for each angular degree ($\ell = 0$ blue squares, $\ell = 2$ green triangles, $\ell = 3$ yellow stars, and resolved $\ell = 1$ mixed modes red circles). Open symbols represent modes with detection probability under the suggested threshold (see Sect.~\ref{sec:test}). The 68\,\% credible intervals for the amplitudes as derived by \diamonds\,\,are shown for each data point. The solid red line represents a polynomial fit to the amplitudes of the $\ell = 1$ mixed modes, included to emphasize the trend with frequency. The dashed vertical line indicates the $\numax$ value listed in Table~\ref{tab:bkg2}.}
    \label{fig:8718745amplitude}
\end{figure}
\clearpage

\begin{figure}
   \centering
   \includegraphics[width=9.0cm]{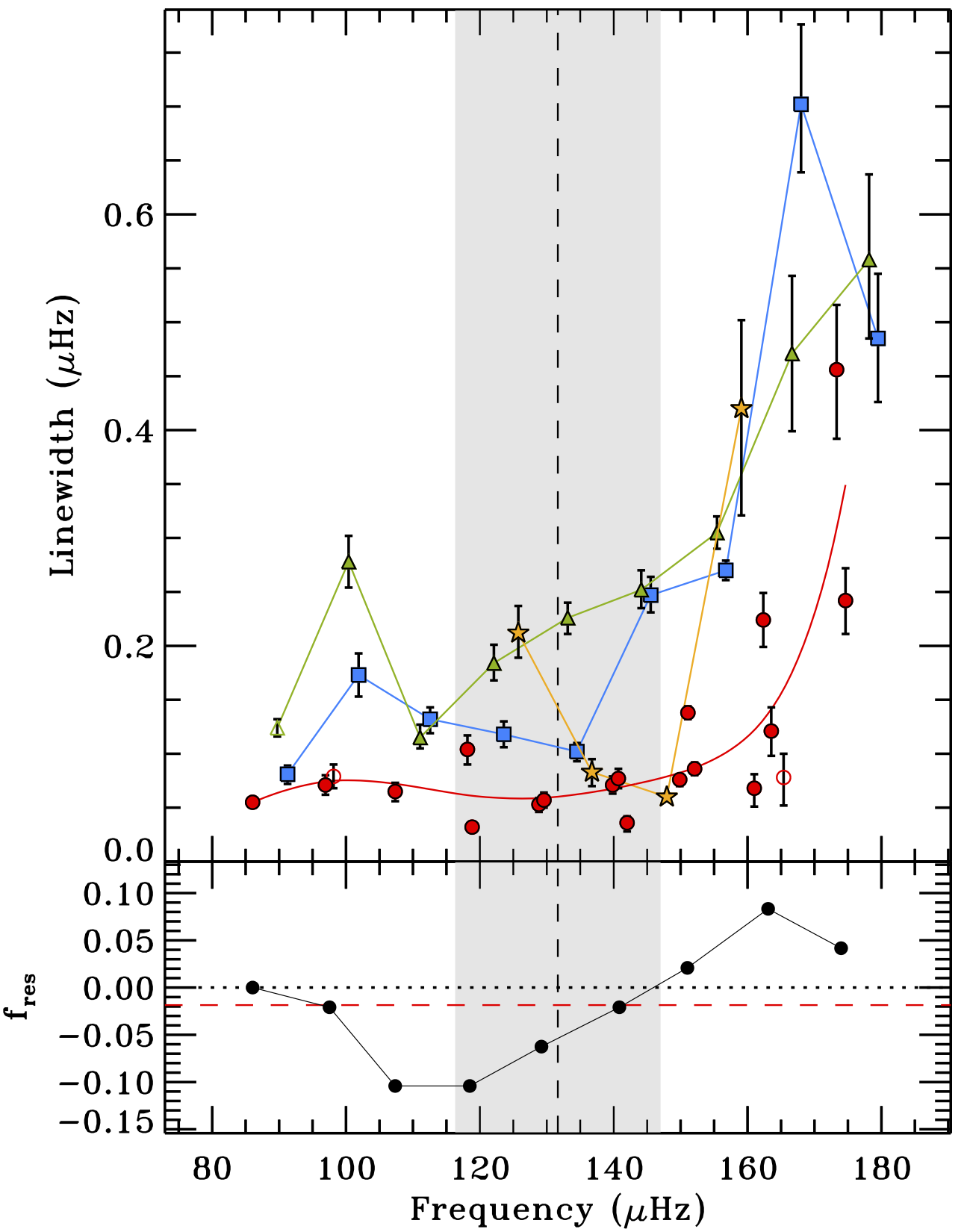}
      \caption{Mode linewidths for KIC~9145955 as a function of the corresponding oscillation frequencies. \textit{Top panel}: linewidth measurements as defined by Eq.~(\ref{eq:resolved_profile}) for each angular degree ($\ell = 0$ blue squares, $\ell = 2$ green triangles, $\ell = 3$ yellow stars, and resolved $\ell = 1$ mixed modes red circles). Open symbols represent modes with detection probability under the suggested threshold (see Sect.~\ref{sec:test}). The 68\,\% credible intervals for the linewidths as derived by \diamonds\,\,are shown for each data point. The red solid line represents a polynomial fit to the linewidths of the $\ell = 1$ mixed modes, included to emphasize the trend with frequency. The shaded region represents the range $\numax \pm \sigma_\mathrm{env}$, with $\numax$ from Table~\ref{tab:bkg2} indicated by the dashed vertical line. \textit{Bottom panel}: the normalized fraction of resolved mixed modes with respect to unresolved ones, $f_\mathrm{res}$ (black dots), defined by Eq.~(\ref{eq:fraction_resolved}). The frequency position of each point is the average frequency of the resolved dipole mixed modes falling in each radial order (or that of the unresolved mixed modes if no resolved mixed modes are present). The horizontal dotted line represents the limit of resolved-dominated regime, as defined in Sect.~\ref{sec:fwhm}, while the horizontal dashed red line marks the average $f_\mathrm{res}$ given by Eq.~(\ref{eq:average_fraction}).}
    \label{fig:9145955fwhm}
\end{figure}

\begin{figure}
   \centering
   \includegraphics[width=9.0cm]{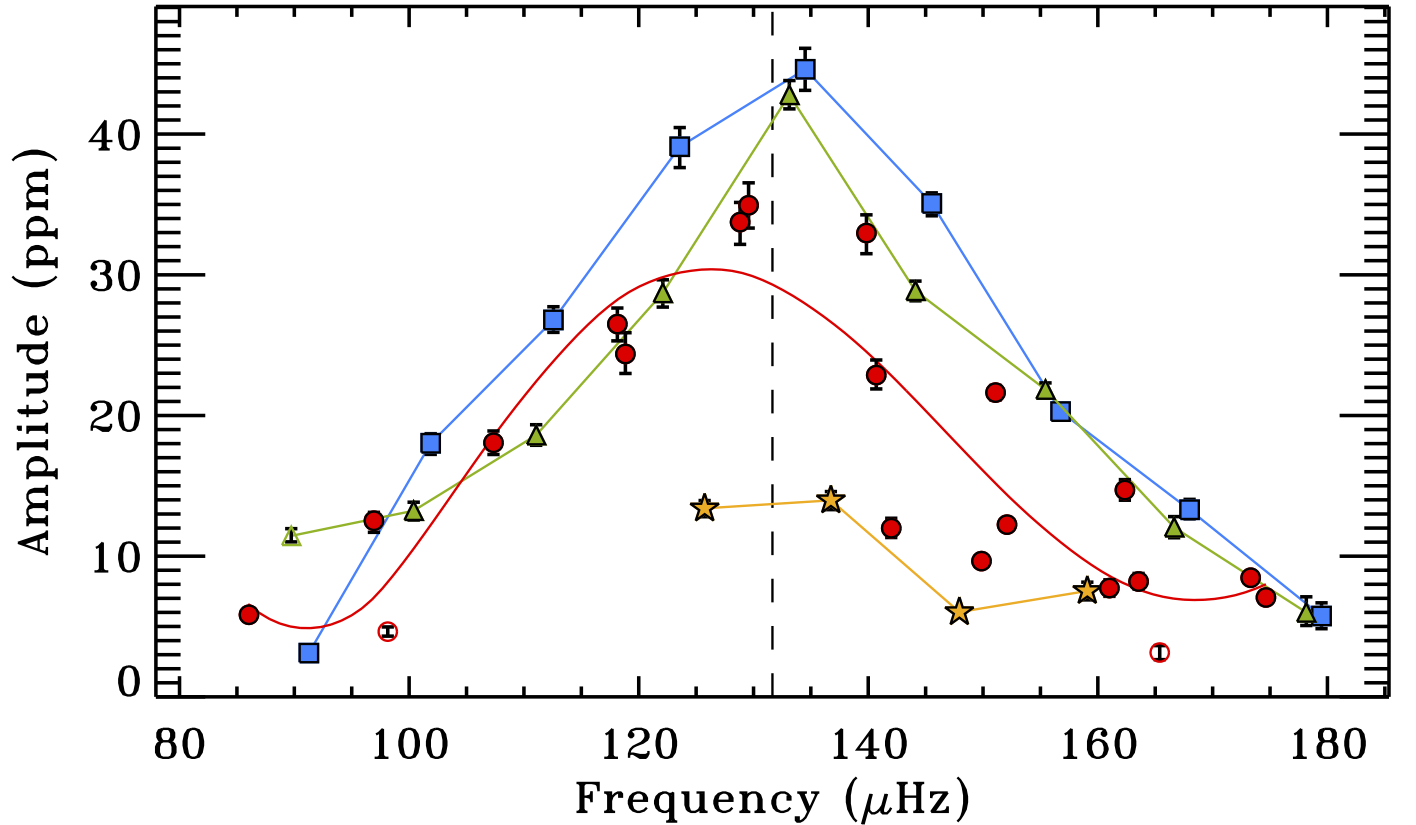}
      \caption{Mode amplitudes for KIC~9145955 as a function of the corresponding oscillation frequencies. Amplitude measurements as defined by Eq.~(\ref{eq:resolved_profile}) for each angular degree ($\ell = 0$ blue squares, $\ell = 2$ green triangles, $\ell = 3$ yellow stars, and resolved $\ell = 1$ mixed modes red circles). Open symbols represent modes with detection probability under the suggested threshold (see Sect.~\ref{sec:test}). The 68\,\% credible intervals for the amplitudes as derived by \diamonds\,\,are shown for each data point. The solid red line represents a polynomial fit to the amplitudes of the $\ell = 1$ mixed modes, included to emphasize the trend with frequency. The dashed vertical line indicates the $\numax$ value listed in Table~\ref{tab:bkg2}.}
    \label{fig:9145955amplitude}
\end{figure}
\clearpage

\begin{figure}
   \centering
   \includegraphics[width=9.0cm]{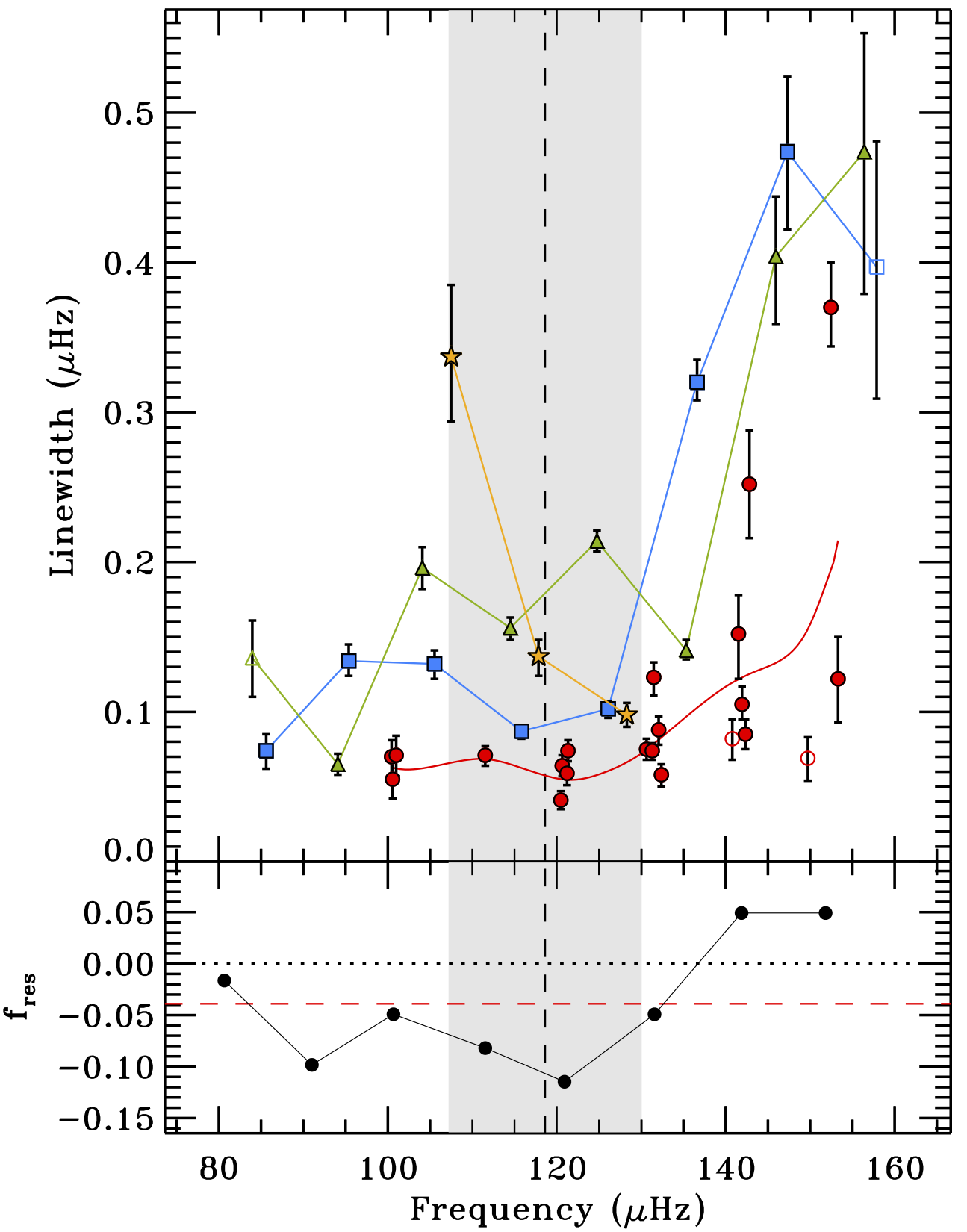}
      \caption{Mode linewidths for KIC~9267654 as a function of the corresponding oscillation frequencies. \textit{Top panel}: linewidth measurements as defined by Eq.~(\ref{eq:resolved_profile}) for each angular degree ($\ell = 0$ blue squares, $\ell = 2$ green triangles, $\ell = 3$ yellow stars, and resolved $\ell = 1$ mixed modes red circles). Open symbols represent modes with detection probability under the suggested threshold (see Sect.~\ref{sec:test}). The 68\,\% credible intervals for the linewidths as derived by \diamonds\,\,are shown for each data point. The red solid line represents a polynomial fit to the linewidths of the $\ell = 1$ mixed modes, included to emphasize the trend with frequency. The shaded region represents the range $\numax \pm \sigma_\mathrm{env}$, with $\numax$ from Table~\ref{tab:bkg2} indicated by the dashed vertical line. \textit{Bottom panel}: the normalized fraction of resolved mixed modes with respect to unresolved ones, $f_\mathrm{res}$ (black dots), defined by Eq.~(\ref{eq:fraction_resolved}). The frequency position of each point is the average frequency of the resolved dipole mixed modes falling in each radial order (or that of the unresolved mixed modes if no resolved mixed modes are present). The horizontal dotted line represents the limit of resolved-dominated regime, as defined in Sect.~\ref{sec:fwhm}, while the horizontal dashed red line marks the average $f_\mathrm{res}$ given by Eq.~(\ref{eq:average_fraction}).}
    \label{fig:9267654fwhm}
\end{figure}

\begin{figure}
   \centering
   \includegraphics[width=9.0cm]{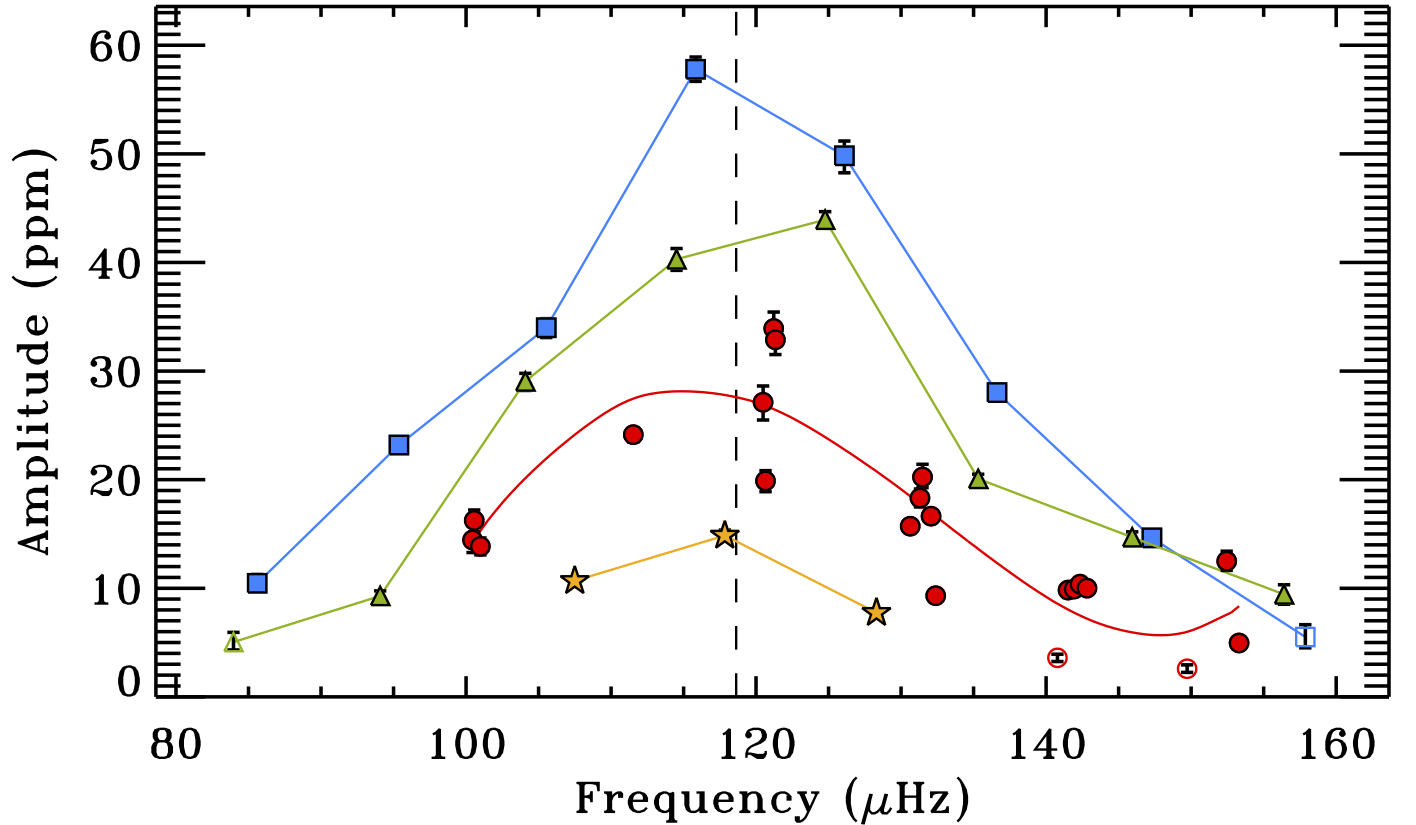}
      \caption{Mode amplitudes for KIC~9267654 as a function of the corresponding oscillation frequencies. Amplitude measurements as defined by Eq.~(\ref{eq:resolved_profile}) for each angular degree ($\ell = 0$ blue squares, $\ell = 2$ green triangles, $\ell = 3$ yellow stars, and resolved $\ell = 1$ mixed modes red circles). Open symbols represent modes with detection probability under the suggested threshold (see Sect.~\ref{sec:test}). The 68\,\% credible intervals for the amplitudes as derived by \diamonds\,\,are shown for each data point. The solid red line represents a polynomial fit to the amplitudes of the $\ell = 1$ mixed modes, included to emphasize the trend with frequency. The dashed vertical line indicates the $\numax$ value listed in Table~\ref{tab:bkg2}.}
    \label{fig:9267654amplitude}
\end{figure}
\clearpage

\begin{figure}
   \centering
   \includegraphics[width=9.0cm]{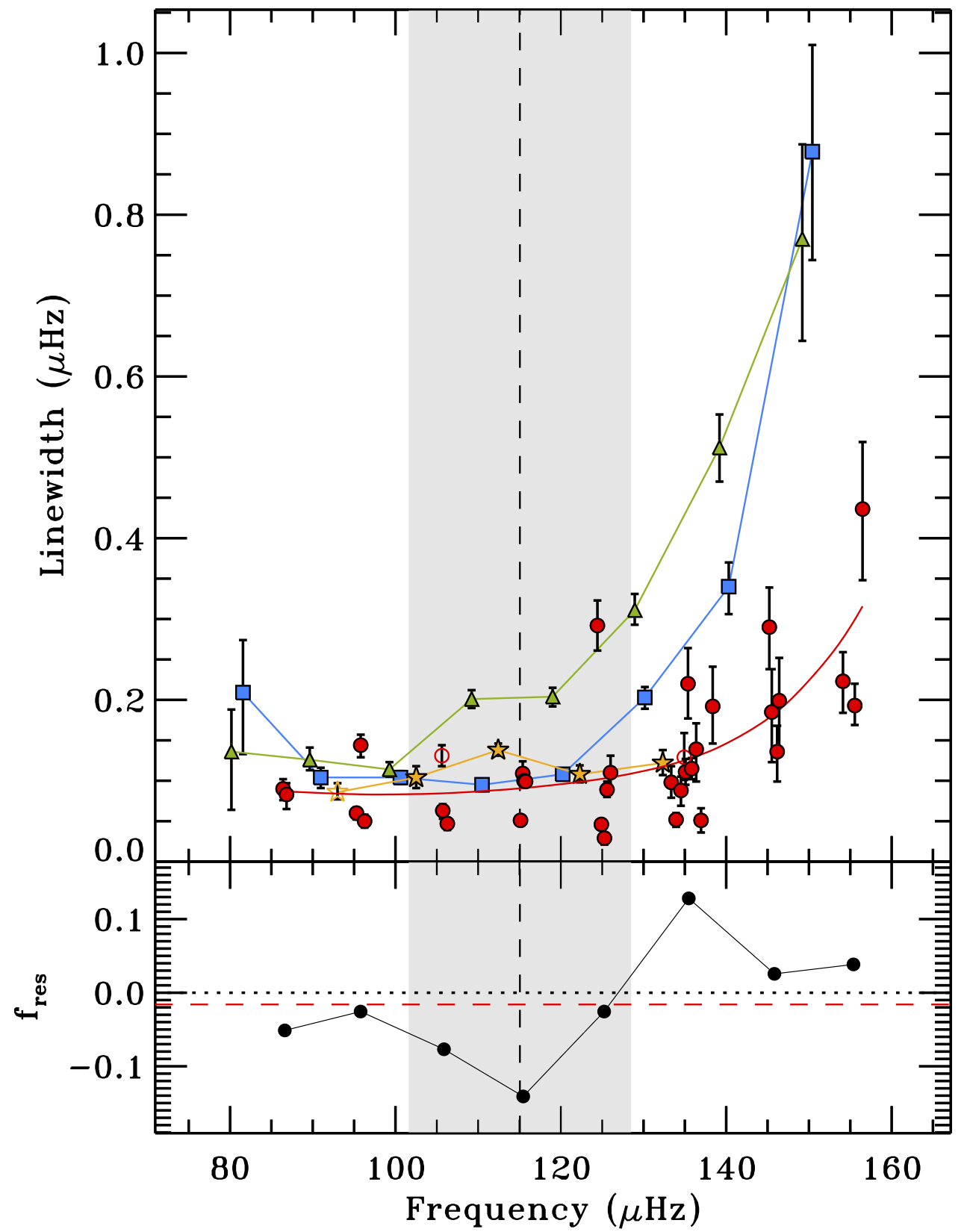}
      \caption{Mode linewidths for KIC~9475697 as a function of the corresponding oscillation frequencies. \textit{Top panel}: linewidth measurements as defined by Eq.~(\ref{eq:resolved_profile}) for each angular degree ($\ell = 0$ blue squares, $\ell = 2$ green triangles, $\ell = 3$ yellow stars, and resolved $\ell = 1$ mixed modes red circles). Open symbols represent modes with detection probability under the suggested threshold (see Sect.~\ref{sec:test}). The 68\,\% credible intervals for the linewidths as derived by \diamonds\,\,are shown for each data point. The red solid line represents a polynomial fit to the linewidths of the $\ell = 1$ mixed modes, included to emphasize the trend with frequency. The shaded region represents the range $\numax \pm \sigma_\mathrm{env}$, with $\numax$ from Table~\ref{tab:bkg2} indicated by the dashed vertical line. \textit{Bottom panel}: the normalized fraction of resolved mixed modes with respect to unresolved ones, $f_\mathrm{res}$ (black dots), defined by Eq.~(\ref{eq:fraction_resolved}). The frequency position of each point is the average frequency of the resolved dipole mixed modes falling in each radial order (or that of the unresolved mixed modes if no resolved mixed modes are present). The horizontal dotted line represents the limit of resolved-dominated regime, as defined in Sect.~\ref{sec:fwhm}, while the horizontal dashed red line marks the average $f_\mathrm{res}$ given by Eq.~(\ref{eq:average_fraction}).}
    \label{fig:9475697fwhm}
\end{figure}

\begin{figure}
   \centering
   \includegraphics[width=9.0cm]{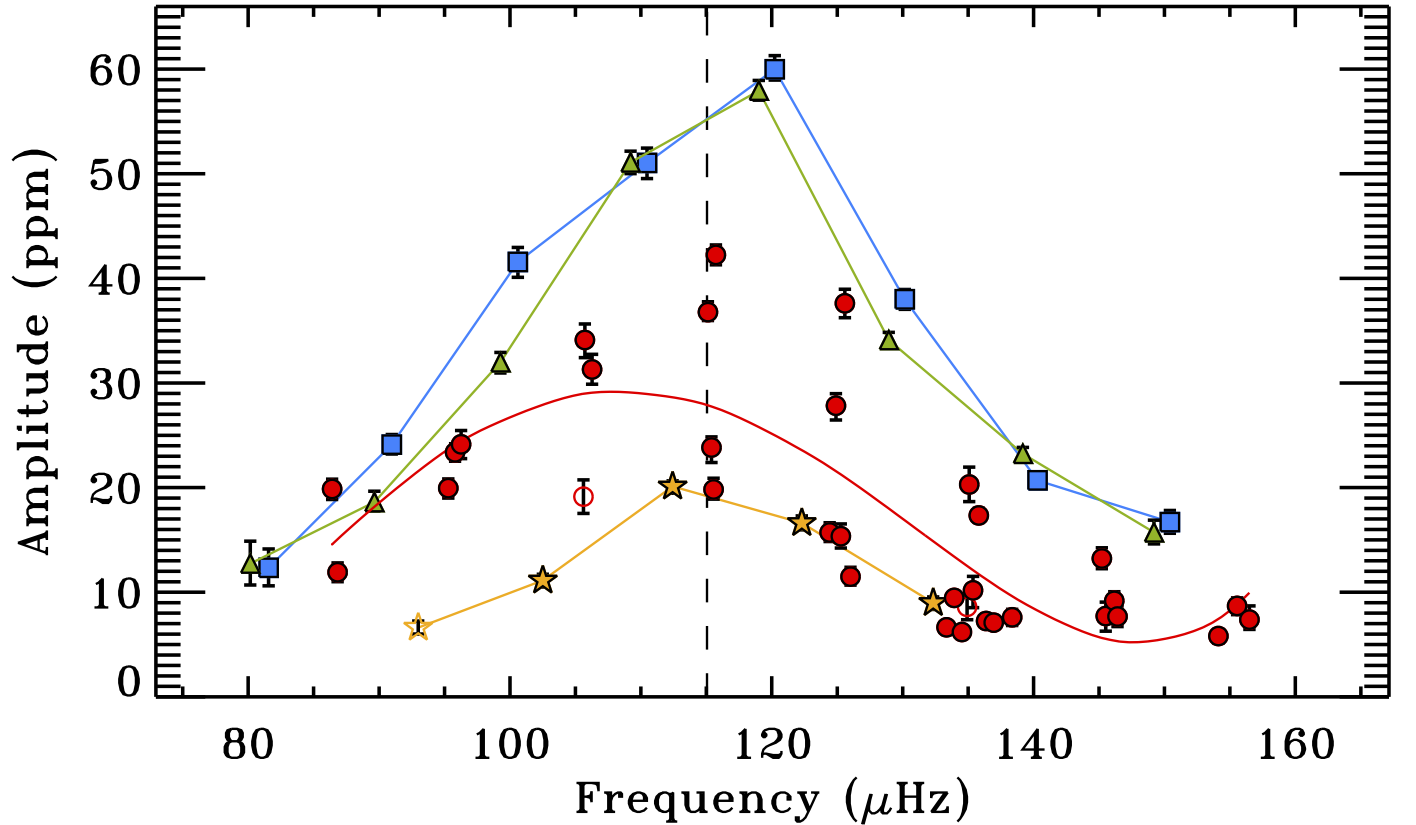}
      \caption{Mode amplitudes for KIC~9475697 as a function of the corresponding oscillation frequencies. Amplitude measurements as defined by Eq.~(\ref{eq:resolved_profile}) for each angular degree ($\ell = 0$ blue squares, $\ell = 2$ green triangles, $\ell = 3$ yellow stars, and resolved $\ell = 1$ mixed modes red circles). Open symbols represent modes with detection probability under the suggested threshold (see Sect.~\ref{sec:test}). The 68\,\% credible intervals for the amplitudes as derived by \diamonds\,\,are shown for each data point. The solid red line represents a polynomial fit to the amplitudes of the $\ell = 1$ mixed modes, included to emphasize the trend with frequency. The dashed vertical line indicates the $\numax$ value listed in Table~\ref{tab:bkg2}.}
    \label{fig:9475697amplitude}
\end{figure}
\clearpage

\begin{figure}
   \centering
   \includegraphics[width=9.0cm]{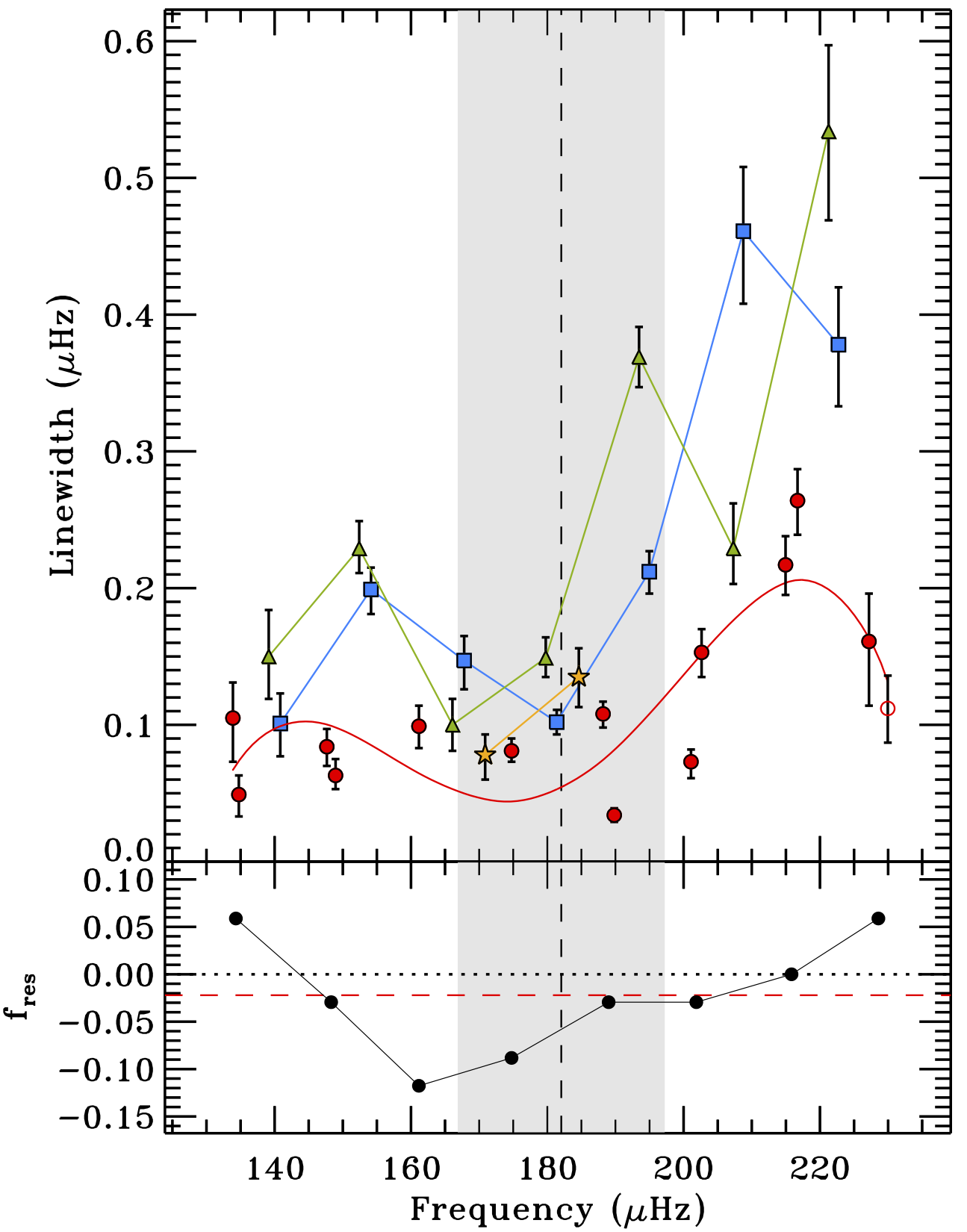}
      \caption{Mode linewidths for KIC~9882316 as a function of the corresponding oscillation frequencies. \textit{Top panel}: linewidth measurements as defined by Eq.~(\ref{eq:resolved_profile}) for each angular degree ($\ell = 0$ blue squares, $\ell = 2$ green triangles, $\ell = 3$ yellow stars, and resolved $\ell = 1$ mixed modes red circles). Open symbols represent modes with detection probability under the suggested threshold (see Sect.~\ref{sec:test}). The 68\,\% credible intervals for the linewidths as derived by \diamonds\,\,are shown for each data point. The red solid line represents a polynomial fit to the linewidths of the $\ell = 1$ mixed modes, included to emphasize the trend with frequency. The shaded region represents the range $\numax \pm \sigma_\mathrm{env}$, with $\numax$ from Table~\ref{tab:bkg2} indicated by the dashed vertical line. \textit{Bottom panel}: the normalized fraction of resolved mixed modes with respect to unresolved ones, $f_\mathrm{res}$ (black dots), defined by Eq.~(\ref{eq:fraction_resolved}). The frequency position of each point is the average frequency of the resolved dipole mixed modes falling in each radial order (or that of the unresolved mixed modes if no resolved mixed modes are present). The horizontal dotted line represents the limit of resolved-dominated regime, as defined in Sect.~\ref{sec:fwhm}, while the horizontal dashed red line marks the average $f_\mathrm{res}$ given by Eq.~(\ref{eq:average_fraction}).}
    \label{fig:9882316fwhm}
\end{figure}

\begin{figure}
   \centering
   \includegraphics[width=9.0cm]{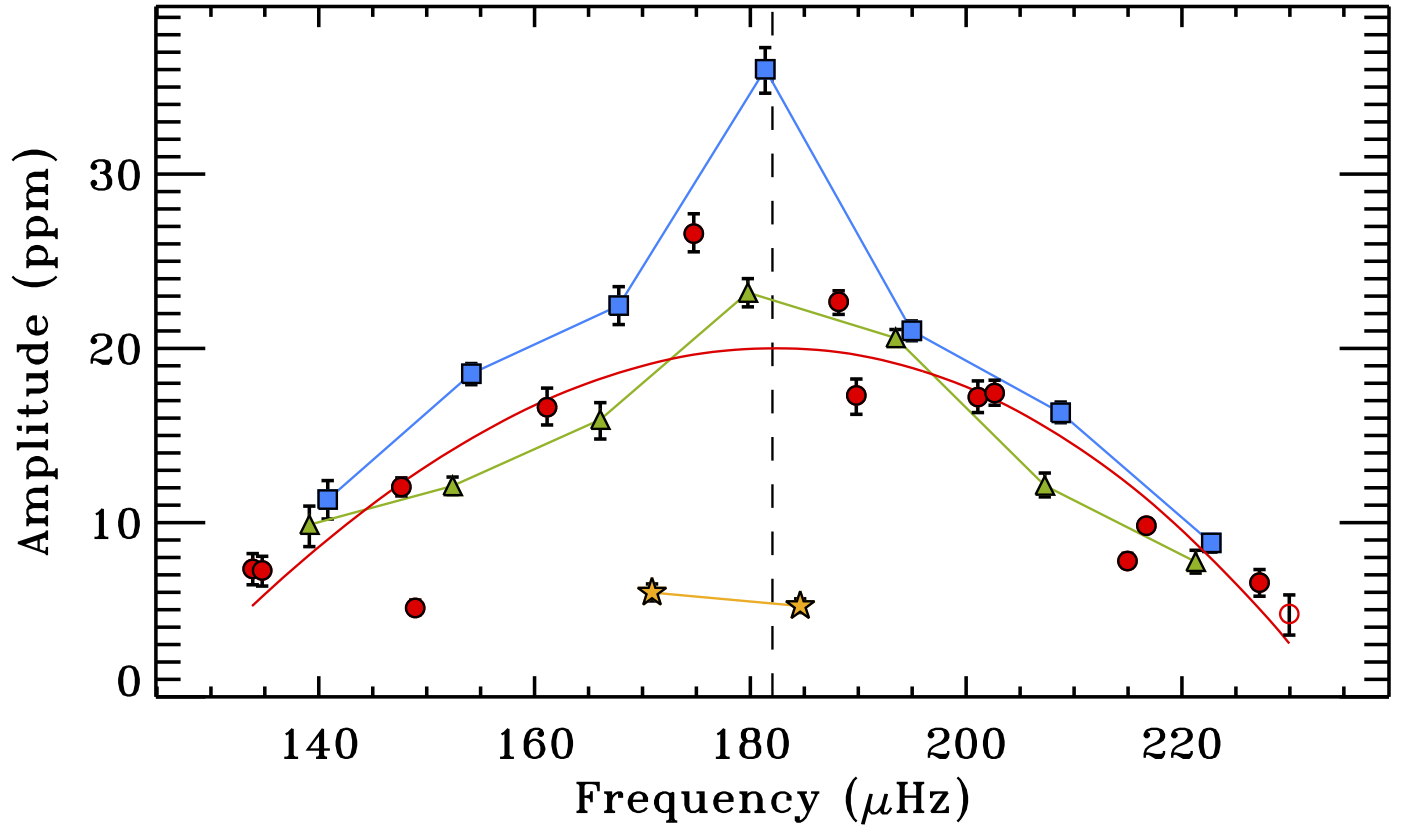}
      \caption{Mode amplitudes for KIC~9882316 as a function of the corresponding oscillation frequencies. Amplitude measurements as defined by Eq.~(\ref{eq:resolved_profile}) for each angular degree ($\ell = 0$ blue squares, $\ell = 2$ green triangles, $\ell = 3$ yellow stars, and resolved $\ell = 1$ mixed modes red circles). Open symbols represent modes with detection probability under the suggested threshold (see Sect.~\ref{sec:test}). The 68\,\% credible intervals for the amplitudes as derived by \diamonds\,\,are shown for each data point. The solid red line represents a polynomial fit to the amplitudes of the $\ell = 1$ mixed modes, included to emphasize the trend with frequency. The dashed vertical line indicates the $\numax$ value listed in Table~\ref{tab:bkg2}.}
    \label{fig:9882316amplitude}
\end{figure}
\clearpage

\begin{figure}
   \centering
   \includegraphics[width=9.0cm]{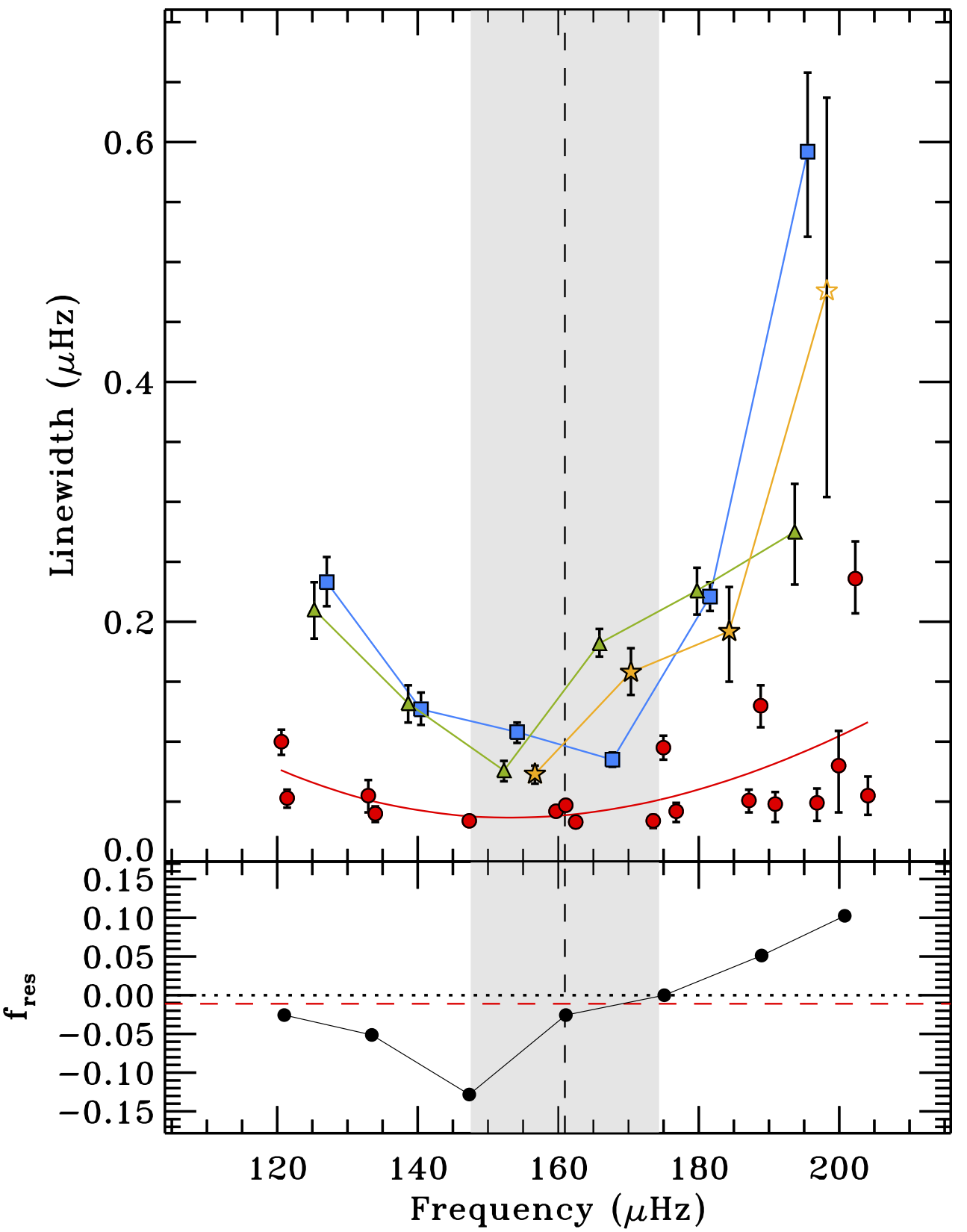}
      \caption{Mode linewidths for KIC~10123207 as a function of the corresponding oscillation frequencies. \textit{Top panel}: linewidth measurements as defined by Eq.~(\ref{eq:resolved_profile}) for each angular degree ($\ell = 0$ blue squares, $\ell = 2$ green triangles, $\ell = 3$ yellow stars, and resolved $\ell = 1$ mixed modes red circles). Open symbols represent modes with detection probability under the suggested threshold (see Sect.~\ref{sec:test}). The 68\,\% credible intervals for the linewidths as derived by \diamonds\,\,are shown for each data point. The red solid line represents a polynomial fit to the linewidths of the $\ell = 1$ mixed modes, included to emphasize the trend with frequency. The shaded region represents the range $\numax \pm \sigma_\mathrm{env}$, with $\numax$ from Table~\ref{tab:bkg2} indicated by the dashed vertical line. \textit{Bottom panel}: the normalized fraction of resolved mixed modes with respect to unresolved ones, $f_\mathrm{res}$ (black dots), defined by Eq.~(\ref{eq:fraction_resolved}). The frequency position of each point is the average frequency of the resolved dipole mixed modes falling in each radial order (or that of the unresolved mixed modes if no resolved mixed modes are present). The horizontal dotted line represents the limit of resolved-dominated regime, as defined in Sect.~\ref{sec:fwhm}, while the horizontal dashed red line marks the average $f_\mathrm{res}$ given by Eq.~(\ref{eq:average_fraction}).}
    \label{fig:10123207fwhm}
\end{figure}

\begin{figure}
   \centering
   \includegraphics[width=9.0cm]{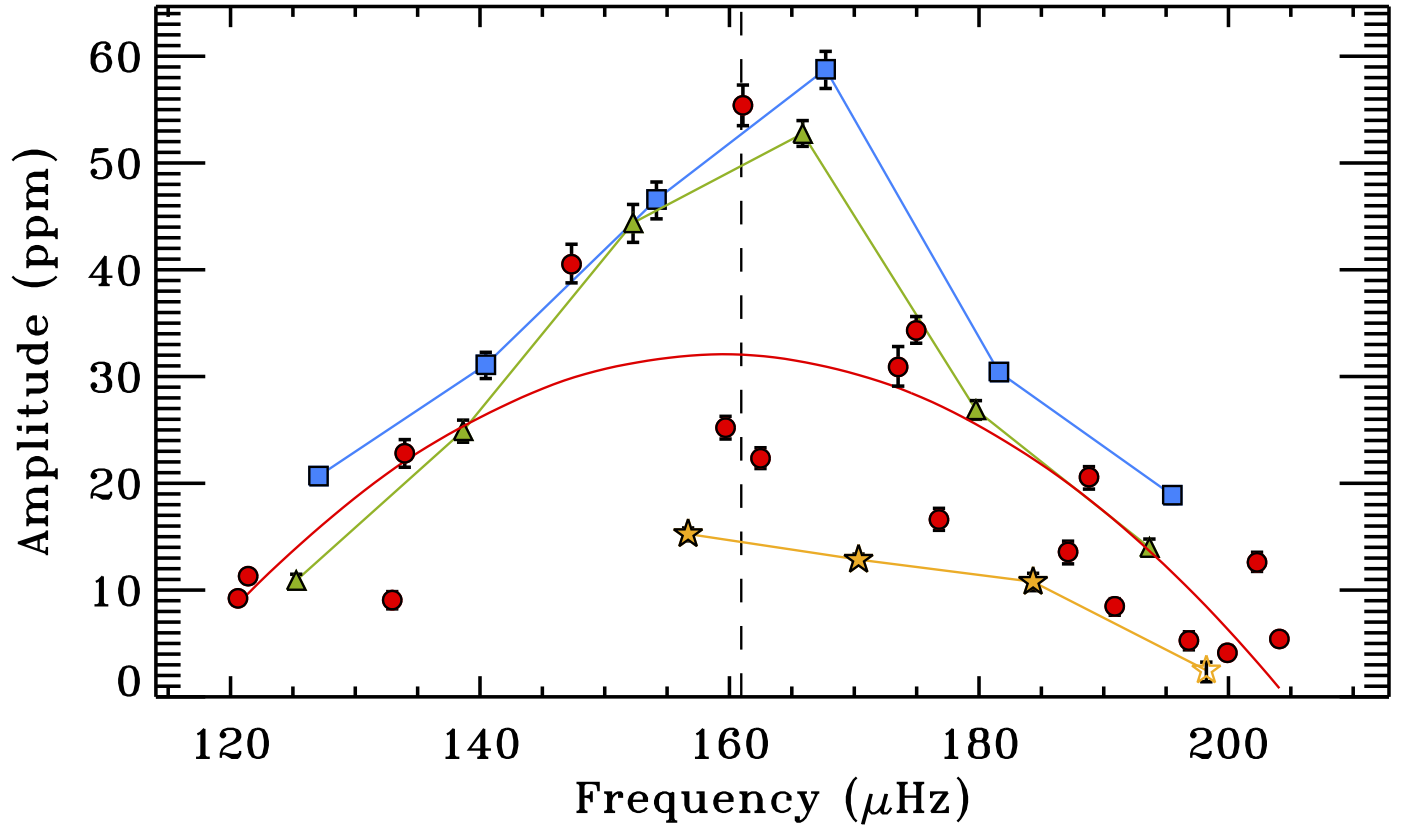}
      \caption{Mode amplitudes for KIC~10123207 as a function of the corresponding oscillation frequencies. Amplitude measurements as defined by Eq.~(\ref{eq:resolved_profile}) for each angular degree ($\ell = 0$ blue squares, $\ell = 2$ green triangles, $\ell = 3$ yellow stars, and resolved $\ell = 1$ mixed modes red circles). Open symbols represent modes with detection probability under the suggested threshold (see Sect.~\ref{sec:test}). The 68\,\% credible intervals for the amplitudes as derived by \diamonds\,\,are shown for each data point. The solid red line represents a polynomial fit to the amplitudes of the $\ell = 1$ mixed modes, included to emphasize the trend with frequency. The dashed vertical line indicates the $\numax$ value listed in Table~\ref{tab:bkg2}.}
    \label{fig:10123207amplitude}
\end{figure}
\clearpage

\begin{figure}
   \centering
   \includegraphics[width=9.0cm]{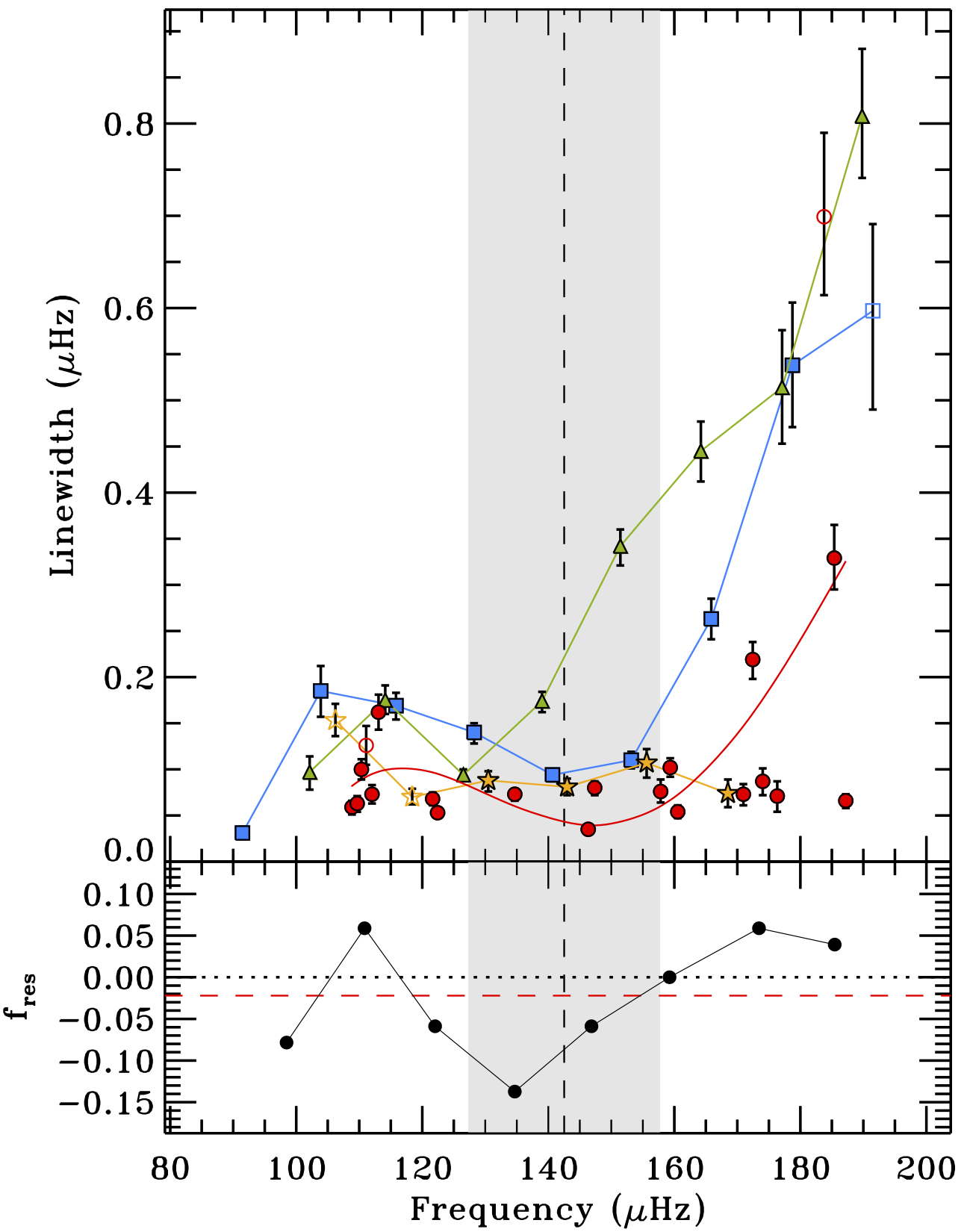}
      \caption{Mode linewidths for KIC~10200377 as a function of the corresponding oscillation frequencies. \textit{Top panel}: linewidth measurements as defined by Eq.~(\ref{eq:resolved_profile}) for each angular degree ($\ell = 0$ blue squares, $\ell = 2$ green triangles, $\ell = 3$ yellow stars, and resolved $\ell = 1$ mixed modes red circles). Open symbols represent modes with detection probability under the suggested threshold (see Sect.~\ref{sec:test}). The 68\,\% credible intervals for the linewidths as derived by \diamonds\,\,are shown for each data point. The red solid line represents a polynomial fit to the linewidths of the $\ell = 1$ mixed modes, included to emphasize the trend with frequency. The shaded region represents the range $\numax \pm \sigma_\mathrm{env}$, with $\numax$ from Table~\ref{tab:bkg2} indicated by the dashed vertical line. \textit{Bottom panel}: the normalized fraction of resolved mixed modes with respect to unresolved ones, $f_\mathrm{res}$ (black dots), defined by Eq.~(\ref{eq:fraction_resolved}). The frequency position of each point is the average frequency of the resolved dipole mixed modes falling in each radial order (or that of the unresolved mixed modes if no resolved mixed modes are present). The horizontal dotted line represents the limit of resolved-dominated regime, as defined in Sect.~\ref{sec:fwhm}, while the horizontal dashed red line marks the average $f_\mathrm{res}$ given by Eq.~(\ref{eq:average_fraction}).}
    \label{fig:10200377fwhm}
\end{figure}

\begin{figure}
   \centering
   \includegraphics[width=9.0cm]{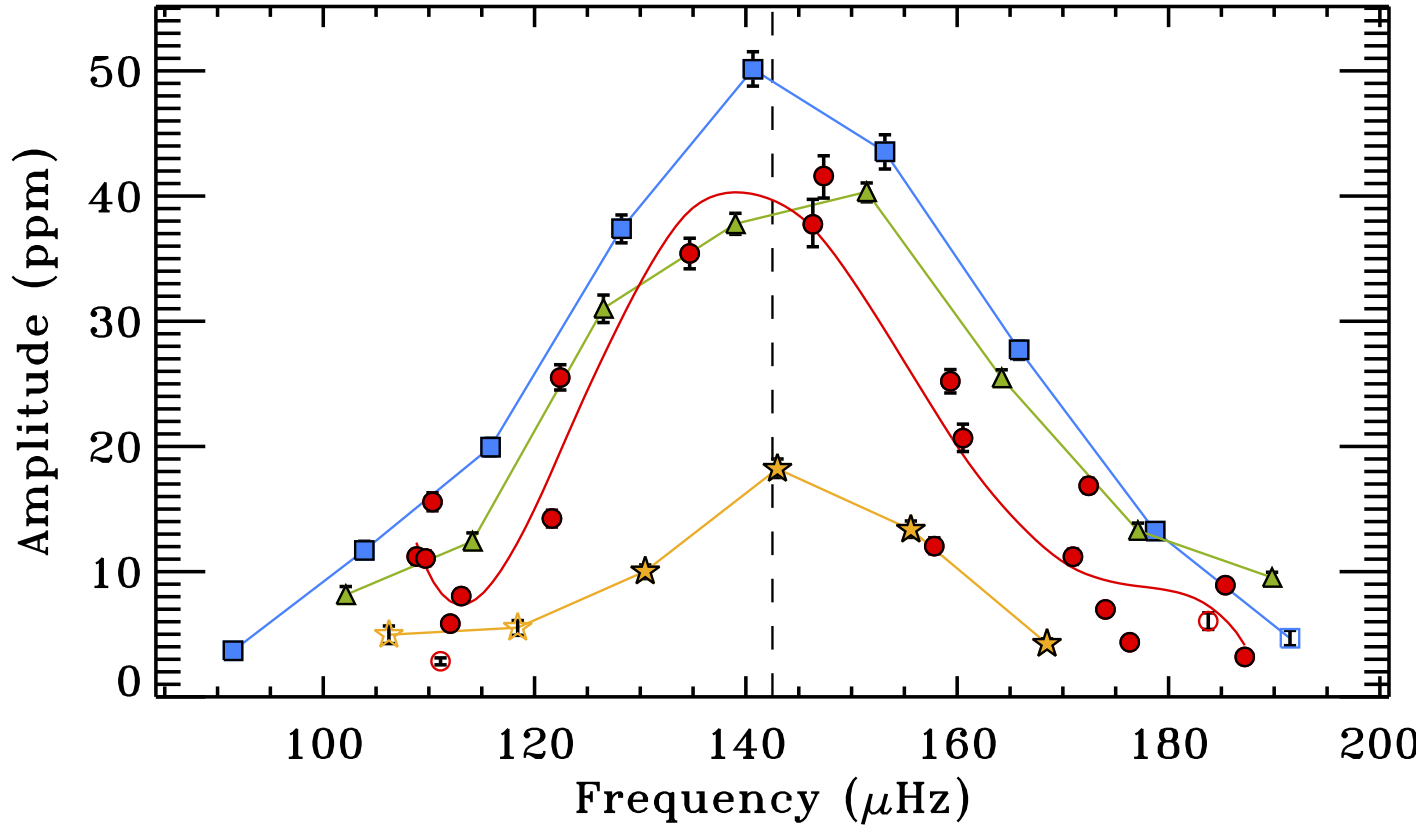}
      \caption{Mode amplitudes for KIC~10200377 as a function of the corresponding oscillation frequencies. Amplitude measurements as defined by Eq.~(\ref{eq:resolved_profile}) for each angular degree ($\ell = 0$ blue squares, $\ell = 2$ green triangles, $\ell = 3$ yellow stars, and resolved $\ell = 1$ mixed modes red circles). Open symbols represent modes with detection probability under the suggested threshold (see Sect.~\ref{sec:test}). The 68\,\% credible intervals for the amplitudes as derived by \diamonds\,\,are shown for each data point. The solid red line represents a polynomial fit to the amplitudes of the $\ell = 1$ mixed modes, included to emphasize the trend with frequency. The dashed vertical line indicates the $\numax$ value listed in Table~\ref{tab:bkg2}.}
    \label{fig:10200377amplitude}
\end{figure}
\clearpage

\begin{figure}
   \centering
   \includegraphics[width=9.0cm]{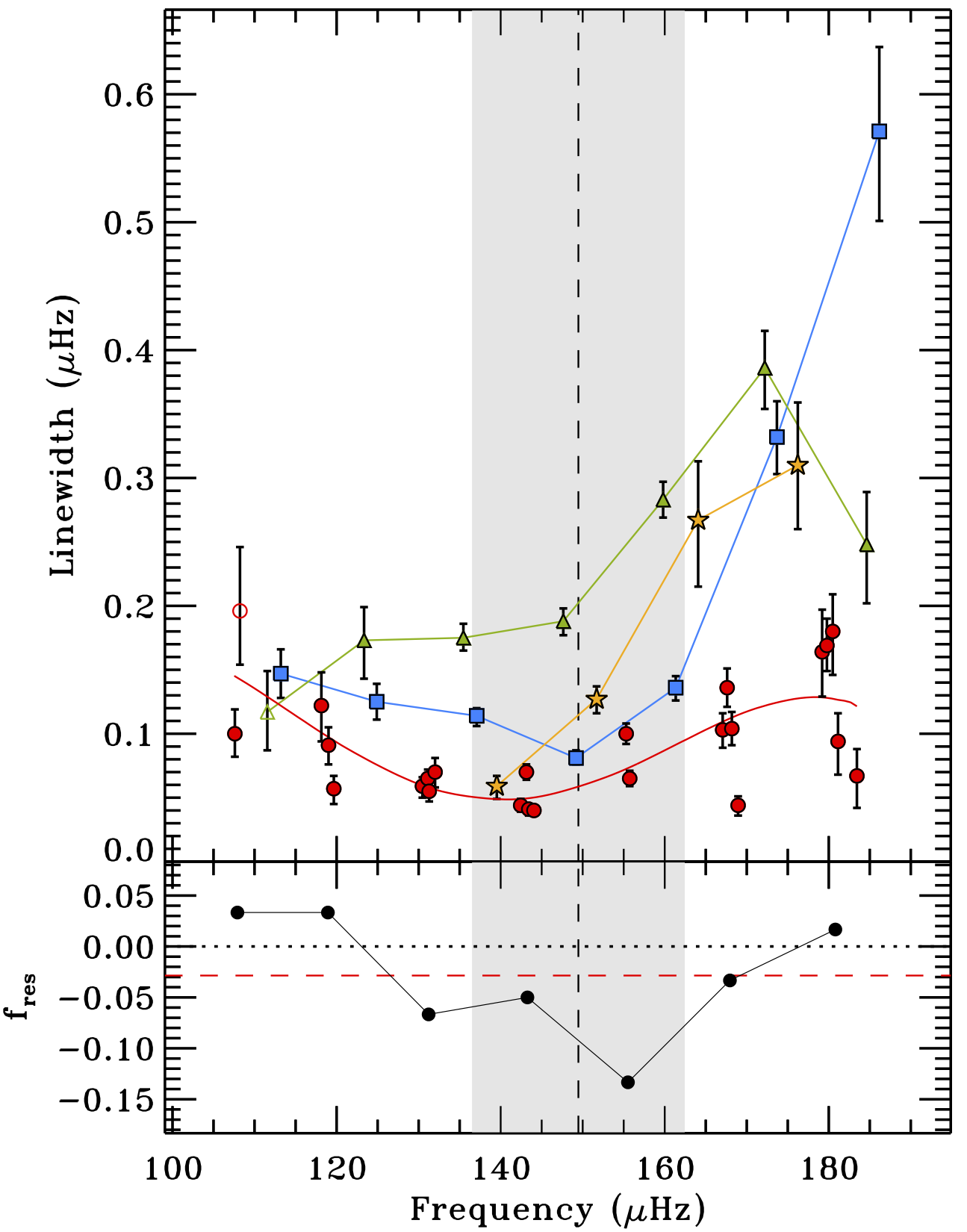}
      \caption{Mode linewidths for KIC~10257278 as a function of the corresponding oscillation frequencies. \textit{Top panel}: linewidth measurements as defined by Eq.~(\ref{eq:resolved_profile}) for each angular degree ($\ell = 0$ blue squares, $\ell = 2$ green triangles, $\ell = 3$ yellow stars, and resolved $\ell = 1$ mixed modes red circles). Open symbols represent modes with detection probability under the suggested threshold (see Sect.~\ref{sec:test}). The 68\,\% credible intervals for the linewidths as derived by \diamonds\,\,are shown for each data point. The red solid line represents a polynomial fit to the linewidths of the $\ell = 1$ mixed modes, included to emphasize the trend with frequency. The shaded region represents the range $\numax \pm \sigma_\mathrm{env}$, with $\numax$ from Table~\ref{tab:bkg2} indicated by the dashed vertical line. \textit{Bottom panel}: the normalized fraction of resolved mixed modes with respect to unresolved ones, $f_\mathrm{res}$ (black dots), defined by Eq.~(\ref{eq:fraction_resolved}). The frequency position of each point is the average frequency of the resolved dipole mixed modes falling in each radial order (or that of the unresolved mixed modes if no resolved mixed modes are present). The horizontal dotted line represents the limit of resolved-dominated regime, as defined in Sect.~\ref{sec:fwhm}, while the horizontal dashed red line marks the average $f_\mathrm{res}$ given by Eq.~(\ref{eq:average_fraction}).}
    \label{fig:10257278fwhm}
\end{figure}

\begin{figure}
   \centering
   \includegraphics[width=9.0cm]{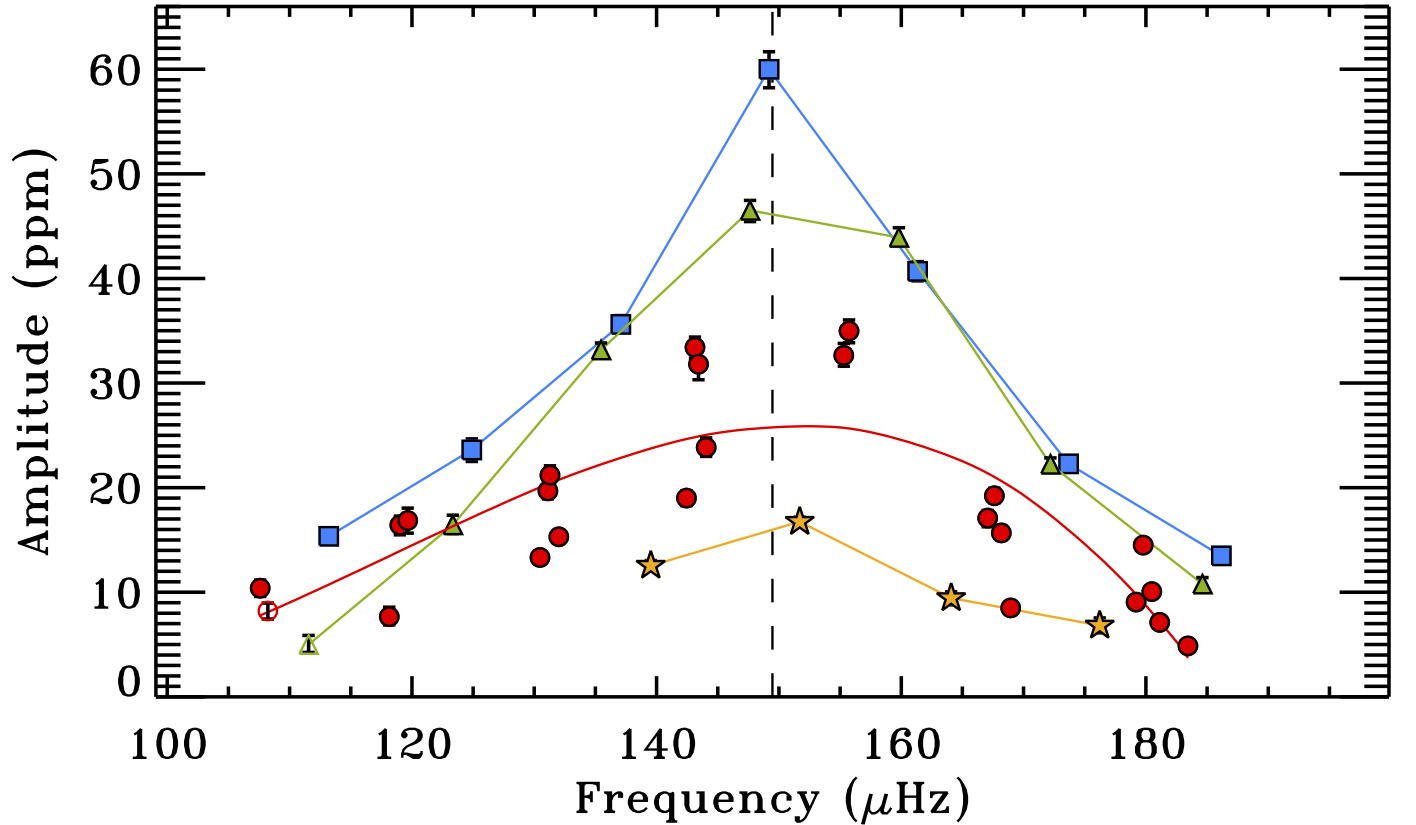}
      \caption{Mode amplitudes for KIC~10257278 as a function of the corresponding oscillation frequencies. Amplitude measurements as defined by Eq.~(\ref{eq:resolved_profile}) for each angular degree ($\ell = 0$ blue squares, $\ell = 2$ green triangles, $\ell = 3$ yellow stars, and resolved $\ell = 1$ mixed modes red circles). Open symbols represent modes with detection probability under the suggested threshold (see Sect.~\ref{sec:test}). The 68\,\% credible intervals for the amplitudes as derived by \diamonds\,\,are shown for each data point. The solid red line represents a polynomial fit to the amplitudes of the $\ell = 1$ mixed modes, included to emphasize the trend with frequency. The dashed vertical line indicates the $\numax$ value listed in Table~\ref{tab:bkg2}.}
    \label{fig:10257278amplitude}
\end{figure}
\clearpage

\begin{figure}
   \centering
   \includegraphics[width=9.0cm]{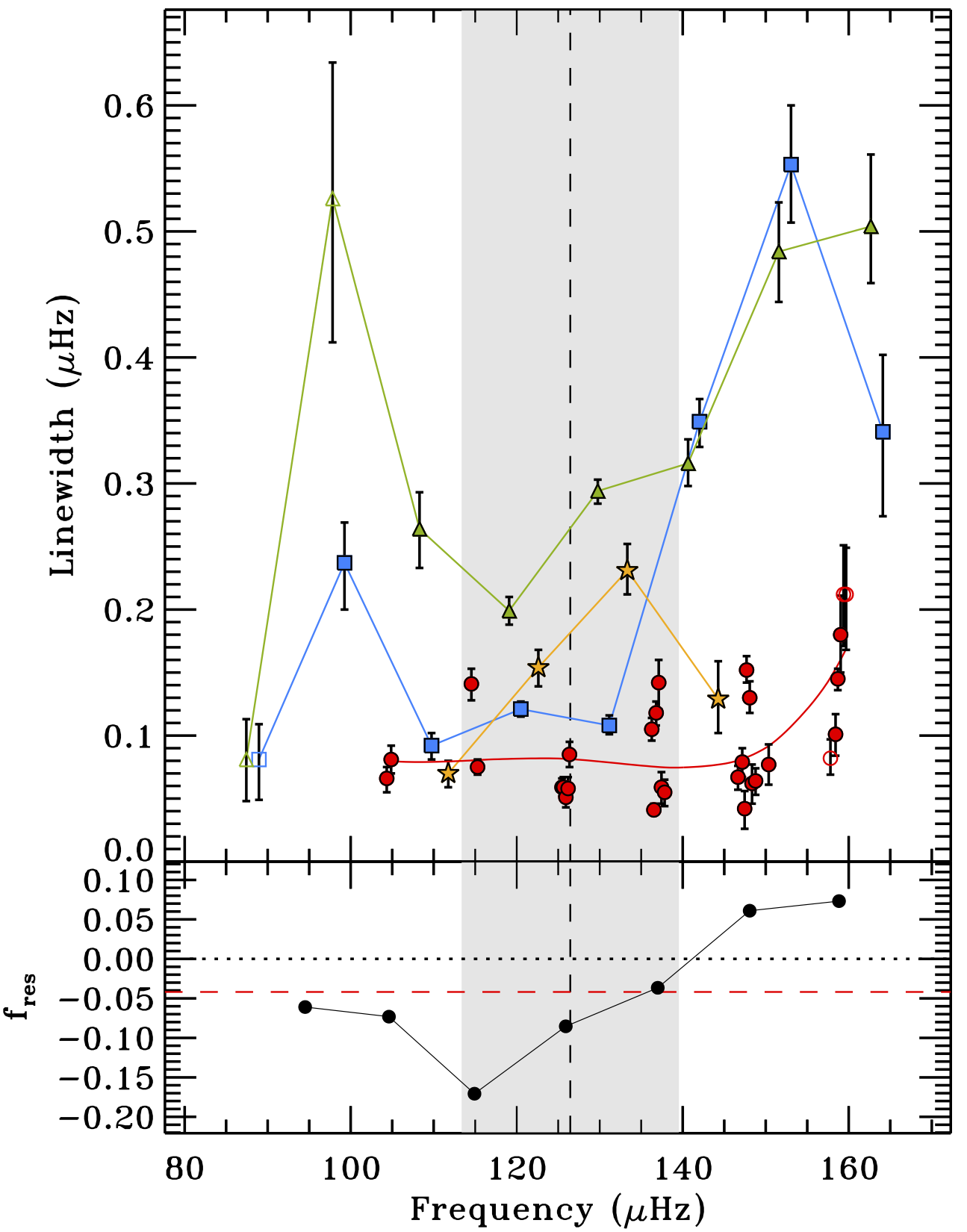}
      \caption{Mode linewidths for KIC~11353313 as a function of the corresponding oscillation frequencies. \textit{Top panel}: linewidth measurements as defined by Eq.~(\ref{eq:resolved_profile}) for each angular degree ($\ell = 0$ blue squares, $\ell = 2$ green triangles, $\ell = 3$ yellow stars, and resolved $\ell = 1$ mixed modes red circles). Open symbols represent modes with detection probability under the suggested threshold (see Sect.~\ref{sec:test}). The 68\,\% credible intervals for the linewidths as derived by \diamonds\,\,are shown for each data point. The red solid line represents a polynomial fit to the linewidths of the $\ell = 1$ mixed modes, included to emphasize the trend with frequency. The shaded region represents the range $\numax \pm \sigma_\mathrm{env}$, with $\numax$ from Table~\ref{tab:bkg2} indicated by the dashed vertical line. \textit{Bottom panel}: the normalized fraction of resolved mixed modes with respect to unresolved ones, $f_\mathrm{res}$ (black dots), defined by Eq.~(\ref{eq:fraction_resolved}). The frequency position of each point is the average frequency of the resolved dipole mixed modes falling in each radial order (or that of the unresolved mixed modes if no resolved mixed modes are present). The horizontal dotted line represents the limit of resolved-dominated regime, as defined in Sect.~\ref{sec:fwhm}, while the horizontal dashed red line marks the average $f_\mathrm{res}$ given by Eq.~(\ref{eq:average_fraction}).}
    \label{fig:11353313fwhm}
\end{figure}

\begin{figure}
   \centering
   \includegraphics[width=9.0cm]{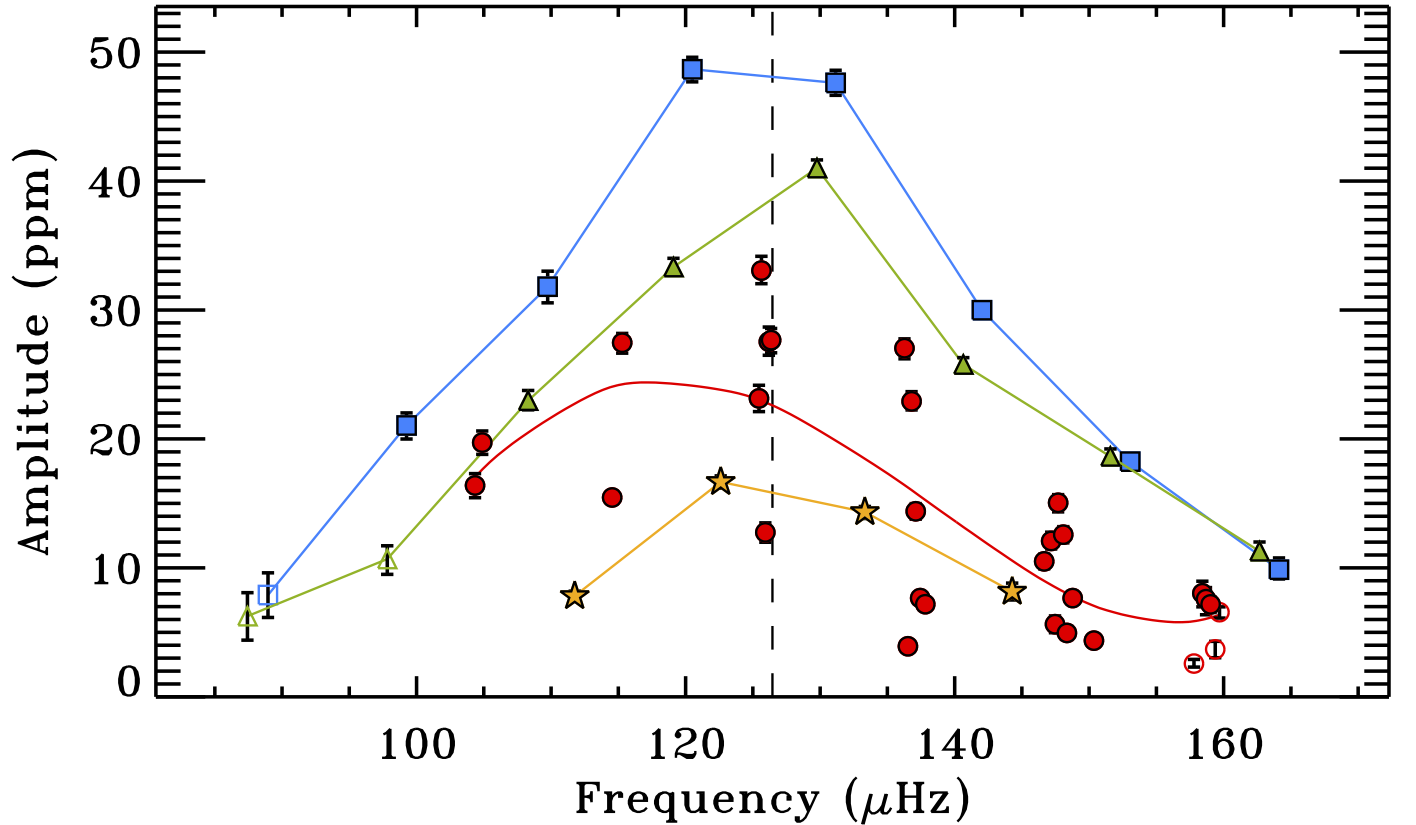}
      \caption{Mode amplitudes for KIC~11353313 as a function of the corresponding oscillation frequencies. Amplitude measurements as defined by Eq.~(\ref{eq:resolved_profile}) for each angular degree ($\ell = 0$ blue squares, $\ell = 2$ green triangles, $\ell = 3$ yellow stars, and resolved $\ell = 1$ mixed modes red circles). Open symbols represent modes with detection probability under the suggested threshold (see Sect.~\ref{sec:test}). The 68\,\% credible intervals for the amplitudes as derived by \diamonds\,\,are shown for each data point. The solid red line represents a polynomial fit to the amplitudes of the $\ell = 1$ mixed modes, included to emphasize the trend with frequency. The dashed vertical line indicates the $\numax$ value listed in Table~\ref{tab:bkg2}.}
    \label{fig:11353313amplitude}
\end{figure}
\clearpage

\begin{figure}
   \centering
   \includegraphics[width=9.0cm]{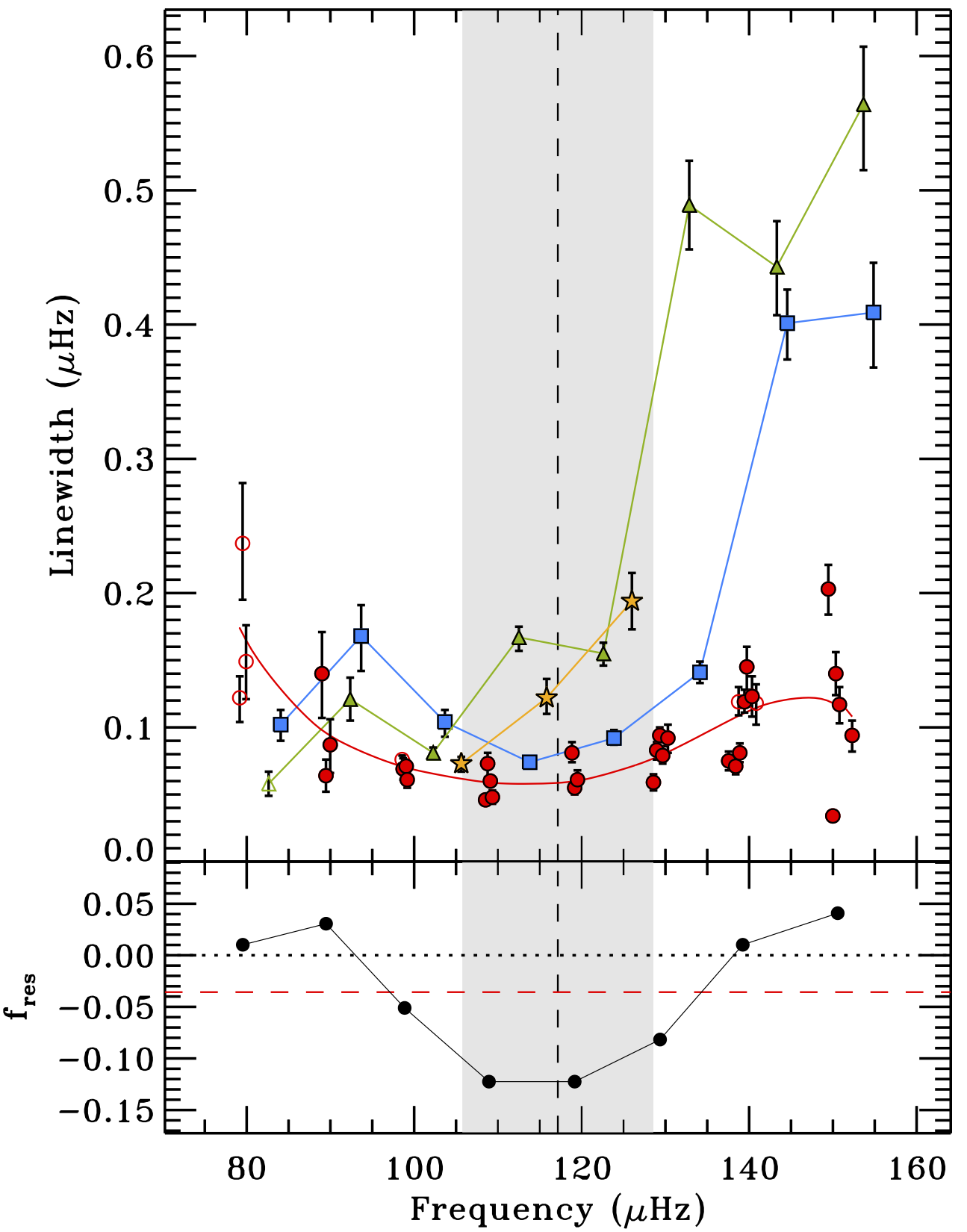}
      \caption{Mode linewidths for KIC~11913545 as a function of the corresponding oscillation frequencies. \textit{Top panel}: linewidth measurements as defined by Eq.~(\ref{eq:resolved_profile}) for each angular degree ($\ell = 0$ blue squares, $\ell = 2$ green triangles, $\ell = 3$ yellow stars, and resolved $\ell = 1$ mixed modes red circles). Open symbols represent modes with detection probability under the suggested threshold (see Sect.~\ref{sec:test}). The 68\,\% credible intervals for the linewidths as derived by \diamonds\,\,are shown for each data point. The red solid line represents a polynomial fit to the linewidths of the $\ell = 1$ mixed modes, included to emphasize the trend with frequency. The shaded region represents the range $\numax \pm \sigma_\mathrm{env}$, with $\numax$ from Table~\ref{tab:bkg2} indicated by the dashed vertical line. \textit{Bottom panel}: the normalized fraction of resolved mixed modes with respect to unresolved ones, $f_\mathrm{res}$ (black dots), defined by Eq.~(\ref{eq:fraction_resolved}). The frequency position of each point is the average frequency of the resolved dipole mixed modes falling in each radial order (or that of the unresolved mixed modes if no resolved mixed modes are present). The horizontal dotted line represents the limit of resolved-dominated regime, as defined in Sect.~\ref{sec:fwhm}, while the horizontal dashed red line marks the average $f_\mathrm{res}$ given by Eq.~(\ref{eq:average_fraction}).}
    \label{fig:11913545fwhm}
\end{figure}

\begin{figure}
   \centering
   \includegraphics[width=9.0cm]{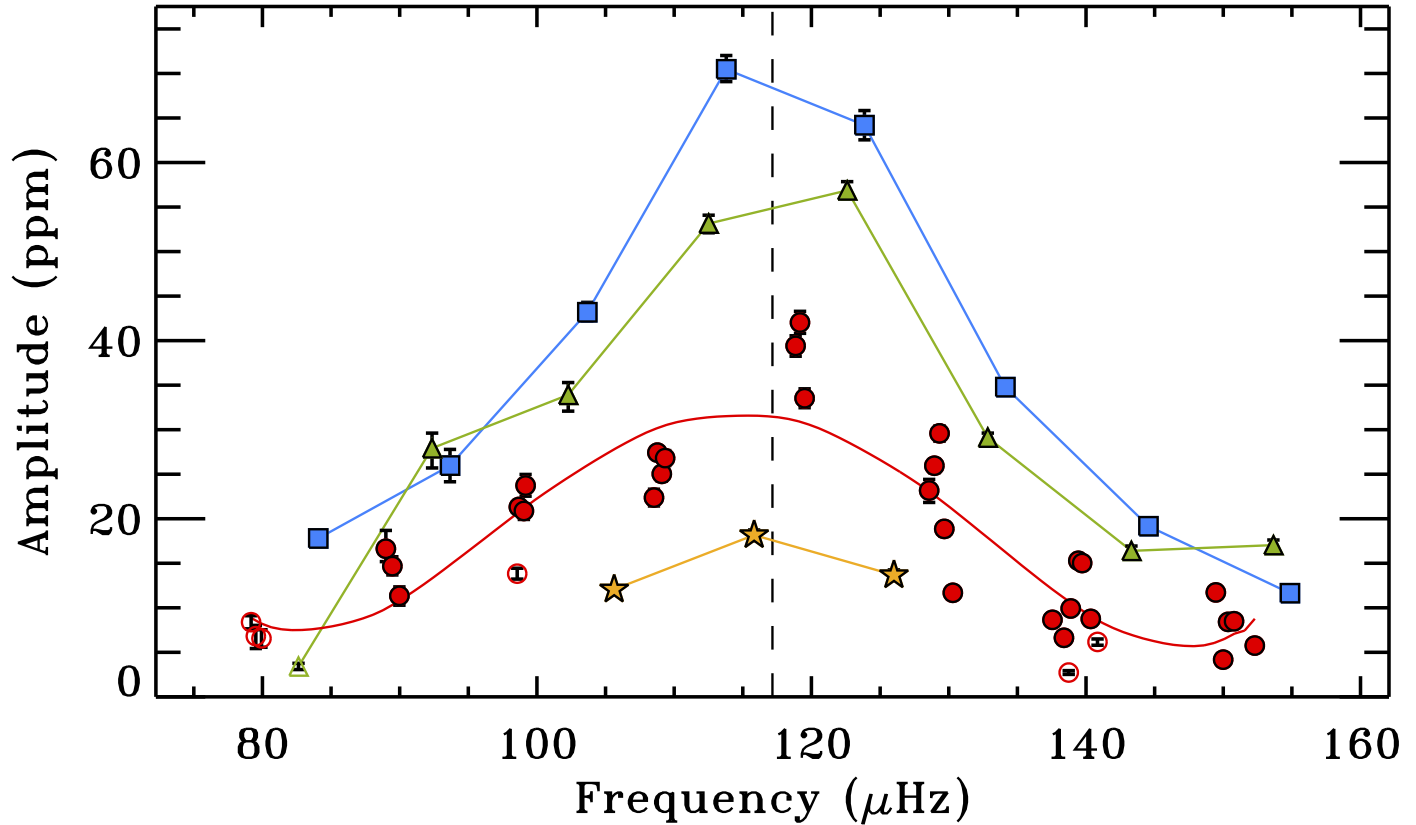}
      \caption{Mode amplitudes for KIC~11913545 as a function of the corresponding oscillation frequencies. Amplitude measurements as defined by Eq.~(\ref{eq:resolved_profile}) for each angular degree ($\ell = 0$ blue squares, $\ell = 2$ green triangles, $\ell = 3$ yellow stars, and resolved $\ell = 1$ mixed modes red circles). Open symbols represent modes with detection probability under the suggested threshold (see Sect.~\ref{sec:test}). The 68\,\% credible intervals for the amplitudes as derived by \diamonds\,\,are shown for each data point. The solid red line represents a polynomial fit to the amplitudes of the $\ell = 1$ mixed modes, included to emphasize the trend with frequency. The dashed vertical line indicates the $\numax$ value listed in Table~\ref{tab:bkg2}.}
    \label{fig:11913545amplitude}
\end{figure}
\clearpage

\begin{figure}
   \centering
   \includegraphics[width=9.0cm]{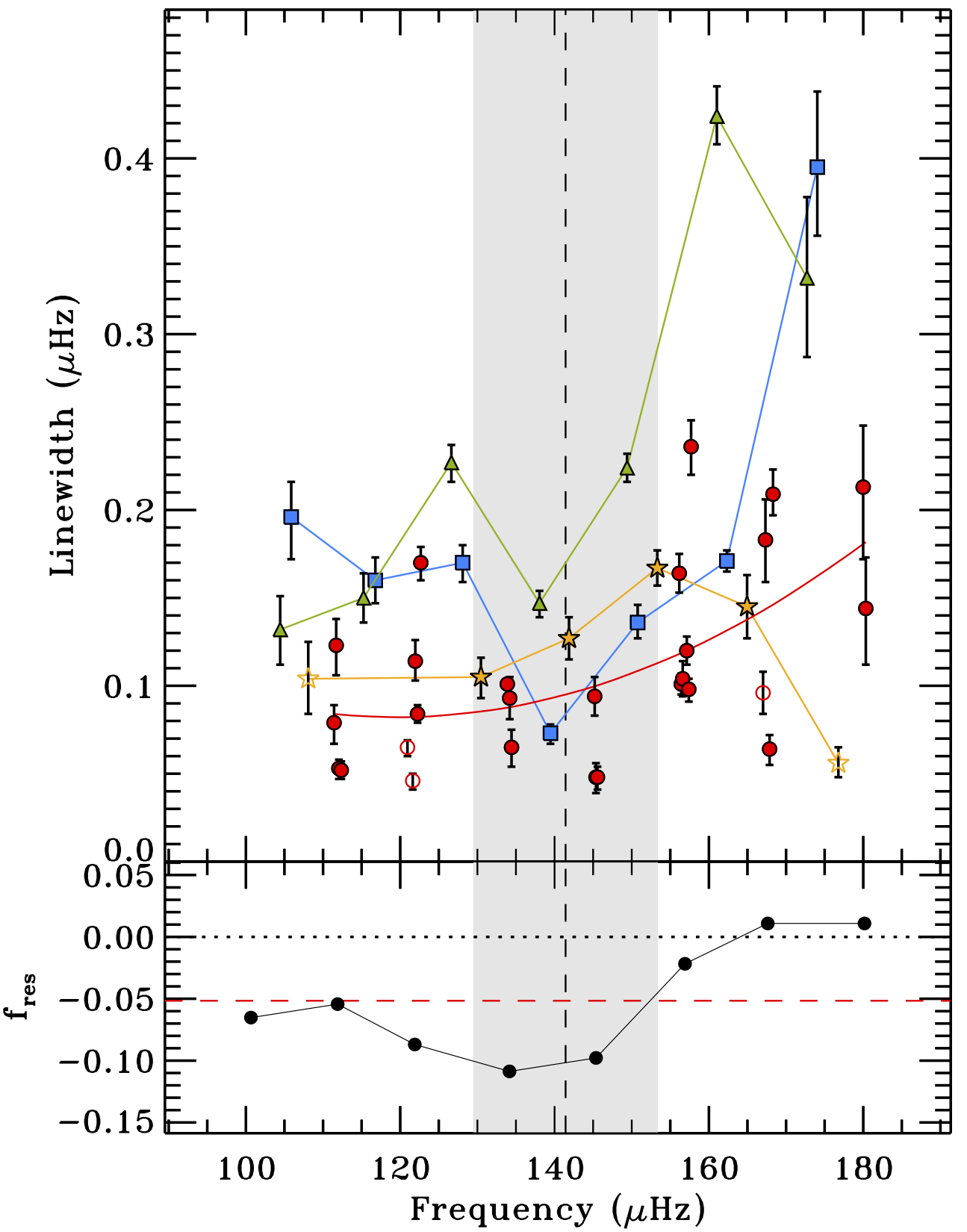}
      \caption{Mode linewidths for KIC~11968334 as a function of the corresponding oscillation frequencies. \textit{Top panel}: linewidth measurements as defined by Eq.~(\ref{eq:resolved_profile}) for each angular degree ($\ell = 0$ blue squares, $\ell = 2$ green triangles, $\ell = 3$ yellow stars, and resolved $\ell = 1$ mixed modes red circles). Open symbols represent modes with detection probability under the suggested threshold (see Sect.~\ref{sec:test}). The 68\,\% credible intervals for the linewidths as derived by \diamonds\,\,are shown for each data point. The red solid line represents a polynomial fit to the linewidths of the $\ell = 1$ mixed modes, included to emphasize the trend with frequency. The shaded region represents the range $\numax \pm \sigma_\mathrm{env}$, with $\numax$ from Table~\ref{tab:bkg2} indicated by the dashed vertical line. \textit{Bottom panel}: the normalized fraction of resolved mixed modes with respect to unresolved ones, $f_\mathrm{res}$ (black dots), defined by Eq.~(\ref{eq:fraction_resolved}). The frequency position of each point is the average frequency of the resolved dipole mixed modes falling in each radial order (or that of the unresolved mixed modes if no resolved mixed modes are present). The horizontal dotted line represents the limit of resolved-dominated regime, as defined in Sect.~\ref{sec:fwhm}, while the horizontal dashed red line marks the average $f_\mathrm{res}$ given by Eq.~(\ref{eq:average_fraction}).}
    \label{fig:11968334fwhm}
\end{figure}

\begin{figure}
   \centering
   \includegraphics[width=9.0cm]{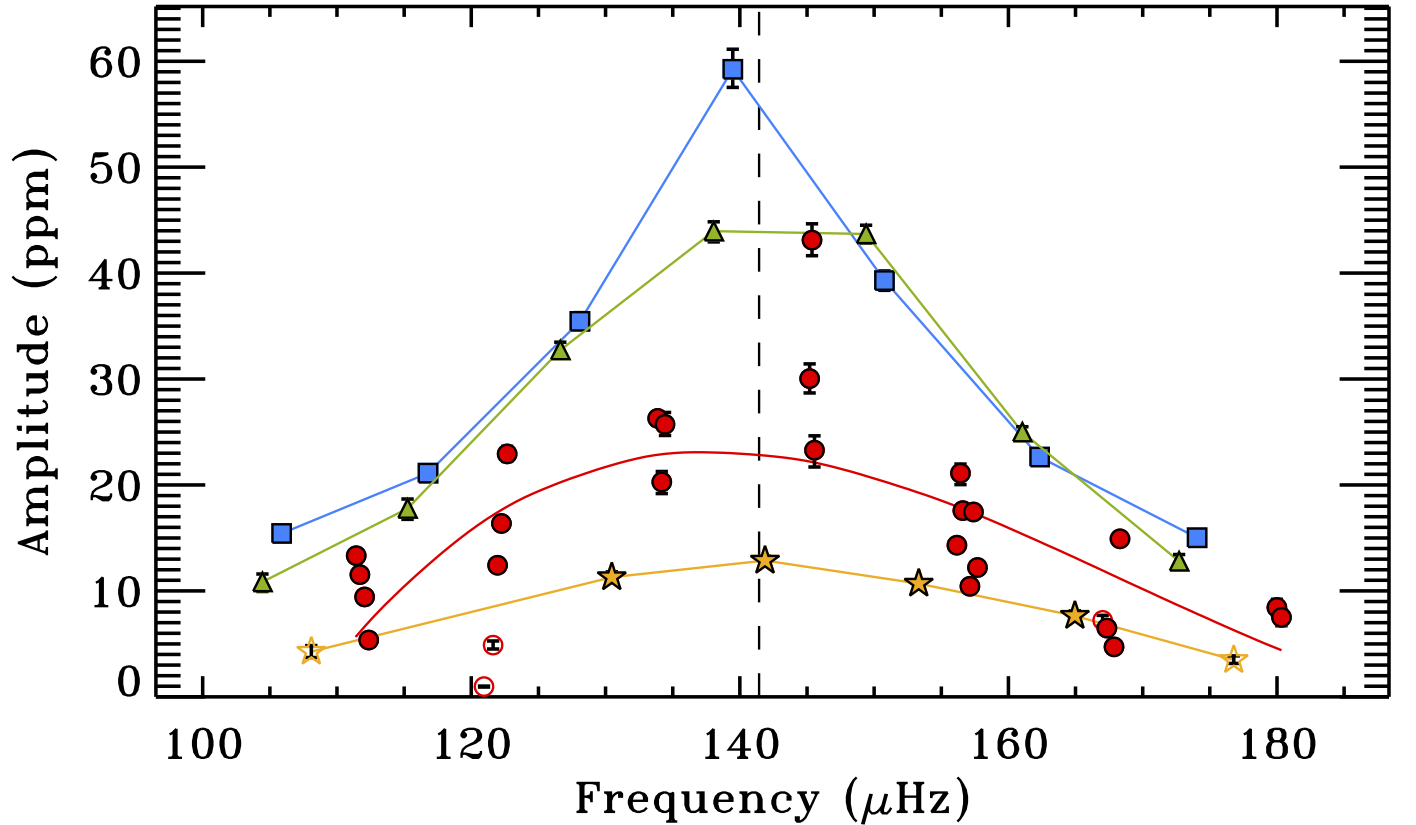}
      \caption{Mode amplitudes for KIC~11968334 as a function of the corresponding oscillation frequencies. Amplitude measurements as defined by Eq.~(\ref{eq:resolved_profile}) for each angular degree ($\ell = 0$ blue squares, $\ell = 2$ green triangles, $\ell = 3$ yellow stars, and resolved $\ell = 1$ mixed modes red circles). Open symbols represent modes with detection probability under the suggested threshold (see Sect.~\ref{sec:test}). The 68\,\% credible intervals for the amplitudes as derived by \diamonds\,\,are shown for each data point. The solid red line represents a polynomial fit to the amplitudes of the $\ell = 1$ mixed modes, included to emphasize the trend with frequency. The dashed vertical line indicates the $\numax$ value listed in Table~\ref{tab:bkg2}.}
    \label{fig:11968334amplitude}
\end{figure}
\clearpage

\begin{table*}[!]
\caption{Median values with corresponding 68.3\,\% shortest credible intervals for the oscillation frequencies, amplitudes, linewidths, and heights of the mixed modes of KIC~3744043, as derived by \diamonds\,\,by using the peak bagging model defined by Eqs.~(\ref{eq:general_pb_model}) and (\ref{eq:pb_model}).}
\label{tab:3744043m}
\centering
\begin{tabular}{llcrrlrc}
\hline\hline
\\[-8pt]          
Peak \# & $\ell$ & $m$ & \multicolumn{1}{c}{Frequency} & \multicolumn{1}{c}{Amplitude} & \multicolumn{1}{c}{Linewidth} & \multicolumn{1}{c}{Height}& $p_\mathrm{B}$\\
 & & & \multicolumn{1}{c}{($\mu$Hz)} & \multicolumn{1}{c}{(ppm)} & \multicolumn{1}{c}{($\mu$Hz)} & \multicolumn{1}{c}{(ppm$^2/\mu$Hz)}\\
\hline \\[-8pt]
0 & 1 & 0 & $     77.3480_{-      0.0116}^{+      0.0103}$ & $       9.343_{-       0.826}^{+       0.913}$ & $       0.093_{-       0.019}^{+       0.017}$ & \multicolumn{1}{c}{\dots} & 0.999\\[1pt]

1  & 1 & 0 & $     82.5174_{-      0.0019}^{+      0.0014}$ & \multicolumn{1}{c}{\dots} & \multicolumn{1}{c}{\dots} & $      865.00_{-       55.89}^{+       49.25}$ & 0.901 \\[1pt]
2 & 1 & 0 & $     85.6808_{-      0.0004}^{+      0.0005}$ & \multicolumn{1}{c}{\dots} & \multicolumn{1}{c}{\dots} & $     3179.07_{-      282.89}^{+      250.26}$ & \dots \\[1pt]
3 & 1 & 0 & $     86.1585_{-      0.0096}^{+      0.0101}$ & $       7.051_{-       0.523}^{+       0.514}$ & $       0.107_{-       0.017}^{+       0.016}$ & \multicolumn{1}{c}{\dots} & 0.984\\
\hline
\end{tabular}
\end{table*}

\begin{table*}[!]
\caption{Table~\ref{tab:3744043m} continued.}
\label{tab:3744043m2}
\centering
\begin{tabular}{llcrrlrc}
\hline\hline
\\[-8pt]          
Peak \# & $\ell$ & $m$ & \multicolumn{1}{c}{Frequency} & \multicolumn{1}{c}{Amplitude} & \multicolumn{1}{c}{Linewidth} & \multicolumn{1}{c}{Height}& $p_\mathrm{B}$\\
 & & & \multicolumn{1}{c}{($\mu$Hz)} & \multicolumn{1}{c}{(ppm)} & \multicolumn{1}{c}{($\mu$Hz)} & \multicolumn{1}{c}{(ppm$^2/\mu$Hz)}\\
\hline \\[-8pt]
51 & 1 & $+1$ & $    115.7813_{-      0.0068}^{+      0.0074}$ & $      21.243_{-       1.363}^{+       1.095}$ & $       0.037_{-       0.008}^{+       0.008}$ & \multicolumn{1}{c}{\dots} & \dots\\[1pt]
52 & 1 & $-1$ & $    116.0955_{-      0.0086}^{+      0.0066}$ & $      25.641_{-       1.068}^{+       1.245}$ & $       0.081_{-       0.005}^{+       0.006}$ & \multicolumn{1}{c}{\dots} & \dots\\[1pt]
53 & 1 & 0 & $    116.2429_{-      0.0001}^{+      0.0001}$ & \multicolumn{1}{c}{\dots} & \multicolumn{1}{c}{\dots} & $    24247.93_{-     1884.42}^{+     3199.00}$ & \dots\\[1pt]
54  & 1 & $+1$ & $    116.4611_{-      0.0070}^{+      0.0068}$ & $       3.737_{-       0.226}^{+       0.236}$ & $       0.057_{-       0.007}^{+       0.006}$ & \multicolumn{1}{c}{\dots} & 0.159\\
\hline
\end{tabular}
\tablefoot{The first column represents the peak number in increasing frequency order and is shown for each angular degree ($\ell$) and azimuthal order ($m$). The last column corresponds to the detection probability introduced by Eq.~(\ref{eq:detection_probability}) and discussed in Sect.~\ref{sec:test}.}
\end{table*}

\begin{table*}[!]
\caption{Median values with corresponding 68.3\,\% shortest credible intervals for the oscillation frequencies, amplitudes, and linewidths of the $p$ modes of KIC~3744043, as derived by \diamonds\,\,by using the peak bagging model defined by Eqs.~(\ref{eq:general_pb_model}) and (\ref{eq:pb_model}).}
\label{tab:3744043p}
\centering
\begin{tabular}{llcrrlrc}
\hline\hline
\\[-8pt]          
Peak \# & $\ell$ & $m$ & \multicolumn{1}{c}{Frequency} & \multicolumn{1}{c}{Amplitude} & \multicolumn{1}{c}{Linewidth} & \multicolumn{1}{c}{Height}& $p_\mathrm{B}$\\
 & & & \multicolumn{1}{c}{($\mu$Hz)} & \multicolumn{1}{c}{(ppm)} & \multicolumn{1}{c}{($\mu$Hz)} & \multicolumn{1}{c}{(ppm$^2/\mu$Hz)}\\
\hline \\[-8pt]
0 & 0 & 0 & $     81.6886_{-      0.0108}^{+      0.0126}$ & $      18.967_{-       1.195}^{+       1.111}$ & $       0.133_{-       0.031}^{+       0.026}$ & \multicolumn{1}{c}{\dots} & 1.000\\[1pt]
1 & 0 & 0 & $     91.1127_{-      0.0080}^{+      0.0074}$ & $      25.098_{-       0.970}^{+       1.056}$ & $       0.090_{-       0.010}^{+       0.010}$ & \multicolumn{1}{c}{\dots} & \dots \\[1pt]
2 & 0 & 0 & $    100.9011_{-      0.0048}^{+      0.0043}$ & $      42.406_{-       1.155}^{+       1.104}$ & $       0.110_{-       0.006}^{+       0.007}$ & \multicolumn{1}{c}{\dots} & \dots \\[1pt]
3 & 0 & 0 & $    110.7508_{-      0.0024}^{+      0.0022}$ & $      62.960_{-       0.964}^{+       0.970}$ & $       0.136_{-       0.006}^{+       0.005}$ & \multicolumn{1}{c}{\dots} & \dots\\
\hline
\end{tabular}
\tablefoot{The first column represents the peak number in increasing frequency order and is shown for each angular degree ($\ell$) and azimuthal order ($m$). The last column corresponds to the detection probability introduced by Eq.~(\ref{eq:detection_probability}) and discussed in Sect.~\ref{sec:test}.}
\end{table*}
\begin{table*}[!]
\caption{Median values with corresponding 68.3\,\% shortest credible intervals for the oscillation frequencies, amplitudes, linewidths, and heights of the mixed modes of KIC~6117517, as derived by \diamonds\,\,by using the peak bagging model defined by Eqs.~(\ref{eq:general_pb_model}) and (\ref{eq:pb_model}).}
\label{tab:6117517m}
\centering
\begin{tabular}{llcrrlrc}
\hline\hline
\\[-8pt]          
Peak \# & $\ell$ & $m$ & \multicolumn{1}{c}{Frequency} & \multicolumn{1}{c}{Amplitude} & \multicolumn{1}{c}{Linewidth} & \multicolumn{1}{c}{Height}& $p_\mathrm{B}$\\
 & & & \multicolumn{1}{c}{($\mu$Hz)} & \multicolumn{1}{c}{(ppm)} & \multicolumn{1}{c}{($\mu$Hz)} & \multicolumn{1}{c}{(ppm$^2/\mu$Hz)}\\
\hline \\[-8pt]
0 & 1 & 0 & $     78.9102_{-      0.0006}^{+      0.0006}$ & \multicolumn{1}{c}{\dots} & \multicolumn{1}{c}{\dots} & $     3899.92_{-      234.43}^{+      350.20}$ & \dots \\[1pt]
1 & 1 & 0 & $     88.6524_{-      0.0061}^{+      0.0059}$ & $      12.641_{-       1.184}^{+       1.096}$ & $       0.056_{-       0.017}^{+       0.014}$ & \multicolumn{1}{c}{\dots} & \dots \\[1pt]
2 & 1 & 0 & $     89.0953_{-      0.0063}^{+      0.0067}$ & $      16.429_{-       1.088}^{+       1.143}$ & $       0.062_{-       0.015}^{+       0.014}$ & \multicolumn{1}{c}{\dots} & \dots \\[1pt]
3 & 1 & 0 & $     89.5298_{-      0.0006}^{+      0.0007}$ & \multicolumn{1}{c}{\dots} & \multicolumn{1}{c}{\dots} & $     5075.76_{-      373.39}^{+      861.42}$ & \dots \\
\hline
\end{tabular}
\tablefoot{The first column represents the peak number in increasing frequency order and is shown for each angular degree ($\ell$) and azimuthal order ($m$). The last column corresponds to the detection probability introduced by Eq.~(\ref{eq:detection_probability}) and discussed in Sect.~\ref{sec:test}.}
\end{table*}

\begin{table*}[!]
\caption{Median values with corresponding 68.3\,\% shortest credible intervals for the oscillation frequencies, amplitudes, and linewidths of the $p$ modes of KIC~6117517, as derived by \diamonds\,\,by using the peak bagging model defined by Eqs.~(\ref{eq:general_pb_model}) and (\ref{eq:pb_model}).}
\label{tab:6117517p}
\centering
\begin{tabular}{llcrrlrc}
\hline\hline
\\[-8pt]          
Peak \# & $\ell$ & $m$ & \multicolumn{1}{c}{Frequency} & \multicolumn{1}{c}{Amplitude} & \multicolumn{1}{c}{Linewidth} & \multicolumn{1}{c}{Height}& $p_\mathrm{B}$\\
 & & & \multicolumn{1}{c}{($\mu$Hz)} & \multicolumn{1}{c}{(ppm)} & \multicolumn{1}{c}{($\mu$Hz)} & \multicolumn{1}{c}{(ppm$^2/\mu$Hz)}\\
\hline \\[-8pt]
0 & 0 & 0 & $     83.9104_{-      0.0111}^{+      0.0103}$ & $      16.100_{-       1.049}^{+       0.935}$ & $       0.175_{-       0.019}^{+       0.014}$ & \multicolumn{1}{c}{\dots} & 1.000\\[1pt]
1 & 0 & 0 & $     93.7208_{-      0.0119}^{+      0.0126}$ & $      26.267_{-       1.309}^{+       1.126}$ & $       0.220_{-       0.044}^{+       0.039}$ & \multicolumn{1}{c}{\dots} & \dots \\[1pt]
2 & 0 & 0 & $    103.4368_{-      0.0064}^{+      0.0060}$ & $      35.832_{-       2.045}^{+       1.827}$ & $       0.089_{-       0.014}^{+       0.014}$ & \multicolumn{1}{c}{\dots} & \dots \\[1pt]
3 & 0 & 0 & $    113.6025_{-      0.0020}^{+      0.0022}$ & $      77.568_{-       2.018}^{+       2.321}$ & $       0.054_{-       0.004}^{+       0.004}$ & \multicolumn{1}{c}{\dots} & \dots\\
\hline
\end{tabular}
\tablefoot{The first column represents the peak number in increasing frequency order and is shown for each angular degree ($\ell$) and azimuthal order ($m$). The last column corresponds to the detection probability introduced by Eq.~(\ref{eq:detection_probability}) and discussed in Sect.~\ref{sec:test}.}
\end{table*}
\begin{table*}[!]
\caption{Median values with corresponding 68.3\,\% shortest credible intervals for the oscillation frequencies, amplitudes, linewidths, and heights of the mixed modes of KIC~6144777, as derived by \diamonds\,\,by using the peak bagging model defined by Eqs.~(\ref{eq:general_pb_model}) and (\ref{eq:pb_model}).}
\label{tab:6144777m}
\centering
\begin{tabular}{llcrrlrc}
\hline\hline
\\[-8pt]          
Peak \# & $\ell$ & $m$ & \multicolumn{1}{c}{Frequency} & \multicolumn{1}{c}{Amplitude} & \multicolumn{1}{c}{Linewidth} & \multicolumn{1}{c}{Height}& $p_\mathrm{B}$\\
 & & & \multicolumn{1}{c}{($\mu$Hz)} & \multicolumn{1}{c}{(ppm)} & \multicolumn{1}{c}{($\mu$Hz)} & \multicolumn{1}{c}{(ppm$^2/\mu$Hz)}\\
\hline \\[-8pt]
0 & 1 & ? & $     85.9962_{-      0.0143}^{+      0.0150}$ & $       6.879_{-       0.864}^{+       0.760}$ & $       0.109_{-       0.019}^{+       0.014}$ & \multicolumn{1}{c}{\dots} & 0.773\\[1pt]
1 & 1 & ? & $     86.1645_{-      0.0114}^{+      0.0108}$ & $       7.415_{-       1.115}^{+       1.428}$ & $       0.067_{-       0.009}^{+       0.008}$ & \multicolumn{1}{c}{\dots} & 0.988\\[1pt]

2 & 1 & ? & $     94.5077_{-      0.0013}^{+      0.0013}$ & \multicolumn{1}{c}{\dots} & \multicolumn{1}{c}{\dots} & $      987.00_{-       67.35}^{+       71.28}$ & 0.980\\[1pt]
3 & 1 & ? & $     95.7274_{-      0.0023}^{+      0.0022}$ & \multicolumn{1}{c}{\dots} & \multicolumn{1}{c}{\dots} & $     1447.13_{-       68.86}^{+       97.77}$ & \dots \\
\hline
\end{tabular}
\end{table*}

\begin{table*}[!]
\caption{Table~\ref{tab:6144777m} continued.}
\label{tab:6144777m2}
\centering
\begin{tabular}{llcrrlrc}
\hline\hline
\\[-8pt]          
Peak \# & $\ell$ & $m$ & \multicolumn{1}{c}{Frequency} & \multicolumn{1}{c}{Amplitude} & \multicolumn{1}{c}{Linewidth} & \multicolumn{1}{c}{Height}& $p_\mathrm{B}$\\
 & & & \multicolumn{1}{c}{($\mu$Hz)} & \multicolumn{1}{c}{(ppm)} & \multicolumn{1}{c}{($\mu$Hz)} & \multicolumn{1}{c}{(ppm$^2/\mu$Hz)}\\
\hline \\[-8pt]
51 & 1 & 0 & $    125.5672_{-      0.0012}^{+      0.0010}$ & \multicolumn{1}{c}{\dots} & \multicolumn{1}{c}{\dots} & $     1250.65_{-       42.29}^{+       29.90}$ & 0.999\\[1pt]
52 & 1 & $-1$ & $    126.5611_{-      0.0012}^{+      0.0019}$ & \multicolumn{1}{c}{\dots} & \multicolumn{1}{c}{\dots} & $     1303.12_{-       38.75}^{+       40.19}$ & \dots\\[1pt]
53 & 1 & 0 & $    126.7983_{-      0.0003}^{+      0.0003}$ & \multicolumn{1}{c}{\dots} & \multicolumn{1}{c}{\dots} & $     2561.85_{-       72.55}^{+       67.68}$ & \dots\\[1pt]
54 & 1 & $+1$ & $    127.0333_{-      0.0003}^{+      0.0003}$ & \multicolumn{1}{c}{\dots} & \multicolumn{1}{c}{\dots} & $     4661.59_{-      158.59}^{+      130.56}$ & \dots\\
\hline
\end{tabular}
\tablefoot{The first column represents the peak number in increasing frequency order and is shown for each angular degree ($\ell$) and azimuthal order ($m$), with question marks placed for m-values that could not be identified. The last column corresponds to the detection probability introduced by Eq.~(\ref{eq:detection_probability}) and discussed in Sect.~\ref{sec:test}.}
\end{table*}

\begin{table*}[!]
\caption{Median values with corresponding 68.3\,\% shortest credible intervals for the oscillation frequencies, amplitudes, and linewidths of the $p$ modes of KIC~6144777, as derived by \diamonds\,\,by using the peak bagging model defined by Eqs.~(\ref{eq:general_pb_model}) and (\ref{eq:pb_model}).}
\label{tab:6144777p}
\centering
\begin{tabular}{llcrrlrc}
\hline\hline
\\[-8pt]          
Peak \# & $\ell$ & $m$ & \multicolumn{1}{c}{Frequency} & \multicolumn{1}{c}{Amplitude} & \multicolumn{1}{c}{Linewidth} & \multicolumn{1}{c}{Height}& $p_\mathrm{B}$\\
 & & & \multicolumn{1}{c}{($\mu$Hz)} & \multicolumn{1}{c}{(ppm)} & \multicolumn{1}{c}{($\mu$Hz)} & \multicolumn{1}{c}{(ppm$^2/\mu$Hz)}\\
\hline \\[-8pt]
0 & 0 & 0 & $     91.3480_{-      0.0123}^{+      0.0135}$ & $      11.466_{-       1.076}^{+       0.966}$ & $       0.152_{-       0.022}^{+       0.019}$ & \multicolumn{1}{c}{\dots} & \dots \\[1pt]
1 & 0 & 0 & $    101.9340_{-      0.0082}^{+      0.0079}$ & $      22.604_{-       0.953}^{+       1.055}$ & $       0.122_{-       0.014}^{+       0.014}$ & \multicolumn{1}{c}{\dots} & \dots\\[1pt]
2 & 0 & 0 & $    112.6211_{-      0.0028}^{+      0.0029}$ & $      36.516_{-       0.706}^{+       0.772}$ & $       0.112_{-       0.007}^{+       0.006}$ & \multicolumn{1}{c}{\dots} & \dots\\[1pt]
3 & 0 & 0 & $    123.6235_{-      0.0022}^{+      0.0021}$ & $      63.663_{-       1.646}^{+       1.604}$ & $       0.072_{-       0.004}^{+       0.004}$ & \multicolumn{1}{c}{\dots} & \dots \\
\hline
\end{tabular}
\tablefoot{The first column represents the peak number in increasing frequency order and is shown for each angular degree ($\ell$) and azimuthal order ($m$). The last column corresponds to the detection probability introduced by Eq.~(\ref{eq:detection_probability}) and discussed in Sect.~\ref{sec:test}.}
\end{table*}
\begin{table*}[!]
\caption{Median values with corresponding 68.3\,\% shortest credible intervals for the oscillation frequencies, amplitudes, linewidths, and heights of the mixed modes of KIC~7060732, as derived by \diamonds\,\,by using the peak bagging model defined by Eqs.~(\ref{eq:general_pb_model}) and (\ref{eq:pb_model}).}
\label{tab:7060732m}
\centering
\begin{tabular}{llcrrlrc}
\hline\hline
\\[-8pt]          
Peak \# & $\ell$ & $m$ & \multicolumn{1}{c}{Frequency} & \multicolumn{1}{c}{Amplitude} & \multicolumn{1}{c}{Linewidth} & \multicolumn{1}{c}{Height}& $p_\mathrm{B}$\\
 & & & \multicolumn{1}{c}{($\mu$Hz)} & \multicolumn{1}{c}{(ppm)} & \multicolumn{1}{c}{($\mu$Hz)} & \multicolumn{1}{c}{(ppm$^2/\mu$Hz)}\\
\hline \\[-8pt]
0 & 1 & ? & $     86.7710_{-      0.0017}^{+      0.0016}$ & \multicolumn{1}{c}{\dots} & \multicolumn{1}{c}{\dots} & $     1376.15_{-      191.05}^{+       96.42}$ & 0.978\\[1pt]

1 & 1 & ? & $     92.3033_{-      0.0010}^{+      0.0009}$ & \multicolumn{1}{c}{\dots} & \multicolumn{1}{c}{\dots} & $      915.55_{-       21.04}^{+       24.60}$ & 0.939\\[1pt]
2 & 1 & ? & $     95.6785_{-      0.0015}^{+      0.0014}$ & \multicolumn{1}{c}{\dots} & \multicolumn{1}{c}{\dots} & $     1318.95_{-       62.12}^{+       56.48}$ & 0.996\\[1pt]
3 & 1 & ? & $     96.2809_{-      0.0052}^{+      0.0051}$ & $       8.602_{-       0.426}^{+       0.427}$ & $       0.075_{-       0.005}^{+       0.005}$ & \multicolumn{1}{c}{\dots} & 1.000\\
\hline
\end{tabular}
\end{table*}

\begin{table*}[!]
\caption{Table~\ref{tab:7060732m} continued.}
\label{tab:7060732m2}
\centering
\begin{tabular}{llcrrlrc}
\hline\hline
\\[-8pt]          
Peak \# & $\ell$ & $m$ & \multicolumn{1}{c}{Frequency} & \multicolumn{1}{c}{Amplitude} & \multicolumn{1}{c}{Linewidth} & \multicolumn{1}{c}{Height}& $p_\mathrm{B}$\\
 & & & \multicolumn{1}{c}{($\mu$Hz)} & \multicolumn{1}{c}{(ppm)} & \multicolumn{1}{c}{($\mu$Hz)} & \multicolumn{1}{c}{(ppm$^2/\mu$Hz)}\\
\hline \\[-8pt]
51 & 1 & $-1$ & $    128.2569_{-      0.0024}^{+      0.0024}$ & $      40.985_{-       1.214}^{+       1.135}$ & $       0.087_{-       0.007}^{+       0.007}$ & \multicolumn{1}{c}{\dots} & \dots \\[1pt]
52 & 1 & $+1$ & $    128.5859_{-      0.0037}^{+      0.0043}$ & $      34.752_{-       1.043}^{+       0.993}$ & $       0.088_{-       0.006}^{+       0.006}$ & \multicolumn{1}{c}{\dots} & \dots \\[1pt]
53 & 1 & $-1$ & $    128.9984_{-      0.0027}^{+      0.0025}$ & $      24.458_{-       0.879}^{+       1.266}$ & $       0.069_{-       0.006}^{+       0.007}$ & \multicolumn{1}{c}{\dots} & \dots \\[1pt]
54 & 1 & 0 & $    129.2379_{-      0.0053}^{+      0.0063}$ & $      17.190_{-       0.748}^{+       0.601}$ & $       0.101_{-       0.007}^{+       0.008}$ & \multicolumn{1}{c}{\dots} & 1.000\\
\hline
\end{tabular}
\tablefoot{The first column represents the peak number in increasing frequency order and is shown for each angular degree ($\ell$) and azimuthal order ($m$), with question marks placed for m-values that could not be identified. The last column corresponds to the detection probability introduced by Eq.~(\ref{eq:detection_probability}) and discussed in Sect.~\ref{sec:test}.}
\end{table*}

\begin{table*}[!]
\caption{Median values with corresponding 68.3\,\% shortest credible intervals for the oscillation frequencies, amplitudes, and linewidths of the $p$ modes of KIC~7060732, as derived by \diamonds\,\,by using the peak bagging model defined by Eqs.~(\ref{eq:general_pb_model}) and (\ref{eq:pb_model}).}
\label{tab:7060732p}
\centering
\begin{tabular}{llcrrlrc}
\hline\hline
\\[-8pt]          
Peak \# & $\ell$ & $m$ & \multicolumn{1}{c}{Frequency} & \multicolumn{1}{c}{Amplitude} & \multicolumn{1}{c}{Linewidth} & \multicolumn{1}{c}{Height}& $p_\mathrm{B}$\\
 & & & \multicolumn{1}{c}{($\mu$Hz)} & \multicolumn{1}{c}{(ppm)} & \multicolumn{1}{c}{($\mu$Hz)} & \multicolumn{1}{c}{(ppm$^2/\mu$Hz)}\\
\hline \\[-8pt]
0 & 0 & 0 & $     90.7578_{-      0.0101}^{+      0.0092}$ & $      11.578_{-       0.826}^{+       0.805}$ & $       0.093_{-       0.014}^{+       0.014}$ & \multicolumn{1}{c}{\dots} & 1.000\\[1pt]
1 & 0 & 0 & $    101.3432_{-      0.0081}^{+      0.0073}$ & $      20.273_{-       0.625}^{+       0.597}$ & $       0.249_{-       0.019}^{+       0.016}$ & \multicolumn{1}{c}{\dots} & \dots \\[1pt]
2 & 0 & 0 & $    111.8478_{-      0.0035}^{+      0.0032}$ & $      29.672_{-       0.591}^{+       0.625}$ & $       0.105_{-       0.005}^{+       0.005}$ & \multicolumn{1}{c}{\dots} & \dots \\[1pt]
3 & 0 & 0 & $    122.7847_{-      0.0046}^{+      0.0048}$ & $      44.961_{-       1.223}^{+       1.097}$ & $       0.108_{-       0.003}^{+       0.003}$ & \multicolumn{1}{c}{\dots} & \dots \\
\hline
\end{tabular}
\tablefoot{The first column represents the peak number in increasing frequency order and is shown for each angular degree ($\ell$) and azimuthal order ($m$). The last column corresponds to the detection probability introduced by Eq.~(\ref{eq:detection_probability}) and discussed in Sect.~\ref{sec:test}.}
\end{table*}
\begin{table*}[!]
\caption{Median values with corresponding 68.3\,\% shortest credible intervals for the oscillation frequencies, amplitudes, linewidths, and heights of the mixed modes of KIC~7619745, as derived by \diamonds\,\,by using the peak bagging model defined by Eqs.~(\ref{eq:general_pb_model}) and (\ref{eq:pb_model}).}
\label{tab:7619745m}
\centering
\begin{tabular}{llcrrlrc}
\hline\hline
\\[-8pt]          
Peak \# & $\ell$ & $m$ & \multicolumn{1}{c}{Frequency} & \multicolumn{1}{c}{Amplitude} & \multicolumn{1}{c}{Linewidth} & \multicolumn{1}{c}{Height}& $p_\mathrm{B}$\\
 & & & \multicolumn{1}{c}{($\mu$Hz)} & \multicolumn{1}{c}{(ppm)} & \multicolumn{1}{c}{($\mu$Hz)} & \multicolumn{1}{c}{(ppm$^2/\mu$Hz)}\\
 \hline \\[-8pt]
0 & 1 & ? & $    127.3272_{-      0.0022}^{+      0.0015}$ & \multicolumn{1}{c}{\dots} & \multicolumn{1}{c}{\dots} & $      802.50_{-       57.42}^{+       61.74}$ & 0.998\\[1pt]
1 & 1 & ? & $    128.4370_{-      0.0115}^{+      0.0107}$ & $       7.429_{-       0.676}^{+       0.682}$ & $       0.097_{-       0.022}^{+       0.021}$ & \multicolumn{1}{c}{\dots} & 1.000\\[1pt]
2 & 1 & ? & $    128.7241_{-      0.0097}^{+      0.0103}$ & $       7.868_{-       0.756}^{+       0.690}$ & $       0.096_{-       0.027}^{+       0.024}$ & \multicolumn{1}{c}{\dots} & \dots \\[1pt]
3 & 1 & ? & $    129.2020_{-      0.0088}^{+      0.0088}$ & $       8.324_{-       0.677}^{+       0.703}$ & $       0.079_{-       0.018}^{+       0.018}$ & \multicolumn{1}{c}{\dots} & 1.000\\
\hline
\end{tabular}
\end{table*}

\begin{table*}[!]
\caption{Table~\ref{tab:7619745m} continued.}
\label{tab:7619745m2}
\centering
\begin{tabular}{llcrrlrc}
\hline\hline
\\[-8pt]          
Peak \# & $\ell$ & $m$ & \multicolumn{1}{c}{Frequency} & \multicolumn{1}{c}{Amplitude} & \multicolumn{1}{c}{Linewidth} & \multicolumn{1}{c}{Height}& $p_\mathrm{B}$\\
 & & & \multicolumn{1}{c}{($\mu$Hz)} & \multicolumn{1}{c}{(ppm)} & \multicolumn{1}{c}{($\mu$Hz)} & \multicolumn{1}{c}{(ppm$^2/\mu$Hz)}\\
\hline \\[-8pt]
51 & 1 & $+1$ & $    192.7203_{-      0.0085}^{+      0.0086}$ & $      10.055_{-       0.722}^{+       0.667}$ & $       0.100_{-       0.021}^{+       0.018}$ & \multicolumn{1}{c}{\dots} & \dots \\[1pt]
52 & 1 & $-1$ & $    193.8389_{-      0.0108}^{+      0.0102}$ & $      14.372_{-       1.013}^{+       1.041}$ & $       0.128_{-       0.024}^{+       0.021}$ & \multicolumn{1}{c}{\dots} & \dots \\[1pt]
53 & 1 & $+1$ & $    194.1314_{-      0.0219}^{+      0.0212}$ & $      12.647_{-       1.018}^{+       1.043}$ & $       0.213_{-       0.041}^{+       0.037}$ & \multicolumn{1}{c}{\dots} & \dots \\[1pt]
54 & 1 & $-1$ & $    195.6776_{-      0.0067}^{+      0.0067}$ & $       6.658_{-       0.612}^{+       0.587}$ & $       0.053_{-       0.014}^{+       0.011}$ & \multicolumn{1}{c}{\dots} & \dots \\
\hline
\end{tabular}
\tablefoot{The first column represents the peak number in increasing frequency order and is shown for each angular degree ($\ell$) and azimuthal order ($m$), with question marks placed for m-values that could not be identified. The last column corresponds to the detection probability introduced by Eq.~(\ref{eq:detection_probability}) and discussed in Sect.~\ref{sec:test}.}
\end{table*}

\begin{table*}[!]
\caption{Median values with corresponding 68.3\,\% shortest credible intervals for the oscillation frequencies, amplitudes, and linewidths of the $p$ modes of KIC~7619745, as derived by \diamonds\,\,by using the peak bagging model defined by Eqs.~(\ref{eq:general_pb_model}) and (\ref{eq:pb_model}).}
\label{tab:7619745p}
\centering
\begin{tabular}{llcrrlrc}
\hline\hline
\\[-8pt]          
Peak \# & $\ell$ & $m$ & \multicolumn{1}{c}{Frequency} & \multicolumn{1}{c}{Amplitude} & \multicolumn{1}{c}{Linewidth} & \multicolumn{1}{c}{Height}& $p_\mathrm{B}$\\
 & & & \multicolumn{1}{c}{($\mu$Hz)} & \multicolumn{1}{c}{(ppm)} & \multicolumn{1}{c}{($\mu$Hz)} & \multicolumn{1}{c}{(ppm$^2/\mu$Hz)}\\
\hline \\[-8pt]
0 & 0 & 0 & $    122.4306_{-      0.0198}^{+      0.0235}$ & $       7.443_{-       0.944}^{+       0.904}$ & $       0.138_{-       0.020}^{+       0.020}$ & \multicolumn{1}{c}{\dots} & 1.000\\[1pt]
1 & 0 & 0 & $    135.0314_{-      0.0122}^{+      0.0131}$ & $      17.804_{-       0.758}^{+       0.757}$ & $       0.255_{-       0.035}^{+       0.029}$ & \multicolumn{1}{c}{\dots} & \dots \\[1pt]
2 & 0 & 0 & $    147.8301_{-      0.0048}^{+      0.0048}$ & $      24.875_{-       1.013}^{+       0.961}$ & $       0.096_{-       0.009}^{+       0.010}$ & \multicolumn{1}{c}{\dots} & \dots \\[1pt]
3 & 0 & 0 & $    160.9626_{-      0.0029}^{+      0.0033}$ & $      39.661_{-       1.564}^{+       1.358}$ & $       0.074_{-       0.007}^{+       0.007}$ & \multicolumn{1}{c}{\dots} & \dots \\
\hline
\end{tabular}
\tablefoot{The first column represents the peak number in increasing frequency order and is shown for each angular degree ($\ell$) and azimuthal order ($m$). The last column corresponds to the detection probability introduced by Eq.~(\ref{eq:detection_probability}) and discussed in Sect.~\ref{sec:test}.}
\end{table*}
\begin{table*}[!]
\caption{Median values with corresponding 68.3\,\% shortest credible intervals for the oscillation frequencies, amplitudes, linewidths, and heights of the mixed modes of KIC~8366239, as derived by \diamonds\,\,by using the peak bagging model defined by Eqs.~(\ref{eq:general_pb_model}) and (\ref{eq:pb_model}).}
\label{tab:8366239m}
\centering
\begin{tabular}{llcrrlrc}
\hline\hline
\\[-8pt]      
Peak \# & $\ell$ & $m$ & \multicolumn{1}{c}{Frequency} & \multicolumn{1}{c}{Amplitude} & \multicolumn{1}{c}{Linewidth} & \multicolumn{1}{c}{Height}& $p_\mathrm{B}$\\
 & & & \multicolumn{1}{c}{($\mu$Hz)} & \multicolumn{1}{c}{(ppm)} & \multicolumn{1}{c}{($\mu$Hz)} & \multicolumn{1}{c}{(ppm$^2/\mu$Hz)}\\
\hline \\[-8pt]

0 & 1 & ? & $    134.2419_{-      0.0020}^{+      0.0024}$ & \multicolumn{1}{c}{\dots} & \multicolumn{1}{c}{\dots} & $      764.26_{-       63.17}^{+       73.97}$ & 0.997\\[1pt]
1 & 1 & ? & $    135.9448_{-      0.0009}^{+      0.0008}$ & \multicolumn{1}{c}{\dots} & \multicolumn{1}{c}{\dots} & $      826.19_{-       78.51}^{+       77.35}$ & 1.000\\[1pt]
2 & 1 & ? & $    136.3370_{-      0.0011}^{+      0.0012}$ & \multicolumn{1}{c}{\dots} & \multicolumn{1}{c}{\dots} & $      782.27_{-       66.23}^{+       72.28}$ & 0.999\\[1pt]

3 & 1 & $-1$ & $    145.7400_{-      0.0011}^{+      0.0011}$ & \multicolumn{1}{c}{\dots} & \multicolumn{1}{c}{\dots} & $      503.53_{-       69.71}^{+       33.03}$ & 0.999\\
\hline
\end{tabular}
\tablefoot{The first column represents the peak number in increasing frequency order and is shown for each angular degree ($\ell$) and azimuthal order ($m$), with question marks placed for m-values that could not be identified. The last column corresponds to the detection probability introduced by Eq.~(\ref{eq:detection_probability}) and discussed in Sect.~\ref{sec:test}.}
\end{table*}

\begin{table*}[!]
\caption{Median values with corresponding 68.3\,\% shortest credible intervals for the oscillation frequencies, amplitudes, and linewidt hs of the $p$ modes of KIC~8366239, as derived by \diamonds\,\,by using the peak bagging model defined by Eqs.~(\ref{eq:general_pb_model}) and (\ref{eq:pb_model}).}
\label{tab:8366239p}
\centering
\begin{tabular}{llcrrlrc}
\hline\hline
\\[-8pt]          
Peak \# & $\ell$ & $m$ & \multicolumn{1}{c}{Frequency} & \multicolumn{1}{c}{Amplitude} & \multicolumn{1}{c}{Linewidth} & \multicolumn{1}{c}{Height}& $p_\mathrm{B}$\\
 & & & \multicolumn{1}{c}{($\mu$Hz)} & \multicolumn{1}{c}{(ppm)} & \multicolumn{1}{c}{($\mu$Hz)} & \multicolumn{1}{c}{(ppm$^2/\mu$Hz)}\\
\hline \\[-8pt]
0 & 0 & 0 & $    128.2646_{-      0.0179}^{+      0.0174}$ & $       7.611_{-       0.957}^{+       0.948}$ & $       0.115_{-       0.032}^{+       0.030}$ & \multicolumn{1}{c}{\dots} & 1.000\\[1pt]
1 & 0 & 0 & $    141.6492_{-      0.0140}^{+      0.0138}$ & $      11.765_{-       0.893}^{+       0.884}$ & $       0.123_{-       0.029}^{+       0.024}$ & \multicolumn{1}{c}{\dots} & \dots \\[1pt]
2 & 0 & 0 & $    154.9071_{-      0.0081}^{+      0.0081}$ & $      20.913_{-       0.808}^{+       0.830}$ & $       0.170_{-       0.019}^{+       0.018}$ & \multicolumn{1}{c}{\dots} & \dots \\[1pt]
3 & 0 & 0 & $    168.5823_{-      0.0077}^{+      0.0075}$ & $      26.274_{-       0.919}^{+       0.984}$ & $       0.149_{-       0.015}^{+       0.015}$ & \multicolumn{1}{c}{\dots} & \dots \\
\hline
\end{tabular}
\tablefoot{The first column represents the peak number in increasing frequency order and is shown for each angular degree ($\ell$) and azimuthal order ($m$). The last column corresponds to the detection probability introduced by Eq.~(\ref{eq:detection_probability}) and discussed in Sect.~\ref{sec:test}.}
\end{table*}
\begin{table*}[!]
\caption{Median values with corresponding 68.3\,\% shortest credible intervals for the oscillation frequencies, amplitudes, linewidths, and heights of the mixed modes of KIC~8475025, as derived by \diamonds\,\,by using the peak bagging model defined by Eqs.~(\ref{eq:general_pb_model}) and (\ref{eq:pb_model}).}
\label{tab:8475025m}
\centering
\begin{tabular}{llcrrlrc}
\hline\hline
\\[-8pt]      
Peak \# & $\ell$ & $m$ & \multicolumn{1}{c}{Frequency} & \multicolumn{1}{c}{Amplitude} & \multicolumn{1}{c}{Linewidth} & \multicolumn{1}{c}{Height}& $p_\mathrm{B}$\\
 & & & \multicolumn{1}{c}{($\mu$Hz)} & \multicolumn{1}{c}{(ppm)} & \multicolumn{1}{c}{($\mu$Hz)} & \multicolumn{1}{c}{(ppm$^2/\mu$Hz)}\\
\hline \\[-8pt]
0 & 1 & ? & $     84.2866_{-      0.0100}^{+      0.0099}$ & $      13.277_{-       0.733}^{+       0.852}$ & $       0.161_{-       0.019}^{+       0.017}$ & \multicolumn{1}{c}{\dots} & 1.000\\[1pt]
1 & 1 & ? & $     84.6562_{-      0.0081}^{+      0.0083}$ & $      15.128_{-       0.915}^{+       0.886}$ & $       0.113_{-       0.013}^{+       0.011}$ & \multicolumn{1}{c}{\dots} & \dots \\[1pt]
2 & 1 & ? & $     85.1655_{-      0.0014}^{+      0.0014}$ & \multicolumn{1}{c}{\dots} & \multicolumn{1}{c}{\dots} & $     1592.16_{-      123.18}^{+      207.86}$ & 0.986\\[1pt]

2 & 1 & ? & $     92.5589_{-      0.0009}^{+      0.0010}$ & \multicolumn{1}{c}{\dots} & \multicolumn{1}{c}{\dots} & $     2021.58_{-      221.71}^{+      245.43}$ & 1.000\\
\hline
\end{tabular}
\end{table*}

\begin{table*}[!]
\caption{Table~\ref{tab:8475025m} continued.}
\label{tab:8475025m2}
\centering
\begin{tabular}{llcrrlrc}
\hline\hline
\\[-8pt]          
Peak \# & $\ell$ & $m$ & \multicolumn{1}{c}{Frequency} & \multicolumn{1}{c}{Amplitude} & \multicolumn{1}{c}{Linewidth} & \multicolumn{1}{c}{Height}& $p_\mathrm{B}$\\
 & & & \multicolumn{1}{c}{($\mu$Hz)} & \multicolumn{1}{c}{(ppm)} & \multicolumn{1}{c}{($\mu$Hz)} & \multicolumn{1}{c}{(ppm$^2/\mu$Hz)}\\
\hline \\[-8pt]
51 & 1 & $-1$ & $    129.3205_{-      0.0007}^{+      0.0008}$ & \multicolumn{1}{c}{\dots} & \multicolumn{1}{c}{\dots} & $     1678.86_{-       49.87}^{+       42.67}$ & \dots \\[1pt]
52 & 1 & $+1$ & $    129.9531_{-      0.0020}^{+      0.0023}$ & \multicolumn{1}{c}{\dots} & \multicolumn{1}{c}{\dots} & $      517.46_{-       15.78}^{+       17.76}$ & 0.902\\[1pt]
53 & 1 & $-1$ & $    130.5478_{-      0.0017}^{+      0.0035}$ & \multicolumn{1}{c}{\dots} & \multicolumn{1}{c}{\dots} & $     3174.40_{-      291.80}^{+      238.77}$ & \dots \\[1pt]
54 & 1 & $+1$ & $    131.1119_{-      0.0099}^{+      0.0094}$ & $       6.540_{-       0.244}^{+       0.249}$ & $       0.110_{-       0.012}^{+       0.010}$ & \multicolumn{1}{c}{\dots} & 0.942\\
\hline
\end{tabular}
\tablefoot{The first column represents the peak number in increasing frequency order and is shown for each angular degree ($\ell$) and azimuthal order ($m$), with question marks placed for m-values that could not be identified. The last column corresponds to the detection probability introduced by Eq.~(\ref{eq:detection_probability}) and discussed in Sect.~\ref{sec:test}.}
\end{table*}

\begin{table*}[!]
\caption{Median values with corresponding 68.3\,\% shortest credible intervals for the oscillation frequencies, amplitudes, and linewidths of the $p$ modes of KIC~8475025, as derived by \diamonds\,\,by using the peak bagging model defined by Eqs.~(\ref{eq:general_pb_model}) and (\ref{eq:pb_model}).}
\label{tab:8475025p}
\centering
\begin{tabular}{llcrrlrc}
\hline\hline
\\[-8pt]          
Peak \# & $\ell$ & $m$ & \multicolumn{1}{c}{Frequency} & \multicolumn{1}{c}{Amplitude} & \multicolumn{1}{c}{Linewidth} & \multicolumn{1}{c}{Height}& $p_\mathrm{B}$\\
 & & & \multicolumn{1}{c}{($\mu$Hz)} & \multicolumn{1}{c}{(ppm)} & \multicolumn{1}{c}{($\mu$Hz)} & \multicolumn{1}{c}{(ppm$^2/\mu$Hz)}\\
\hline \\[-8pt]
0 & 0 & 0 & $     79.6595_{-      0.0074}^{+      0.0066}$ & $      18.576_{-       0.884}^{+       0.917}$ & $       0.114_{-       0.014}^{+       0.013}$ & \multicolumn{1}{c}{\dots} & \dots \\[1pt]
1 & 0 & 0 & $     88.8818_{-      0.0096}^{+      0.0114}$ & $      23.299_{-       0.777}^{+       0.839}$ & $       0.297_{-       0.031}^{+       0.030}$ & \multicolumn{1}{c}{\dots} & \dots \\[1pt]
2 & 0 & 0 & $     98.2762_{-      0.0121}^{+      0.0122}$ & $      28.403_{-       1.592}^{+       1.417}$ & $       0.160_{-       0.025}^{+       0.023}$ & \multicolumn{1}{c}{\dots} & \dots \\[1pt]
3 & 0 & 0 & $    107.9356_{-      0.0028}^{+      0.0026}$ & $      61.524_{-       1.535}^{+       1.431}$ & $       0.084_{-       0.005}^{+       0.004}$ & \multicolumn{1}{c}{\dots} & \dots \\
\hline
\end{tabular}
\tablefoot{The first column represents the peak number in increasing frequency order and is shown for each angular degree ($\ell$) and azimuthal order ($m$). The last column corresponds to the detection probability introduced by Eq.~(\ref{eq:detection_probability}) and discussed in Sect.~\ref{sec:test}.}
\end{table*}
\begin{table*}[!]
\caption{Median values with corresponding 68.3\,\% shortest credible intervals for the oscillation frequencies, amplitudes, linewidths, and heights of the mixed modes of KIC~8718745, as derived by \diamonds\,\,by using the peak bagging model defined by Eqs.~(\ref{eq:general_pb_model}) and (\ref{eq:pb_model}).}
\label{tab:8718745m}
\centering
\begin{tabular}{llcrrlrc}
\hline\hline
\\[-8pt]      
Peak \# & $\ell$ & $m$ & \multicolumn{1}{c}{Frequency} & \multicolumn{1}{c}{Amplitude} & \multicolumn{1}{c}{Linewidth} & \multicolumn{1}{c}{Height}& $p_\mathrm{B}$\\
 & & & \multicolumn{1}{c}{($\mu$Hz)} & \multicolumn{1}{c}{(ppm)} & \multicolumn{1}{c}{($\mu$Hz)} & \multicolumn{1}{c}{(ppm$^2/\mu$Hz)}\\
\hline \\[-8pt]

0 & 1 & 0 & $     96.3463_{-      0.0013}^{+      0.0011}$ & \multicolumn{1}{c}{\dots} & \multicolumn{1}{c}{\dots} & $     1235.51_{-       95.60}^{+       94.35}$ & 0.994\\[1pt]
1 & 1 & ? & $     96.7350_{-      0.0012}^{+      0.0012}$ & \multicolumn{1}{c}{\dots} & \multicolumn{1}{c}{\dots} & $     1221.90_{-       82.60}^{+       85.46}$  & 0.999\\[1pt]
2 & 1 & 0 & $     98.4801_{-      0.0010}^{+      0.0010}$ & \multicolumn{1}{c}{\dots} & \multicolumn{1}{c}{\dots} & $     3417.79_{-      252.03}^{+      309.03}$ & \dots \\[1pt]
3 & 1 & 0 & $     99.2110_{-      0.0007}^{+      0.0006}$ & \multicolumn{1}{c}{\dots} & \multicolumn{1}{c}{\dots} & $     2444.75_{-      179.24}^{+      188.74}$ & \dots \\
\hline
\end{tabular}
\end{table*}

\begin{table*}[!]
\caption{Table~\ref{tab:8718745m} continued.}
\label{tab:8718745m2}
\centering
\begin{tabular}{llcrrlrc}
\hline\hline
\\[-8pt]          
Peak \# & $\ell$ & $m$ & \multicolumn{1}{c}{Frequency} & \multicolumn{1}{c}{Amplitude} & \multicolumn{1}{c}{Linewidth} & \multicolumn{1}{c}{Height}& $p_\mathrm{B}$\\
 & & & \multicolumn{1}{c}{($\mu$Hz)} & \multicolumn{1}{c}{(ppm)} & \multicolumn{1}{c}{($\mu$Hz)} & \multicolumn{1}{c}{(ppm$^2/\mu$Hz)}\\
\hline \\[-8pt]

51 & 1 & 0 & $    146.0053_{-      0.0042}^{+      0.0040}$ & $      19.944_{-       0.495}^{+       0.596}$ & $       0.084_{-       0.009}^{+       0.009}$ & \multicolumn{1}{c}{\dots} & \dots \\[1pt]
52 & 1 & $+1$ & $    146.3312_{-      0.0044}^{+      0.0046}$ & $       6.962_{-       0.408}^{+       0.351}$ & $       0.060_{-       0.006}^{+       0.006}$ & \multicolumn{1}{c}{\dots} & 0.995 \\[1pt]
53 & 1 & 0 & $    147.5242_{-      0.0055}^{+      0.0049}$ & $      10.033_{-       0.391}^{+       0.405}$ & $       0.105_{-       0.009}^{+       0.008}$ & \multicolumn{1}{c}{\dots} & 1.000 \\[1pt]

54 & 1 & 0 & $    154.5704_{-      0.0045}^{+      0.0053}$ & $       8.208_{-       0.399}^{+       0.419}$ & $       0.074_{-       0.009}^{+       0.009}$ & \multicolumn{1}{c}{\dots} & \dots \\
\hline
\end{tabular}
\tablefoot{The first column represents the peak number in increasing frequency order and is shown for each angular degree ($\ell$) and azimuthal order ($m$), with question marks placed for m-values that could not be identified. The last column corresponds to the detection probability introduced by Eq.~(\ref{eq:detection_probability}) and discussed in Sect.~\ref{sec:test}.}
\end{table*}

\begin{table*}[!]
\caption{Median values with corresponding 68.3\,\% shortest credible intervals for the oscillation frequencies, amplitudes, and linewidths of the $p$ modes of KIC~8718745, as derived by \diamonds\,\,by using the peak bagging model defined by Eqs.~(\ref{eq:general_pb_model}) and (\ref{eq:pb_model}).}
\label{tab:8718745p}
\centering
\begin{tabular}{llcrrlrc}
\hline\hline
\\[-8pt]          
Peak \# & $\ell$ & $m$ & \multicolumn{1}{c}{Frequency} & \multicolumn{1}{c}{Amplitude} & \multicolumn{1}{c}{Linewidth} & \multicolumn{1}{c}{Height}& $p_\mathrm{B}$\\
 & & & \multicolumn{1}{c}{($\mu$Hz)} & \multicolumn{1}{c}{(ppm)} & \multicolumn{1}{c}{($\mu$Hz)} & \multicolumn{1}{c}{(ppm$^2/\mu$Hz)}\\
\hline \\[-8pt]
0 & 0 & 0 & $     94.4523_{-      0.0108}^{+      0.0101}$ & $      16.216_{-       2.087}^{+       1.720}$ & $       0.093_{-       0.022}^{+       0.018}$ & \multicolumn{1}{c}{\dots} & \dots \\[1pt]
1 & 0 & 0 & $    105.2881_{-      0.0060}^{+      0.0061}$ & $      27.554_{-       1.427}^{+       1.371}$ & $       0.087_{-       0.012}^{+       0.012}$ & \multicolumn{1}{c}{\dots} & \dots \\[1pt]
2 & 0 & 0 & $    116.5777_{-      0.0046}^{+      0.0043}$ & $      42.850_{-       1.398}^{+       1.397}$ & $       0.108_{-       0.011}^{+       0.009}$ & \multicolumn{1}{c}{\dots} & \dots \\[1pt]
3 & 0 & 0 & $    127.9038_{-      0.0021}^{+      0.0022}$ & $      60.299_{-       2.089}^{+       1.705}$ & $       0.074_{-       0.005}^{+       0.005}$ & \multicolumn{1}{c}{\dots} & \dots \\
\hline
\end{tabular}
\tablefoot{The first column represents the peak number in increasing frequency order and is shown for each angular degree ($\ell$) and azimuthal order ($m$). The last column corresponds to the detection probability introduced by Eq.~(\ref{eq:detection_probability}) and discussed in Sect.~\ref{sec:test}.}
\end{table*}
\begin{table*}[!]
\caption{Median values with corresponding 68.3\,\% shortest credible intervals for the oscillation frequencies, amplitudes, linewidths, and heights of the mixed modes of KIC~9145955, as derived by \diamonds\,\,by using the peak bagging model defined by Eqs.~(\ref{eq:general_pb_model}) and (\ref{eq:pb_model}).}
\label{tab:9145955m}
\centering
\begin{tabular}{llcrrlrc}
\hline\hline
\\[-8pt]      
Peak \# & $\ell$ & $m$ & \multicolumn{1}{c}{Frequency} & \multicolumn{1}{c}{Amplitude} & \multicolumn{1}{c}{Linewidth} & \multicolumn{1}{c}{Height}& $p_\mathrm{B}$\\
 & & & \multicolumn{1}{c}{($\mu$Hz)} & \multicolumn{1}{c}{(ppm)} & \multicolumn{1}{c}{($\mu$Hz)} & \multicolumn{1}{c}{(ppm$^2/\mu$Hz)}\\
\hline \\[-8pt]
0 & 1 & 0 & $     85.6211_{-      0.0012}^{+      0.0013}$ & \multicolumn{1}{c}{\dots} & \multicolumn{1}{c}{\dots} & $      950.46_{-       61.73}^{+       40.07}$ & 0.972\\[1pt]
1 & 1 & 0 & $     86.0435_{-      0.0058}^{+      0.0057}$ & $       5.835_{-       0.294}^{+       0.353}$ & $       0.055_{-       0.004}^{+       0.004}$ & \multicolumn{1}{c}{\dots} & 1.000\\[1pt]

2 & 1 & 0 & $     95.7987_{-      0.0013}^{+      0.0014}$ & \multicolumn{1}{c}{\dots} & \multicolumn{1}{c}{\dots} & $     1600.71_{-       74.24}^{+       58.25}$ & \dots \\[1pt]
3 & 1 & 0 & $     96.3989_{-      0.0012}^{+      0.0015}$ & \multicolumn{1}{c}{\dots} & \multicolumn{1}{c}{\dots} & $     1423.21_{-       59.61}^{+       60.33}$ & \dots \\
\hline
\end{tabular}
\tablefoot{The first column represents the peak number in increasing frequency order and is shown for each angular degree ($\ell$) and azimuthal order ($m$), with question marks placed for m-values that could not be identified. The last column corresponds to the detection probability introduced by Eq.~(\ref{eq:detection_probability}) and discussed in Sect.~\ref{sec:test}.}
\end{table*}

\begin{table*}[!]
\caption{Median values with corresponding 68.3\,\% shortest credible intervals for the oscillation frequencies, amplitudes, and linewidths of the $p$ modes of KIC~9145955, as derived by \diamonds\,\,by using the peak bagging model defined by Eqs.~(\ref{eq:general_pb_model}) and (\ref{eq:pb_model}).}
\label{tab:9145955p}
\centering
\begin{tabular}{llcrrlrc}
\hline\hline
\\[-8pt]          
Peak \# & $\ell$ & $m$ & \multicolumn{1}{c}{Frequency} & \multicolumn{1}{c}{Amplitude} & \multicolumn{1}{c}{Linewidth} & \multicolumn{1}{c}{Height}& $p_\mathrm{B}$\\
 & & & \multicolumn{1}{c}{($\mu$Hz)} & \multicolumn{1}{c}{(ppm)} & \multicolumn{1}{c}{($\mu$Hz)} & \multicolumn{1}{c}{(ppm$^2/\mu$Hz)}\\
\hline \\[-8pt]
0 & 0 & 0 & $     91.2635_{-      0.0109}^{+      0.0103}$ & $       3.113_{-       0.168}^{+       0.193}$ & $       0.081_{-       0.009}^{+       0.008}$ & \multicolumn{1}{c}{\dots} & \dots \\[1pt]
1 & 0 & 0 & $    101.8793_{-      0.0087}^{+      0.0092}$ & $      18.024_{-       0.791}^{+       0.680}$ & $       0.173_{-       0.020}^{+       0.020}$ & \multicolumn{1}{c}{\dots} & \dots \\[1pt]
2 & 0 & 0 & $    112.5691_{-      0.0051}^{+      0.0058}$ & $      26.790_{-       0.885}^{+       0.940}$ & $       0.132_{-       0.013}^{+       0.011}$ & \multicolumn{1}{c}{\dots} & \dots \\[1pt]
3 & 0 & 0 & $    123.5742_{-      0.0056}^{+      0.0051}$ & $      39.099_{-       1.481}^{+       1.365}$ & $       0.118_{-       0.012}^{+       0.012}$ & \multicolumn{1}{c}{\dots} & \dots \\
\hline
\end{tabular}
\tablefoot{The first column represents the peak number in increasing frequency order and is shown for each angular degree ($\ell$) and azimuthal order ($m$). The last column corresponds to the detection probability introduced by Eq.~(\ref{eq:detection_probability}) and discussed in Sect.~\ref{sec:test}.}
\end{table*}
\begin{table*}[!]
\caption{Median values with corresponding 68.3\,\% shortest credible intervals for the oscillation frequencies, amplitudes, linewidths, and heights of the mixed modes of KIC~9267654, as derived by \diamonds\,\,by using the peak bagging model defined by Eqs.~(\ref{eq:general_pb_model}) and (\ref{eq:pb_model}).}
\label{tab:9267654m}
\centering
\begin{tabular}{llcrrlrc}
\hline\hline
\\[-8pt]      
Peak \# & $\ell$ & $m$ & \multicolumn{1}{c}{Frequency} & \multicolumn{1}{c}{Amplitude} & \multicolumn{1}{c}{Linewidth} & \multicolumn{1}{c}{Height}& $p_\mathrm{B}$\\
 & & & \multicolumn{1}{c}{($\mu$Hz)} & \multicolumn{1}{c}{(ppm)} & \multicolumn{1}{c}{($\mu$Hz)} & \multicolumn{1}{c}{(ppm$^2/\mu$Hz)}\\
\hline \\[-8pt]
0 & 1 & ? & $     80.6484_{-      0.0020}^{+      0.0020}$ & \multicolumn{1}{c}{\dots} & \multicolumn{1}{c}{\dots} & $     1610.79_{-       95.31}^{+      260.38}$ & 0.998\\[1pt]
 
1 & 1 & ? & $     90.2169_{-      0.0014}^{+      0.0013}$ & \multicolumn{1}{c}{\dots} & \multicolumn{1}{c}{\dots} & $     1320.69_{-       41.34}^{+       46.79}$ & 0.997\\[1pt]
2 & 1 & ? & $     90.4656_{-      0.0004}^{+      0.0004}$ & \multicolumn{1}{c}{\dots} & \multicolumn{1}{c}{\dots} & $     1666.71_{-       93.73}^{+       77.61}$ & 1.000\\[1pt]
3 & 1 & ? & $     90.7017_{-      0.0005}^{+      0.0005}$ & \multicolumn{1}{c}{\dots} & \multicolumn{1}{c}{\dots} & $     1391.60_{-       44.57}^{+       46.99}$ & 1.000\\
\hline
\end{tabular}
\end{table*}

\begin{table*}[!]
\caption{Table~\ref{tab:9267654m} continued.}
\label{tab:9267654m2}
\centering
\begin{tabular}{llcrrlrc}
\hline\hline
\\[-8pt]          
Peak \# & $\ell$ & $m$ & \multicolumn{1}{c}{Frequency} & \multicolumn{1}{c}{Amplitude} & \multicolumn{1}{c}{Linewidth} & \multicolumn{1}{c}{Height}& $p_\mathrm{B}$\\
 & & & \multicolumn{1}{c}{($\mu$Hz)} & \multicolumn{1}{c}{(ppm)} & \multicolumn{1}{c}{($\mu$Hz)} & \multicolumn{1}{c}{(ppm$^2/\mu$Hz)}\\
\hline \\[-8pt]
51 & 1 & $-1$ & $    139.3404_{-      0.0013}^{+      0.0014}$ & \multicolumn{1}{c}{\dots} & \multicolumn{1}{c}{\dots} & $      431.22_{-       26.14}^{+       25.46}$ & 0.964\\[1pt]
52 & 1 & $+1$ & $    140.2373_{-      0.0023}^{+      0.0029}$ & \multicolumn{1}{c}{\dots} & \multicolumn{1}{c}{\dots} & $      797.40_{-       91.92}^{+       76.13}$ & 1.000 \\[1pt]
53 & 1 & $-1$ & $    140.7795_{-      0.0137}^{+      0.0123}$ & $       3.594_{-       0.345}^{+       0.335}$ & $       0.082_{-       0.014}^{+       0.013}$ & \multicolumn{1}{c}{\dots} & 0.960 \\[1pt]
54 & 1 & $+1$ & $    141.5134_{-      0.0122}^{+      0.0107}$ & $       9.830_{-       0.673}^{+       0.594}$ & $       0.152_{-       0.030}^{+       0.026}$ & \multicolumn{1}{c}{\dots} & \dots \\
\hline
\end{tabular}
\tablefoot{The first column represents the peak number in increasing frequency order and is shown for each angular degree ($\ell$) and azimuthal order ($m$), with question marks placed for m-values that could not be identified. The last column corresponds to the detection probability introduced by Eq.~(\ref{eq:detection_probability}) and discussed in Sect.~\ref{sec:test}.}
\end{table*}

\begin{table*}[!]
\caption{Median values with corresponding 68.3\,\% shortest credible intervals for the oscillation frequencies, amplitudes, and linewidths of the $p$ modes of KIC~9267654, as derived by \diamonds\,\,by using the peak bagging model defined by Eqs.~(\ref{eq:general_pb_model}) and (\ref{eq:pb_model}).}
\label{tab:9267654p}
\centering
\begin{tabular}{llcrrlrc}
\hline\hline
\\[-8pt]
Peak \# & $\ell$ & $m$ & \multicolumn{1}{c}{Frequency} & \multicolumn{1}{c}{Amplitude} & \multicolumn{1}{c}{Linewidth} & \multicolumn{1}{c}{Height}& $p_\mathrm{B}$\\
 & & & \multicolumn{1}{c}{($\mu$Hz)} & \multicolumn{1}{c}{(ppm)} & \multicolumn{1}{c}{($\mu$Hz)} & \multicolumn{1}{c}{(ppm$^2/\mu$Hz)}\\
\hline \\[-8pt]
0 & 0 & 0 & $     85.5971_{-      0.0082}^{+      0.0081}$ & $      10.456_{-       0.641}^{+       0.816}$ & $       0.074_{-       0.012}^{+       0.011}$ & \multicolumn{1}{c}{\dots} & 1.000\\[1pt]
1 & 0 & 0 & $     95.3733_{-      0.0053}^{+      0.0051}$ & $      23.188_{-       0.704}^{+       0.646}$ & $       0.134_{-       0.010}^{+       0.011}$ & \multicolumn{1}{c}{\dots} & \dots \\[1pt]
2 & 0 & 0 & $    105.5316_{-      0.0048}^{+      0.0044}$ & $      33.985_{-       0.905}^{+       0.838}$ & $       0.132_{-       0.010}^{+       0.009}$ & \multicolumn{1}{c}{\dots} & \dots \\[1pt]
3 & 0 & 0 & $    115.8315_{-      0.0023}^{+      0.0023}$ & $      57.790_{-       1.120}^{+       1.127}$ & $       0.087_{-       0.005}^{+       0.004}$ & \multicolumn{1}{c}{\dots} & \dots \\
\hline
\end{tabular}
\tablefoot{The first column represents the peak number in increasing frequency order and is shown for each angular degree ($\ell$) and azimuthal order ($m$). The last column corresponds to the detection probability introduced by Eq.~(\ref{eq:detection_probability}) and discussed in Sect.~\ref{sec:test}.}
\end{table*}

\begin{table*}[!]
\caption{Median values with corresponding 68.3\,\% shortest credible intervals for the oscillation frequencies, amplitudes, linewidths, and heights of the mixed modes of KIC~9475697, as derived by \diamonds\,\,by using the peak bagging model defined by Eqs.~(\ref{eq:general_pb_model}) and (\ref{eq:pb_model}).}
\label{tab:9475697m}
\centering
\begin{tabular}{llcrrlrc}
\hline\hline
\\[-8pt]      
Peak \# & $\ell$ & $m$ & \multicolumn{1}{c}{Frequency} & \multicolumn{1}{c}{Amplitude} & \multicolumn{1}{c}{Linewidth} & \multicolumn{1}{c}{Height}& $p_\mathrm{B}$\\
 & & & \multicolumn{1}{c}{($\mu$Hz)} & \multicolumn{1}{c}{(ppm)} & \multicolumn{1}{c}{($\mu$Hz)} & \multicolumn{1}{c}{(ppm$^2/\mu$Hz)}\\
\hline \\[-8pt]
0 & 1 & 0 & $     83.8974_{-      0.0019}^{+      0.0019}$ & \multicolumn{1}{c}{\dots} & \multicolumn{1}{c}{\dots} & $     1441.21_{-       82.84}^{+       72.87}$ & 0.995\\[1pt]
1 & 1 & 0 & $     85.4863_{-      0.0016}^{+      0.0016}$ & \multicolumn{1}{c}{\dots} & \multicolumn{1}{c}{\dots} & $     1244.56_{-       72.14}^{+       65.74}$ & 0.986\\[1pt]
2 & 1 & 0 & $     85.9867_{-      0.0004}^{+      0.0004}$ & \multicolumn{1}{c}{\dots} & \multicolumn{1}{c}{\dots} & $     3380.44_{-      283.74}^{+      377.79}$ & \dots \\[1pt]
3 & 1 & 0 & $     86.4125_{-      0.0065}^{+      0.0054}$ & $      19.869_{-       1.009}^{+       0.934}$ & $       0.090_{-       0.014}^{+       0.012}$ & \multicolumn{1}{c}{\dots} & \dots \\
\hline
\end{tabular}
\end{table*}

\begin{table*}[!]
\caption{Table~\ref{tab:9475697m} continued.}
\label{tab:9475697m2}
\centering
\begin{tabular}{llcrrlrc}
\hline\hline
\\[-8pt]          
Peak \# & $\ell$ & $m$ & \multicolumn{1}{c}{Frequency} & \multicolumn{1}{c}{Amplitude} & \multicolumn{1}{c}{Linewidth} & \multicolumn{1}{c}{Height}& $p_\mathrm{B}$\\
 & & & \multicolumn{1}{c}{($\mu$Hz)} & \multicolumn{1}{c}{(ppm)} & \multicolumn{1}{c}{($\mu$Hz)} & \multicolumn{1}{c}{(ppm$^2/\mu$Hz)}\\
\hline \\[-8pt]
51 & 1 & 0 & $    123.9250_{-      0.0003}^{+      0.0003}$ & \multicolumn{1}{c}{\dots} & \multicolumn{1}{c}{\dots} & $     5561.82_{-      309.07}^{+      363.86}$ & \dots \\[1pt]
52 & 1 & ? & $    124.4225_{-      0.0168}^{+      0.0167}$ & $      15.712_{-       0.864}^{+       0.925}$ & $       0.292_{-       0.031}^{+       0.031}$ & \multicolumn{1}{c}{\dots} & 1.000\\[1pt]
53 & 1 & 0 & $    124.9012_{-      0.0028}^{+      0.0030}$ & $      27.827_{-       1.360}^{+       1.149}$ & $       0.046_{-       0.006}^{+       0.006}$ & \multicolumn{1}{c}{\dots} & \dots \\[1pt]
54 & 1 & ? & $    125.2748_{-      0.0031}^{+      0.0030}$ & $      15.352_{-       1.122}^{+       1.173}$ & $       0.029_{-       0.008}^{+       0.007}$ & \multicolumn{1}{c}{\dots} & \dots \\
\hline
\end{tabular}
\tablefoot{The first column represents the peak number in increasing frequency order and is shown for each angular degree ($\ell$) and azimuthal order ($m$), with question marks placed for m-values that could not be identified. The last column corresponds to the detection probability introduced by Eq.~(\ref{eq:detection_probability}) and discussed in Sect.~\ref{sec:test}.}
\end{table*}

\begin{table*}[!]
\caption{Median values with corresponding 68.3\,\% shortest credible intervals for the oscillation frequencies, amplitudes, and linewidths of the $p$ modes of KIC~9475697, as derived by \diamonds\,\,by using the peak bagging model defined by Eqs.~(\ref{eq:general_pb_model}) and (\ref{eq:pb_model}).}
\label{tab:9475697p}
\centering
\begin{tabular}{llcrrlrc}
\hline\hline
\\[-8pt]
Peak \# & $\ell$ & $m$ & \multicolumn{1}{c}{Frequency} & \multicolumn{1}{c}{Amplitude} & \multicolumn{1}{c}{Linewidth} & \multicolumn{1}{c}{Height}& $p_\mathrm{B}$\\
 & & & \multicolumn{1}{c}{($\mu$Hz)} & \multicolumn{1}{c}{(ppm)} & \multicolumn{1}{c}{($\mu$Hz)} & \multicolumn{1}{c}{(ppm$^2/\mu$Hz)}\\
\hline \\[-8pt]
0 & 0 & 0 & $     81.5614_{-      0.0284}^{+      0.0303}$ & $      12.337_{-       1.723}^{+       1.801}$ & $       0.209_{-       0.076}^{+       0.065}$ & \multicolumn{1}{c}{\dots} & 0.999\\[1pt]
1 & 0 & 0 & $     90.9681_{-      0.0060}^{+      0.0055}$ & $      24.110_{-       0.902}^{+       0.963}$ & $       0.104_{-       0.013}^{+       0.012}$ & \multicolumn{1}{c}{\dots} & \dots \\[1pt]
2 & 0 & 0 & $    100.6017_{-      0.0041}^{+      0.0040}$ & $      41.567_{-       1.472}^{+       1.390}$ & $       0.104_{-       0.008}^{+       0.008}$ & \multicolumn{1}{c}{\dots} & \dots \\[1pt]
3 & 0 & 0 & $    110.4707_{-      0.0034}^{+      0.0032}$ & $      51.033_{-       1.493}^{+       1.423}$ & $       0.095_{-       0.008}^{+       0.007}$ & \multicolumn{1}{c}{\dots} & \dots \\
\hline
\end{tabular}
\tablefoot{The first column represents the peak number in increasing frequency order and is shown for each angular degree ($\ell$) and azimuthal order ($m$). The last column corresponds to the detection probability introduced by Eq.~(\ref{eq:detection_probability}) and discussed in Sect.~\ref{sec:test}.}
\end{table*}
\begin{table*}[!]
\caption{Median values with corresponding 68.3\,\% shortest credible intervals for the oscillation frequencies, amplitudes, linewidths, and heights of the mixed modes of KIC~9882316, as derived by \diamonds\,\,by using the peak bagging model defined by Eqs.~(\ref{eq:general_pb_model}) and (\ref{eq:pb_model}).}
\label{tab:9882316m}
\centering
\begin{tabular}{llcrrlrc}
\hline\hline
\\[-8pt]      
Peak \# & $\ell$ & $m$ & \multicolumn{1}{c}{Frequency} & \multicolumn{1}{c}{Amplitude} & \multicolumn{1}{c}{Linewidth} & \multicolumn{1}{c}{Height}& $p_\mathrm{B}$\\
 & & & \multicolumn{1}{c}{($\mu$Hz)} & \multicolumn{1}{c}{(ppm)} & \multicolumn{1}{c}{($\mu$Hz)} & \multicolumn{1}{c}{(ppm$^2/\mu$Hz)}\\
\hline \\[-8pt]
0 & 1 & 0 & $    133.8746_{-      0.0188}^{+      0.0186}$ & $       7.342_{-       0.907}^{+       0.881}$ & $       0.105_{-       0.032}^{+       0.026}$ & \multicolumn{1}{c}{\dots} & 1.000\\[1pt]
1 & 1 & 0 & $    134.7506_{-      0.0086}^{+      0.0084}$ & $       7.253_{-       0.894}^{+       0.817}$ & $       0.049_{-       0.016}^{+       0.014}$ & \multicolumn{1}{c}{\dots} & 1.000\\[1pt]

2 & 1 & 0 & $    145.1127_{-      0.0010}^{+      0.0011}$ & \multicolumn{1}{c}{\dots} & \multicolumn{1}{c}{\dots} & $      653.64_{-       30.75}^{+       29.20}$ & 1.000\\[1pt]
3 & 1 & 0 & $    146.6253_{-      0.0002}^{+      0.0002}$ & \multicolumn{1}{c}{\dots} & \multicolumn{1}{c}{\dots} & $     2759.47_{-      268.22}^{+      236.73}$ & \dots \\
\hline
\end{tabular}
\tablefoot{The first column represents the peak number in increasing frequency order and is shown for each angular degree ($\ell$) and azimuthal order ($m$). The last column corresponds to the detection probability introduced by Eq.~(\ref{eq:detection_probability}) and discussed in Sect.~\ref{sec:test}.}
\end{table*}

\begin{table*}[!]
\caption{Median values with corresponding 68.3\,\% shortest credible intervals for the oscillation frequencies, amplitudes, and linewidths of the $p$ modes of KIC~9882316, as derived by \diamonds\,\,by using the peak bagging model defined by Eqs.~(\ref{eq:general_pb_model}) and (\ref{eq:pb_model}).}
\label{tab:9882316p}
\centering
\begin{tabular}{llcrrlrc}
\hline\hline
\\[-8pt]
Peak \# & $\ell$ & $m$ & \multicolumn{1}{c}{Frequency} & \multicolumn{1}{c}{Amplitude} & \multicolumn{1}{c}{Linewidth} & \multicolumn{1}{c}{Height}& $p_\mathrm{B}$\\
 & & & \multicolumn{1}{c}{($\mu$Hz)} & \multicolumn{1}{c}{(ppm)} & \multicolumn{1}{c}{($\mu$Hz)} & \multicolumn{1}{c}{(ppm$^2/\mu$Hz)}\\
\hline \\[-8pt]
0 & 0 & 0 & $    140.8260_{-      0.0117}^{+      0.0107}$ & $      11.324_{-       1.114}^{+       1.085}$ & $       0.101_{-       0.024}^{+       0.022}$ & \multicolumn{1}{c}{\dots} & 1.000 \\[1pt] 
1 & 0 & 0 & $    154.1351_{-      0.0084}^{+      0.0082}$ & $      18.544_{-       0.630}^{+       0.589}$ & $       0.199_{-       0.018}^{+       0.016}$ & \multicolumn{1}{c}{\dots} & \dots \\[1pt]
2 & 0 & 0 & $    167.7997_{-      0.0099}^{+      0.0093}$ & $      22.456_{-       1.091}^{+       1.084}$ & $       0.147_{-       0.021}^{+       0.018}$ & \multicolumn{1}{c}{\dots} & \dots \\[1pt]
3 & 0 & 0 & $    181.3776_{-      0.0046}^{+      0.0051}$ & $      36.020_{-       1.377}^{+       1.241}$ & $       0.102_{-       0.009}^{+       0.009}$ & \multicolumn{1}{c}{\dots} & \dots \\
\hline
\end{tabular}
\tablefoot{The first column represents the peak number in increasing frequency order and is shown for each angular degree ($\ell$) and azimuthal order ($m$). The last column corresponds to the detection probability introduced by Eq.~(\ref{eq:detection_probability}) and discussed in Sect.~\ref{sec:test}.}
\end{table*}
\begin{table*}[!]
\caption{Median values with corresponding 68.3\,\% shortest credible intervals for the oscillation frequencies, amplitudes, linewidths, and heights of the mixed modes of KIC~10123207, as derived by \diamonds\,\,by using the peak bagging model defined by Eqs.~(\ref{eq:general_pb_model}) and (\ref{eq:pb_model}).}
\label{tab:10123207m}
\centering
\begin{tabular}{llcrrlrc}
\hline\hline
\\[-8pt]      
Peak \# & $\ell$ & $m$ & \multicolumn{1}{c}{Frequency} & \multicolumn{1}{c}{Amplitude} & \multicolumn{1}{c}{Linewidth} & \multicolumn{1}{c}{Height}& $p_\mathrm{B}$\\
 & & & \multicolumn{1}{c}{($\mu$Hz)} & \multicolumn{1}{c}{(ppm)} & \multicolumn{1}{c}{($\mu$Hz)} & \multicolumn{1}{c}{(ppm$^2/\mu$Hz)}\\
\hline \\[-8pt]

0 & 1 & 0 & $    119.6874_{-      0.0011}^{+      0.0011}$ & \multicolumn{1}{c}{\dots} & \multicolumn{1}{c}{\dots} & $     1484.15_{-      127.88}^{+       59.84}$ & 0.998\\[1pt]
1 & 1 & 0 & $    120.5964_{-      0.0085}^{+      0.0087}$ & $       9.227_{-       0.595}^{+       0.585}$ & $       0.100_{-       0.011}^{+       0.010}$ & \multicolumn{1}{c}{\dots} & 1.000\\[1pt]
2 & 1 & 0 & $    121.4157_{-      0.0038}^{+      0.0033}$ & $      11.302_{-       0.661}^{+       0.616}$ & $       0.053_{-       0.008}^{+       0.007}$ & \multicolumn{1}{c}{\dots} & \dots \\[1pt]
3 & 1 & 0 & $    122.4780_{-      0.0012}^{+      0.0011}$ & \multicolumn{1}{c}{\dots} & \multicolumn{1}{c}{\dots} & $     1076.73_{-       46.94}^{+       47.33}$ & 0.999\\
\hline
\end{tabular}
\tablefoot{The first column represents the peak number in increasing frequency order and is shown for each angular degree ($\ell$) and azimuthal order ($m$). The last column corresponds to the detection probability introduced by Eq.~(\ref{eq:detection_probability}) and discussed in Sect.~\ref{sec:test}.}
\end{table*}

\begin{table*}[!]
\caption{Median values with corresponding 68.3\,\% shortest credible intervals for the oscillation frequencies, amplitudes, and linewidths of the $p$ modes of KIC~10123207, as derived by \diamonds\,\,by using the peak bagging model defined by Eqs.~(\ref{eq:general_pb_model}) and (\ref{eq:pb_model}).}
\label{tab:10123207p}
\centering
\begin{tabular}{llcrrlrc}
\hline\hline
\\[-8pt]
Peak \# & $\ell$ & $m$ & \multicolumn{1}{c}{Frequency} & \multicolumn{1}{c}{Amplitude} & \multicolumn{1}{c}{Linewidth} & \multicolumn{1}{c}{Height}& $p_\mathrm{B}$\\
 & & & \multicolumn{1}{c}{($\mu$Hz)} & \multicolumn{1}{c}{(ppm)} & \multicolumn{1}{c}{($\mu$Hz)} & \multicolumn{1}{c}{(ppm$^2/\mu$Hz)}\\
\hline \\[-8pt]
0 & 0 & 0 & $    127.0529_{-      0.0109}^{+      0.0110}$ & $      20.677_{-       0.727}^{+       0.659}$ & $       0.233_{-       0.020}^{+       0.021}$ & \multicolumn{1}{c}{\dots} & \dots \\[1pt]
1 & 0 & 0 & $    140.4660_{-      0.0065}^{+      0.0064}$ & $      31.104_{-       1.294}^{+       1.167}$ & $       0.127_{-       0.013}^{+       0.014}$ & \multicolumn{1}{c}{\dots} & \dots \\[1pt]
2 & 0 & 0 & $    154.1486_{-      0.0040}^{+      0.0040}$ & $      46.584_{-       1.814}^{+       1.640}$ & $       0.108_{-       0.009}^{+       0.008}$ & \multicolumn{1}{c}{\dots} & \dots \\[1pt]
3 & 0 & 0 & $    167.7083_{-      0.0031}^{+      0.0031}$ & $      58.781_{-       1.802}^{+       1.675}$ & $       0.085_{-       0.006}^{+       0.006}$ & \multicolumn{1}{c}{\dots} & \dots \\
\hline
\end{tabular}
\tablefoot{The first column represents the peak number in increasing frequency order and is shown for each angular degree ($\ell$) and azimuthal order ($m$). The last column corresponds to the detection probability introduced by Eq.~(\ref{eq:detection_probability}) and discussed in Sect.~\ref{sec:test}.}
\end{table*}
\begin{table*}[!]
\caption{Median values with corresponding 68.3\,\% shortest credible intervals for the oscillation frequencies, amplitudes, linewidths, and heights of the mixed modes of KIC~10200377, as derived by \diamonds\,\,by using the peak bagging model defined by Eqs.~(\ref{eq:general_pb_model}) and (\ref{eq:pb_model}).}
\label{tab:10200377m}
\centering
\begin{tabular}{llcrrlrc}
\hline\hline
\\[-8pt]      
Peak \# & $\ell$ & $m$ & \multicolumn{1}{c}{Frequency} & \multicolumn{1}{c}{Amplitude} & \multicolumn{1}{c}{Linewidth} & \multicolumn{1}{c}{Height}& $p_\mathrm{B}$\\
 & & & \multicolumn{1}{c}{($\mu$Hz)} & \multicolumn{1}{c}{(ppm)} & \multicolumn{1}{c}{($\mu$Hz)} & \multicolumn{1}{c}{(ppm$^2/\mu$Hz)}\\
\hline \\[-8pt]
0 & 1 & 0 & $     96.1205_{-      0.0014}^{+      0.0017}$ & \multicolumn{1}{c}{\dots} & \multicolumn{1}{c}{\dots} & $      833.21_{-       50.41}^{+       49.46}$ & 0.960\\[1pt]
1 & 1 & 0 & $     98.0810_{-      0.0021}^{+      0.0021}$ & \multicolumn{1}{c}{\dots} & \multicolumn{1}{c}{\dots} & $     1226.71_{-       72.60}^{+       78.40}$ & 0.997\\[1pt]
2 & 1 & 0 & $     98.7103_{-      0.0015}^{+      0.0014}$ & \multicolumn{1}{c}{\dots} & \multicolumn{1}{c}{\dots} & $     1037.12_{-       83.50}^{+       60.32}$ & 0.987\\[1pt]
3 & 1 & 0 & $    100.9617_{-      0.0015}^{+      0.0016}$ & \multicolumn{1}{c}{\dots} & \multicolumn{1}{c}{\dots} & $     1083.03_{-       63.67}^{+       79.39}$ & 0.994\\
\hline
\end{tabular}
\tablefoot{The first column represents the peak number in increasing frequency order and is shown for each angular degree ($\ell$) and azimuthal order ($m$). The last column corresponds to the detection probability introduced by Eq.~(\ref{eq:detection_probability}) and discussed in Sect.~\ref{sec:test}.}
\end{table*}

\begin{table*}[!]
\caption{Median values with corresponding 68.3\,\% shortest credible intervals for the oscillation frequencies, amplitudes, and linewidths of the $p$ modes of KIC~10200377, as derived by \diamonds\,\,by using the peak bagging model defined by Eqs.~(\ref{eq:general_pb_model}) and (\ref{eq:pb_model}).}
\label{tab:10200377p}
\centering
\begin{tabular}{llcrrlrc}
\hline\hline
\\[-8pt]
Peak \# & $\ell$ & $m$ & \multicolumn{1}{c}{Frequency} & \multicolumn{1}{c}{Amplitude} & \multicolumn{1}{c}{Linewidth} & \multicolumn{1}{c}{Height}& $p_\mathrm{B}$\\
 & & & \multicolumn{1}{c}{($\mu$Hz)} & \multicolumn{1}{c}{(ppm)} & \multicolumn{1}{c}{($\mu$Hz)} & \multicolumn{1}{c}{(ppm$^2/\mu$Hz)}\\
\hline \\[-8pt]
0 & 0 & 0 & $     91.4633_{-      0.0062}^{+      0.0064}$ & $       3.680_{-       0.341}^{+       0.393}$ & $       0.031_{-       0.007}^{+       0.007}$ & \multicolumn{1}{c}{\dots} & 0.990\\[1pt]
1 & 0 & 0 & $    103.8863_{-      0.0148}^{+      0.0144}$ & $      11.696_{-       0.627}^{+       0.730}$ & $       0.185_{-       0.028}^{+       0.027}$ & \multicolumn{1}{c}{\dots} & 1.000\\[1pt]
2 & 0 & 0 & $    115.8351_{-      0.0075}^{+      0.0078}$ & $      19.944_{-       0.705}^{+       0.710}$ & $       0.169_{-       0.015}^{+       0.014}$ & \multicolumn{1}{c}{\dots} & \dots \\[1pt]
3 & 0 & 0 & $    128.2301_{-      0.0044}^{+      0.0049}$ & $      37.392_{-       1.129}^{+       1.098}$ & $       0.140_{-       0.012}^{+       0.010}$ & \multicolumn{1}{c}{\dots} & \dots \\
\hline
\end{tabular}
\tablefoot{The first column represents the peak number in increasing frequency order and is shown for each angular degree ($\ell$) and azimuthal order ($m$). The last column corresponds to the detection probability introduced by Eq.~(\ref{eq:detection_probability}) and discussed in Sect.~\ref{sec:test}.}
\end{table*} 
\begin{table*}[!]
\caption{Median values with corresponding 68.3\,\% shortest credible intervals for the oscillation frequencies, amplitudes, linewidths, and heights of the mixed modes of KIC~10257278, as derived by \diamonds\,\,by using the peak bagging model defined by Eqs.~(\ref{eq:general_pb_model}) and (\ref{eq:pb_model}).}
\label{tab:10257278m}
\centering
\begin{tabular}{llcrrlrc}
\hline\hline
\\[-8pt]      
Peak \# & $\ell$ & $m$ & \multicolumn{1}{c}{Frequency} & \multicolumn{1}{c}{Amplitude} & \multicolumn{1}{c}{Linewidth} & \multicolumn{1}{c}{Height}& $p_\mathrm{B}$\\
 & & & \multicolumn{1}{c}{($\mu$Hz)} & \multicolumn{1}{c}{(ppm)} & \multicolumn{1}{c}{($\mu$Hz)} & \multicolumn{1}{c}{(ppm$^2/\mu$Hz)}\\
\hline \\[-8pt]
0 & 1 & ? & $    107.5976_{-      0.0098}^{+      0.0101}$ & $      10.388_{-       0.792}^{+       0.777}$ & $       0.100_{-       0.018}^{+       0.019}$ & \multicolumn{1}{c}{\dots} & 1.000\\[1pt]
1 & 1 & ? & $    108.2197_{-      0.0219}^{+      0.0208}$ & $       8.209_{-       0.767}^{+       0.754}$ & $       0.196_{-       0.042}^{+       0.050}$ & \multicolumn{1}{c}{\dots} & 0.941\\[1pt]

2 & 1 & ? & $    117.1499_{-      0.0017}^{+      0.0013}$ & \multicolumn{1}{c}{\dots} & \multicolumn{1}{c}{\dots} & $     1467.83_{-      218.93}^{+      145.21}$ & 1.000 \\[1pt]
3 & 1 & ? & $    118.1459_{-      0.0179}^{+      0.0186}$ & $       7.661_{-       0.788}^{+       0.912}$ & $       0.122_{-       0.028}^{+       0.026}$ & \multicolumn{1}{c}{\dots} & 0.998\\
\hline
\end{tabular}
\tablefoot{The first column represents the peak number in increasing frequency order and is shown for each angular degree ($\ell$) and azimuthal order ($m$), with question marks placed for m-values that could not be identified. The last column corresponds to the detection probability introduced by Eq.~(\ref{eq:detection_probability}) and discussed in Sect.~\ref{sec:test}.}
\end{table*}

\begin{table*}[!]
\caption{Median values with corresponding 68.3\,\% shortest credible intervals for the oscillation frequencies, amplitudes, and linewidths of the $p$ modes of KIC~10257278, as derived by \diamonds\,\,by using the peak bagging model defined by Eqs.~(\ref{eq:general_pb_model}) and (\ref{eq:pb_model}).}
\label{tab:10257278p}
\centering
\begin{tabular}{llcrrlrc}
\hline\hline
\\[-8pt]
Peak \# & $\ell$ & $m$ & \multicolumn{1}{c}{Frequency} & \multicolumn{1}{c}{Amplitude} & \multicolumn{1}{c}{Linewidth} & \multicolumn{1}{c}{Height}& $p_\mathrm{B}$\\
 & & & \multicolumn{1}{c}{($\mu$Hz)} & \multicolumn{1}{c}{(ppm)} & \multicolumn{1}{c}{($\mu$Hz)} & \multicolumn{1}{c}{(ppm$^2/\mu$Hz)}\\
\hline \\[-8pt]
0 & 0 & 0 & $    113.1994_{-      0.0093}^{+      0.0094}$ & $      15.351_{-       0.726}^{+       0.811}$ & $       0.147_{-       0.019}^{+       0.019}$ & \multicolumn{1}{c}{\dots} & \dots \\[1pt]
1 & 0 & 0 & $    124.8924_{-      0.0073}^{+      0.0076}$ & $      23.606_{-       1.099}^{+       1.055}$ & $       0.125_{-       0.014}^{+       0.014}$ & \multicolumn{1}{c}{\dots} & \dots \\[1pt]
2 & 0 & 0 & $    137.0780_{-      0.0035}^{+      0.0034}$ & $      35.608_{-       0.759}^{+       0.836}$ & $       0.114_{-       0.008}^{+       0.006}$ & \multicolumn{1}{c}{\dots} & \dots \\[1pt]
3 & 0 & 0 & $    149.1918_{-      0.0029}^{+      0.0026}$ & $      60.000_{-       1.764}^{+       1.672}$ & $       0.081_{-       0.005}^{+       0.006}$ & \multicolumn{1}{c}{\dots} & \dots \\
\hline
\end{tabular}
\tablefoot{The first column represents the peak number in increasing frequency order and is shown for each angular degree ($\ell$) and azimuthal order ($m$). The last column corresponds to the detection probability introduced by Eq.~(\ref{eq:detection_probability}) and discussed in Sect.~\ref{sec:test}.}
\end{table*} 
\begin{table*}[!]
\caption{Median values with corresponding 68.3\,\% shortest credible intervals for the oscillation frequencies, amplitudes, linewidths, and heights of the mixed modes of KIC~11353313, as derived by \diamonds\,\,by using the peak bagging model defined by Eqs.~(\ref{eq:general_pb_model}) and (\ref{eq:pb_model}).}
\label{tab:11353313m}
\centering
\begin{tabular}{llcrrlrc}
\hline\hline
\\[-8pt]      
Peak \# & $\ell$ & $m$ & \multicolumn{1}{c}{Frequency} & \multicolumn{1}{c}{Amplitude} & \multicolumn{1}{c}{Linewidth} & \multicolumn{1}{c}{Height}& $p_\mathrm{B}$\\
 & & & \multicolumn{1}{c}{($\mu$Hz)} & \multicolumn{1}{c}{(ppm)} & \multicolumn{1}{c}{($\mu$Hz)} & \multicolumn{1}{c}{(ppm$^2/\mu$Hz)}\\
\hline \\[-8pt]
0 & 1 & ? & $     93.1208_{-      0.0021}^{+      0.0019}$ & \multicolumn{1}{c}{\dots} & \multicolumn{1}{c}{\dots} & $      978.00_{-       71.43}^{+       73.32}$ & 0.983\\[1pt]
1 & 1 & ? & $     94.3103_{-      0.0020}^{+      0.0021}$ & \multicolumn{1}{c}{\dots} & \multicolumn{1}{c}{\dots} & $     1221.95_{-      116.44}^{+       92.39}$ & 0.996\\[1pt]
2 & 1 & ? & $     94.4490_{-      0.0010}^{+      0.0009}$ & \multicolumn{1}{c}{\dots} & \multicolumn{1}{c}{\dots} & $     1361.51_{-      105.24}^{+      106.56}$ & 1.000\\[1pt]
3 & 1 & ? & $     95.1427_{-      0.0018}^{+      0.0017}$ & \multicolumn{1}{c}{\dots} & \multicolumn{1}{c}{\dots} & $     1031.11_{-       75.78}^{+       79.48}$ & 0.995\\
\hline
\end{tabular}
\end{table*}

\begin{table*}[!]
\caption{Table~\ref{tab:11353313m} continued.}
\label{tab:11353313m2}
\centering
\begin{tabular}{llcrrlrc}
\hline\hline
\\[-8pt]          
Peak \# & $\ell$ & $m$ & \multicolumn{1}{c}{Frequency} & \multicolumn{1}{c}{Amplitude} & \multicolumn{1}{c}{Linewidth} & \multicolumn{1}{c}{Height}& $p_\mathrm{B}$\\
 & & & \multicolumn{1}{c}{($\mu$Hz)} & \multicolumn{1}{c}{(ppm)} & \multicolumn{1}{c}{($\mu$Hz)} & \multicolumn{1}{c}{(ppm$^2/\mu$Hz)}\\
\hline \\[-8pt]
51 & 1 & $-1$  & $    133.8120_{-      0.0007}^{+      0.0006}$ & \multicolumn{1}{c}{\dots} & \multicolumn{1}{c}{\dots} & $      644.72_{-       31.78}^{+       30.32}$ & 1.000 \\[1pt]
52 & 1 & 0 & $    134.2241_{-      0.0010}^{+      0.0011}$ & \multicolumn{1}{c}{\dots} & \multicolumn{1}{c}{\dots} & $      553.95_{-       24.54}^{+       25.78}$ & 0.997 \\[1pt]
53 & 1 & $+1$ & $    134.6929_{-      0.0003}^{+      0.0003}$ & \multicolumn{1}{c}{\dots} & \multicolumn{1}{c}{\dots} & $      976.61_{-       43.64}^{+       44.82}$ & \dots \\[1pt]
54 & 1 & $+1$ & $    135.9288_{-      0.0001}^{+      0.0001}$ & \multicolumn{1}{c}{\dots} & \multicolumn{1}{c}{\dots} & $    14185.56_{-     1435.50}^{+     1417.70}$ & \dots \\
\hline
\end{tabular}
\tablefoot{The first column represents the peak number in increasing frequency order and is shown for each angular degree ($\ell$) and azimuthal order ($m$), with question marks placed for m-values that could not be identified. The last column corresponds to the detection probability introduced by Eq.~(\ref{eq:detection_probability}) and discussed in Sect.~\ref{sec:test}.}
\end{table*}

\begin{table*}[!]
\caption{Median values with corresponding 68.3\,\% shortest credible intervals for the oscillation frequencies, amplitudes, and linewidths of the $p$ modes of KIC~11353313, as derived by \diamonds\,\,by using the peak bagging model defined by Eqs.~(\ref{eq:general_pb_model}) and (\ref{eq:pb_model}).}
\label{tab:11353313p}
\centering
\begin{tabular}{llcrrlrc}
\hline\hline
\\[-8pt]
Peak \# & $\ell$ & $m$ & \multicolumn{1}{c}{Frequency} & \multicolumn{1}{c}{Amplitude} & \multicolumn{1}{c}{Linewidth} & \multicolumn{1}{c}{Height}& $p_\mathrm{B}$\\
 & & & \multicolumn{1}{c}{($\mu$Hz)} & \multicolumn{1}{c}{(ppm)} & \multicolumn{1}{c}{($\mu$Hz)} & \multicolumn{1}{c}{(ppm$^2/\mu$Hz)}\\
\hline \\[-8pt]
0 & 0 & 0 & $     88.9532_{-      0.0185}^{+      0.0193}$ & $       7.897_{-       1.746}^{+       1.717}$ & $       0.081_{-       0.032}^{+       0.028}$ & \multicolumn{1}{c}{\dots} & 0.981 \\[1pt]
1 & 0 & 0 & $     99.2568_{-      0.0148}^{+      0.0152}$ & $      21.045_{-       1.052}^{+       0.978}$ & $       0.237_{-       0.037}^{+       0.032}$ & \multicolumn{1}{c}{\dots} & \dots \\[1pt]
2 & 0 & 0 & $    109.7428_{-      0.0049}^{+      0.0047}$ & $      31.806_{-       1.252}^{+       1.201}$ & $       0.092_{-       0.011}^{+       0.010}$ & \multicolumn{1}{c}{\dots} & \dots \\[1pt]
3 & 0 & 0 & $    120.4942_{-      0.0030}^{+      0.0031}$ & $      48.669_{-       0.975}^{+       0.928}$ & $       0.121_{-       0.006}^{+       0.006}$ & \multicolumn{1}{c}{\dots} & \dots \\
\hline
\end{tabular}
\tablefoot{The first column represents the peak number in increasing frequency order and is shown for each angular degree ($\ell$) and azimuthal order ($m$). The last column corresponds to the detection probability introduced by Eq.~(\ref{eq:detection_probability}) and discussed in Sect.~\ref{sec:test}.}
\end{table*}
\begin{table*}[!]
\caption{Median values with corresponding 68.3\,\% shortest credible intervals for the oscillation frequencies, amplitudes, linewidths, and heights of the mixed modes of KIC~11913545, as derived by \diamonds\,\,by using the peak bagging model defined by Eqs.~(\ref{eq:general_pb_model}) and (\ref{eq:pb_model}).}
\label{tab:11913545m}
\centering

\end{table*}

\begin{table*}[!]
\caption{Table~\ref{tab:11913545m} continued.}
\label{tab:11913545m2}
\centering
%
\tablefoot{The first column represents the peak number in increasing frequency order and is shown for each angular degree ($\ell$) and azimuthal order ($m$), with question marks placed for m-values that could not be identified. The last column corresponds to the detection probability introduced by Eq.~(\ref{eq:detection_probability}) and discussed in Sect.~\ref{sec:test}.}
\end{table*}

\begin{table*}[!]
\caption{Median values with corresponding 68.3\,\% shortest credible intervals for the oscillation frequencies, amplitudes, and linewidths of the $p$ modes of KIC~11913545, as derived by \diamonds\,\,by using the peak bagging model defined by Eqs.~(\ref{eq:general_pb_model}) and (\ref{eq:pb_model}).}
\label{tab:11913545p}
\centering

\tablefoot{The first column represents the peak number in increasing frequency order and is shown for each angular degree ($\ell$) and azimuthal order ($m$). The last column corresponds to the detection probability introduced by Eq.~(\ref{eq:detection_probability}) and discussed in Sect.~\ref{sec:test}.}
\end{table*}
\begin{table*}[!]
\caption{Median values with corresponding 68.3\,\% shortest credible intervals for the oscillation frequencies, amplitudes, linewidths, and heights of the mixed modes of KIC~11968334, as derived by \diamonds\,\,by using the peak bagging model defined by Eqs.~(\ref{eq:general_pb_model}) and (\ref{eq:pb_model}).}
\label{tab:11968334m}
\centering

\end{table*}

\begin{table*}[!]
\caption{Table~\ref{tab:11968334m} continued.}
\label{tab:11968334m2}
\centering
%
\tablefoot{The first column represents the peak number in increasing frequency order and is shown for each angular degree ($\ell$) and azimuthal order ($m$), with question marks placed for m-values that could not be identified. The last column corresponds to the detection probability introduced by Eq.~(\ref{eq:detection_probability}) and discussed in Sect.~\ref{sec:test}.}
\end{table*}

\begin{table*}[!]
\caption{Median values with corresponding 68.3\,\% shortest credible intervals for the oscillation frequencies, amplitudes, and linewidths of the $p$ modes of KIC~11968334, as derived by \diamonds\,\,by using the peak bagging model defined by Eqs.~(\ref{eq:general_pb_model}) and (\ref{eq:pb_model}).}
\label{tab:11968334p}
\centering

\tablefoot{The first column represents the peak number in increasing frequency order and is shown for each angular degree ($\ell$) and azimuthal order ($m$). The last column corresponds to the detection probability introduced by Eq.~(\ref{eq:detection_probability}) and discussed in Sect.~\ref{sec:test}.}
\end{table*}
\begin{table*}[!]
\caption{Median values with corresponding 68.3\,\% shortest credible intervals for the oscillation frequencies, amplitudes, linewidths, and heights of the mixed modes of KIC~12008916, as derived by \diamonds\,\,by using the peak bagging model defined by Eqs.~(\ref{eq:general_pb_model}) and (\ref{eq:pb_model}).}
\label{tab:12008916m}
\centering

\tablefoot{The first column represents the peak number in increasing frequency order and is shown for each angular degree ($\ell$) and azimuthal order ($m$). The last column corresponds to the detection probability introduced by Eq.~(\ref{eq:detection_probability}) and discussed in Sect.~\ref{sec:test}.}
\end{table*}

\begin{table*}[!]
\caption{Median values with corresponding 68.3\,\% shortest credible intervals for the oscillation frequencies, amplitudes, and linewidths of the $p$ modes of KIC~12008916, as derived by \diamonds\,\,by using the peak bagging model defined by Eqs.~(\ref{eq:general_pb_model}) and (\ref{eq:pb_model}).}
\label{tab:12008916p}
\centering

\tablefoot{The first column represents the peak number in increasing frequency order and is shown for each angular degree ($\ell$) and azimuthal order ($m$). The last column corresponds to the detection probability introduced by Eq.~(\ref{eq:detection_probability}) and discussed in Sect.~\ref{sec:test}.}
\end{table*}

\end{document}